\documentclass[twoside,12pt]{article}
\usepackage{epsfig}
\usepackage{amssymb}
\usepackage{mathtools}

\def\Journal#1#2#3#4{{#1} {#2} (#4) #3 }

\def\NPA{{\em Nucl. Phys.} A}

\def\PRO{{\em Prog. Theor. Phys.}}
\def\NPB{{\em Nucl. Phys.} B}

\def\PLB{{\em Phys. Lett.} B}

\def\PL{{\em Phys. Lett.}}
\def\PRL{\em Phys. Rev. Lett.}

\def\PREP{\em Phys. Rep.}

\def\PRD{{\em Phys. Rev.} D}
\def\PRC{{\em Phys. Rev.} C}

\def\ZPC{{\em Z. Phys.} C}
\def\ZPA{{\em Z. Phys.} A}

\def\RMP{{\em Rev. Mod. Phys.}}

\def\etal{{\em et al.}}

\newcommand{\be}{\begin{equation}}
\newcommand{\ee}{\end{equation}}
\newcommand{\bea}{\begin{eqnarray}}
\newcommand{\eea}{\end{eqnarray}}

\begin{document}

\title{ \vspace{1cm}  Hadron Collider Searches for Diboson Resonances}
\author{Tommaso Dorigo$^1$ \\
$^1$ INFN, Sezione di Padova, Italy}
\maketitle

\begin{abstract} This review covers results of searches for new elementary particles that decay into boson pairs (dibosons), performed at the CERN Large Hadron Collider in proton-proton collision data collected by the ATLAS and CMS experiments at 7-, 8-, and 13-TeV center-of-mass energy until the year 2017. The available experimental results of the analysis of final states including most of the possible two-object combinations of $W$ and $Z$ bosons, photons, Higgs bosons, and gluons place stringent constraints on  a variety of theoretical ideas that extend the standard model, pushing into the multi-TeV region the scale of allowed new physics phenomena.\par
\vspace{1cm}

\noindent \textcopyright  2018. This manuscript version is made available under the CC-BY-NC-ND 4.0 license \par 
http://creativecommons.org/licenses/by-nc-nd/4.0/
\end{abstract}



\section{Introduction}
\label{s:intro}

Our current understanding of subnuclear physics is enshrined in the standard model (SM) \label{s:sm}, a theoretical construction based on the $SU(3)_C \times SU(2)_L \times U(1)_Y$ gauge group. The SM describes in detail electromagnetic, weak, and strong interactions of elementary fermions, and allows for the self-interaction of some of the interaction carriers, according to the structure of the relative gauge groups. The model, first constructed in the sixties of the past century~\cite{higgs1,higgs2,higgs3,higgs4,higgs5,higgs6}, includes a scalar field that enables the spontaneous breaking of electroweak symmetry, thus allowing for non-zero mass terms for a triplet of weak bosons in the Lagrangian density~\cite{higgs7,higgs8,higgs9}. After the proof of its renormalizability~\cite{thooft} and the discovery of weak neutral currents~\cite{weakcurrents}, the model was accepted as the standard theory of electroweak interactions. Following a string of increasingly accurate verifications of theory predictions in the description of the physics of subnuclear collisions
, which confirmed its validity, the SM was finally crowned by the 2012 discovery of a heavy scalar particle by the ATLAS~\cite{higgsatlas} and CMS~\cite{higgscms} experiments at the CERN Large Hadron Collider (LHC) \label{s:lhc}. That particle, now commonly called the Higgs boson~\footnote{ We adopt in this review the common term ``Higgs boson'' in place of the correct one for the particle, BEH boson (for Brout-Englert-Higgs, an acronym which honors the three theorist who gave crucial contributions to its prediction).}, has since then been the subject of a broad range of experimental studies which have so far confirmed all SM predictions for its intrinsic properties (see {\em e.g.}~\cite{higgspropatlas1} and references therein).

Despite its huge successes in explaining the phenomenology of subnuclear physics from low-energy interactions all the way to the TeV scale, the SM is today considered at most an effective theory, one which should eventually break down when higher-energy reactions are studied, to become a low-energy approximation of a larger and more comprehensive theory~\cite{davidpassarino}. There are multiple reasons for this belief. First of all, the SM manifestly fails to include a description of gravity. Secondly, it does not yield a description of the properties of non-massless neutrinos, in particular an explanation of the smallness of their masses. Yet the most compelling argument demanding some more fundamental theory to replace the SM is the fact that the size of quantum corrections to the Higgs boson mass arising from the contribution of virtual loops involving SM particles exceed the observed value by many orders of magnitude in the absence of new physics below the Planck scale: the physical mass of the Higgs boson, which results from the addition and subtraction of those large contributions, appears exceptionally fine-tuned, ending up to be unnaturally light and close to the electroweak scale~\cite{smshortcomings1,smshortcomings2,smshortcomings4}.
To the above list of clear shortcomings one could add the absence in the SM of other desirable features of a final theory -- an explanation, e.g., of the hierarchy of fermion masses, of the weakness of gravity with respect to the other forces, of the observed abundance of dark matter, and of the observed matter-antimatter asymmetry in the universe; or the lack of a common unification scale of running couplings at high energy.

A number of theoretical ideas have been put forth in the past few decades to extend the SM in ways that either offer a direct solution to some of the above issues, or attempt to enlarge the model thus providing it with the elasticity needed to accommodate {\em ad hoc} solution to others. Notable examples of the former kind are supersymmetry (SUSY) \label{s:susy} models~\cite{susy1,susy2,susy3,susy4,susy5,susy6,susy7,susy8,susy9,SUSY} which radically solve the fine-tuning of the Higgs mass by introducing an automatic cancellation of its quantum corrections through a symmetry of bosons and fermions; and large extra dimension theories~\cite{RandallSundrum1,ADD1,ADD2} which directly explain the weakness of gravity by allowing for its dispersion outside of the (3+1)-dimensional subspace where the SM lives. Examples of the latter include technicolour~\cite{weinbergtechnicolor} and other ideas like theories with a composite Higgs~\cite{strongsector1,strongsector2,strongsector3,strongsector4,strongsector5,strongsector6,strongsector7}, and grand-unification theories based on larger gauge groups that include the SM one~\cite{georgiglashow,langacker1984,robinettrosner1,robinettrosner2}.

A common trait of new physics scenarios that extend the SM is their inclusion of new gauge bosons or resonances that couple to the SM fields, and may thus decay into pairs of the respective fermions and bosons. In many cases the mass scale at which these particles should make their presence felt cannot much exceed the one accessible by the LHC experiments, lest the involved models become less compelling as an explanation of the observed shortcomings of the SM at the electroweak scale. This is the case, {\em e.g.}, of  new $U(1)'$ groups occurring in models of dynamical symmetry breaking, Little Higgs (LH) \label{s:lh} models, or the above mentioned models with large extra dimensions at scales of the order of ${\cal{O}}(1/$ TeV).  This has led the LHC experiments to carry out an extensive campaign of searches for localized excesses of observed event rates over SM predictions in all the accessible final states of heavy resonance decays. Among those searches, the ones focusing on pairs of gauge bosons have recently polarized the attention of experimentalists, thanks to two independent encouragements of different origin: first and foremost, the successful observation of three different diboson final states ($WW$, $ZZ$, $\gamma \gamma$) of the SM Higgs scalar, which gave additional motivation to the development of new experimental techniques to improve signal detection and to tame the relevant backgrounds; and second, the recent breakthroughs in the identification of highly-boosted hadronic decays of $W$, $Z$, and Higgs bosons within individual ``fat jets''. Furthermore, the inclusion of the Higgs as a confirmed elementary boson has brought new entries in the list of final states considered in the class of diboson searches: ones including a vector boson and a Higgs boson, and the one consisting in Higgs boson pairs. These have widened the experimental reach in the parameter space of some of the new physics models, especially ones extending the Higgs sector. The null results obtained by all LHC searches performed so far have been turned into severe bounds on the phase space allowed to models predicting the existence of resonances with calculated cross sections above the observed limits, which in some cases extend down into the sub-femtobarn range. 

The present review covers searches for heavy resonances that extend the SM, performed by the ATLAS and CMS experiments in diboson final states. The bosons we consider are elementary, and we somewhat limit the range of the related theoretical models by only focusing on SM bosons: the four electroweak vectors, the 125 GeV Higgs scalar, and the gluon. We however make an exception or two by including in our review the search for $H \to aa$ decays and $H \to ZA$ decays arising in extended Higgs sector theories, because of their connection to related signatures of interest here. The specified criterion excludes as a studied resonance signal the 125 GeV scalar discovered in 2012, as we consider it to genuinely belong to the SM. Although it may feel strange to not include in this report the one and only resonant signal we have so far been able to discover in diboson final states, a careful treatment of that topic would require too much space; there are a number of excellent reviews that cover the topic in detail (see {\em e.g.}~\cite{higgsrev1,higgsrev2}), to which we refer the interested reader. However, we do discuss results of ATLAS and CMS searches for Higgs-like resonances in their mass spectra of $WW$, $ZZ$, and $\gamma \gamma$ pairs, as those searches are sensitive to additional Higgs bosons arising in Two-Higgs-Doublet Models (2HDM) \label{s:2hdm} and other SM extensions: besides the intrinsic interest of their target, they have directed a large experimental effort into the optimization of the necessary techniques, as discussed in detail in Sec.~\ref{s:exptechniques}. Those techniques have found application in a host of other studies.  

Given the steep rise of the total integrated luminosity that is being made available for analysis by the LHC, we make no attempt at including in this review unpublished preliminary results appearing hot off the press at the time of finalizing this report. The rigorous internal review process that is routinely performed by the ATLAS and CMS collaborations before publishing conference notes, physics analysis summaries, and preprints  guarantees that those results are fully trustable, and they usually end up in refereed journal publications with little or no substantive modifications; however, we see no reason for attempting to capture the latest ephemeral detail in a quickly evolving picture. Rather, the emphasis of this review will be on the techniques more than on the actual search results, especially given that the latter are only constituted by exclusion limits, and there are no exciting hints of possible new discoveries awaiting to be confirmed. Exceptions are made when discussing analysis methods or techniques used in the published results for which only preliminary reports are available.

The datasets on which the reported searches are based come from the data-taking phases of LHC operation called Run 1 and Run 2. Run 1 took place in the years 2010-2012, when the experiments collected up to 5 inverse femtobarns of integrated luminosity from 7-TeV proton-proton collisions (in 2010-2011) and then over 20 inverse femtobarns of integrated luminosity from 8-TeV collisions (in 2012). Following a two-year shutdown to allow for important upgrades to the accelerating complex and to the detectors, Run 2 started in 2015 and is currently still ongoing. The center-of-mass energy of proton-proton collisions has been increased to 13 TeV, providing large increases in the experimental sensitivity to massive new particles. While a full picture of the reach for new resonances from the full Run 2 statistics will only be drawn from the year 2019 onwards, when a new long shutdown of the LHC is planned, a few experimental results based on up to 40 inverse femtobarns of integrated luminosity collected until recently at 13 TeV are already available in refereed journals at the time of writing this review. Those results and the relative information on the remaining range of validity of a broad range of new physics models will be carefully discussed.

This document is organized as follows. In Section~\ref{s:theory} we offer a brief overview of the theoretical models that may be successfully probed by searches for diboson resonances. Section 3 deals with a description of the experimental facilities. In Section 4 there follows a discussion of the state-of-the-art of the reconstruction procedures employed to extract the most possible information from the detectors on the identity and kinematics of individual particles (electrons, muons, $\tau$ leptons, photons) or collective effects (hadronic jets, $b$-tags, missing transverse energy) that allow the identification and measurement of SM bosons and their decay products. Section 5 offers an overview of the experimental techniques and the statistical methods employed in the extraction of results. After a brief summary of diboson resonance searches performed at past accelerator facilities, Section 6 presents the LHC results, divided in subsections that deal with four non-overlapping categories of final state signatures: ones involving only $W$ and $Z$ bosons, ones including at least one photon, ones including at least one Higgs boson, and ones involving gluon pairs. Section~\ref{s:outlook} finally offers a discussion of the general picture resulting from the described studies, as well as a brief overview of the future prospects of diboson searches. In addition, Appendix A offers a decoding table for all the acronyms used in the text.


\section {Theoretical  Models}
\label{s:theory}

\noindent
Full-fledged theoretical models that predict the phenomenology of new physics extending the SM without making a number of assumptions, or relying on a large number of tunable parameters, do not currently exist. Hence the experimental investigation of SM extensions faces the exploration of a vast and multi-dimensional space. This usually proceeds via the definition of suitable benchmark points that may capture the different identity and kinematic signatures of the possible final states of new theoretical models. An attractive alternative to this {\em modus operandi} is to consider simplified models which make generic predictions on specific observable processes. Heavy resonances are particularly handy within such a program, as their general features can be used to probe new physics scenarios with minimal assumptions. 

In this section we offer an overview of the classes of new physics models that predict the existence of heavy resonances, when they allow for a significant branching ratio of decays into SM boson pairs. We mention simplified models whenever they are used by the experiments in their investigations.

\subsection{\it $W'$ and $Z'$ Bosons from New Gauge Groups}

\noindent
Many new physics scenarios imply the existence of symmetry groups extending the SM structure. These result in the addition of new gauge bosons, conventionally labeled as $W'$ and $Z'$. Some of those theories are discussed in separate sections below; here we consider their general features. The simplest extensions involve the appearance of an extra $U(1)'$ group in the gauge structure of the model, which is spontaneously broken. While not solving directly any of the shortcomings of the SM, this is one of its best motivated extensions, as extra $U(1)'$ symmetries arise very commonly from the breaking of larger non-Abelian groups~\cite{langacker}.  In supersymmetric versions of grand-unification, for example, the breaking scale of an extra $U(1)'$ arises in connection to the soft supersymmetry breaking scale~\cite{cveticlangacker1,cveticlangacker2}. 

The relevant part of the group structure involves the decomposition\par

\begin{equation}
SU(3)_C \times SU(2)_L \times U(1)_Y \times U(1)' .
\end{equation}

\noindent
A massive spin-1 $Z'$ arises from the breaking at the TeV scale of the $U(1)'$ group. The new boson has couplings to SM fermions given by the coefficients $g_V^{f}$ and $g_A^f$ of the Lagrangian interaction term\par

\begin{equation}
\frac{g'}{2}Z'_\mu \bar{f} \gamma^{\mu} (g_V^f-g_A^f \gamma^5)f,
\end{equation}

\noindent
where $f$ is the fermion field, and $g'$ is the $U(1)_Y$ coupling constant. Flavour-universal couplings of the $Z'$ are often assumed in simplified scenarios, to avoid the problem of possible large Flavour-Changing Neutral Currents (FCNC) \label{s:fcnc}. In such a framework one may consider the following three classes of models:\par

\begin{itemize}
\item Models based on the group E6 at the Grand-Unification (GUT) \label{s:gut} scale~\cite{hewettrizzo1989,langacker1984,robinettrosner1}: the E6 group is broken into $SO(10) \times U(1)_\psi$, and SO(10) is further broken into $SO(5) \times U(1)_\chi$; $SO(5)$ then generates the SM gauge groups. All breakings take place at the GUT scale, while a linear combination of the two $U(1)$ groups may survive to the TeV scale. The resulting $Z'$ model depends on the mixing of the two $U(1)$ generators.
\item Generalized left-right symmetric models (GLR)\label{s:glr}, motivated by the symmetry breaking of the group $SU(2)_L \times SU(2)_R \times U(1)_{B-L}$ into $SU(2)_L \times U(1)_Y$~\cite{mohapatrareview}. 
\item Generalized sequential models (GSM)\label{s:gsm},  which extend the sequential standard model (SSM) \label{s:ssm}  idea of a new $Z'$ boson with couplings equal to those of the $Z$~\cite{ssm1,langacker}. 
\end{itemize}


\noindent
The three classes of models above generally depend on a continuous angle, and thus define an infinite set of possible theories. A number of different benchmarks have been examined in detail, and the resulting literature is quite extended.
One model considered in a number of LHC searches, in particular by the ATLAS collaboration, is the ``extended gauge model'' (EGM)~\cite{altarellimele}, \label{s:egm} where the coupling of a heavy $W'$ boson to SM $W$ and $Z$ bosons is equal to the $WWZ$ coupling strength in the SM. This can be generalized by scaling the new particle coupling by a factor $c_{EGM} \times (m_W/m_{W'})^2$, which has the consequence that the $W'$ will exhibit a partial width scaling linearly with $m_{W'}$, in contrast with the SSM where the width has a steep $m_{W'}^5$ dependence. For both SSM and EGM, the produced $W$ and $Z$ bosons emitted in the $W'$ decay are expected to be longitudinally polarized.
Direct experimental searches at the Tevatron, LEP, and LEP-II, as well as the more recent LHC ones which are the real focus of this review, have specifically targeted some of those scenarios, placing limits to the allowed regions of their parameter space.
Indirect constraints on the models arise instead essentially from the implication on mass and couplings of the extra resonance of LEP precision measurements at the $Z$ pole and LEP-II off-pole measurements~\cite{continobounds}.

Similarly to the $Z'$ models mentioned above, any model that includes a massive $W'$ boson associates it with a spontaneously broken gauge symmetry. This is true even when the $W'$ is composite (provided that has a mass much smaller than the compositeness scale), or when it propagates in the bulk in extra-dimension theories~\cite{pdgrevW'}. The simplest versions of gauge extensions including $W'$ bosons contain the structure $SU(2)_1 \times SU(2)_2 \times U(1)$; they imply the existence of an accompanying neutral $Z'$. Perhaps the most important feature of the new resonance, besides its mass and width, is the helicity of its couplings to SM fermions, as this is either purely left- or right-handed in most scenarios, with small mixing effects~\cite{rizzo2007}. The classical examples are the left-right symmetric model first discussed in~\cite{ref4inpdgw'rev,ref4binpdgw'rev}, and the alternate left-right model~\cite{ref5inpdgw'rev}, where the  new $W'$ boson only couples to SM fermion--new fermion pairs. In general, $W'$ bosons are a common product of a wide class of new physics models, including the Randall-Sundrum (RS) \label{s:rs} model with bulk gauge fields~\cite{RandallSundrum2,ref3inrizzo2007a,ref3inrizzo2007b,ref3inrizzo2007c,ref3inrizzo2007d,ref3inrizzo2007e,ref3inrizzo2007f,ref3inrizzo2007g,ref3inrizzo2007h,ref3inrizzo2007i}, theories with universal extra dimensions~\cite{ref4inrizzo2007a,ref4inrizzo2007b}, and LH models~\cite{ref1inrizzo2007a,ref1inrizzo2007b}. 

A useful simplified model for the interpretation of experimental searches for heavy new gauge bosons $W'$ and $Z'$, which summarizes many different versions of similar constructions, is the Heavy Vector Triplet (HVT) \label{s:hvt} model proposed in~\cite{HVT}. The HVT model can be described with a small number of parameters which determine the main observable features of heavy vector boson production. The $g_V$ parameter governs the strength of the new vector boson interaction; the coupling of the new bosons to SM fermions is determined by the coefficient $c_F$, and their coupling to the Higgs boson and to longitudinally-polarized $W$ and $Z$ bosons is determined by the coefficient $c_H$. The phenomenology ends up being almost completely described in terms of the combination $g^2 c_F/g_V$ and $g_V c_H$ (where $g$ is the SU(2) coupling), along with the mass scale $M_V$ of the practically degenerate heavy vector triplet. 


An often considered benchmark for the HVT model is the one called ``model A'', which describes the vector triplet produced when the $SU(2)_1 \times SU(2)_2 \times U(1)_Y$ group breaks into the SM electroweak gauge group through a linear $\sigma$ model~\cite{ref17inHVT}.  In model A the coupling strength $g_V$ is set to 1.0, and the branching fractions of the new resonances to SM fermions and bosons are comparable. The model describes heavy vectors emerging from an underlying weakly-coupled extensions of the SM gauge group~\cite{ref17inHVT}, such as in the sequential standard model~\cite{altarellimele}. 
A second model (``model B'') describes the triplet resulting from a non-linearly-realized SO(5)/SO(4) global symmetry, where $g_V$ is set to 3.0. In this case the new bosons have dominant branching fractions into SM bosons; however, they are still produced mainly by the Drell-Yan mechanism, as the vector-boson fusion (VBF)\label{s:vbf} production process remains sub-dominant for reasonable values of the parameters. Also, large values of $g_V$ lead to broad resonances, with a width-to-mass ratio exceeding 0.1, which cannot be easily constrained by experimental searches for localized bumps in mass distributions. Model B corresponds to a strongly-coupled composite Higgs scenario~\cite{strongsector4}. The ATLAS and CMS collaborations have started to use the HVT model benchmarks, especially model B, in reporting their search results for resonances sought in $WW$ and $WZ$ final states relevant to this review. Those results are reported in Sec.~\ref{s:VV}.

\subsection {\it Large Extra Dimension Models and Gravitons}

\noindent
The first exploration of the idea that we live in a subspace of a larger-dimensional universe dates back to a 1921 article by Kaluza~\cite{kaluzaklein1} and by its quantum interpretation by Klein~\cite{kaluzaklein2,kaluzaklein3}. That idea was resuscitated 20 years ago to provide a possible explanation of the so-called hierarchy problem of the SM. The hierarchy of scales -- the large disparity between the electroweak scale, ${\cal{O}}$(TeV), and the Planck scale, $M_{Pl} \sim {\cal{O}}(10^{16}$TeV)-- is the origin of the unnatural fine-tuning of the Higgs boson mass, and is arguably the most compelling argument calling for a SM extension. The model put forth in 1999 by Randall and Sundrum~\cite{RandallSundrum1} attempts to generate a large difference between the electroweak and the Planck scale by introducing a fifth space-time dimension between two subspaces (``branes'', a shorthand for membranes), separated by a 5-dimensional ``bulk''. SM particles are confined in the TeV brane, but gravity is produced at the Planck scale and can extend into the bulk. This automatically explains the weakness of gravity in the TeV brane, as the wavefunction of its interaction carrier has a small overlap with the region where matter fields live. The Planck brane generates the scale of physical phenomena on the TeV brane by a warp factor \par

\begin{equation}
\Lambda_{Pl} = {\overline{M}}_{Pl} e^{-k \pi R_c} .
\end{equation}

\noindent
Above, ${\overline{M}}_{Pl}=M_{Pl}/\sqrt{8 \pi}$, $R_c$ is the compactification radius, and $k$ is the curvature scale of the extra dimension. For $k R_c \sim 12$ one obtains the wanted hierarchy of the two mass scales. A small compactification radius produces a KK tower of spin-2 gravitons, with universal couplings to SM fields and masses \par

\begin{equation}
M_n = x_n k / {\overline{M}}_{pl} \Lambda_\pi 
\end{equation}

\noindent
where $x_n$ is the n-th root of the first-order Bessel function, and the parameter $\Lambda_\pi$ describes the inverse coupling of the excitations. In order to explore the phenomenology of the RS model, it is advantageous to consider two direct observables which are a function of the $R_c$ and $k$ parameters: the mass of the lightest Kaluza-Klein (KK) \label{s:kk} excitation $M_1$, and the dimensionless coupling parameter $\tilde{k} = k / {\overline{M}}_{Pl}$. The fine-tuning of the Higgs mass is then eliminated if $0.01 < \tilde{k} <0.1$ and $M_1$ is at the TeV scale; larger values of $\tilde{k}$ would make the theory non-perturbative~\cite{nohierarchy}. 

RS graviton states are predicted to have s-wave decays to photon pairs and p-wave decays to lepton pairs; the former have a favourable branching ratio~\cite{gravitondecays}. They can also have a significant decay rate into gluon pairs~\cite{RSgg},
in which case they can be searched for in high-mass jet-pair (dijet) events. 
In a version generically called ``bulk RS graviton'' model, the constraints coming from the flavour structure of the SM are addressed by a localization of fermion fields in the warped extra dimension. 
In the bulk graviton model the decay $G^*_{bulk} \to \gamma \gamma$ and the one to light fermions are suppressed by a factor proportional to the volume of the extra dimension~\cite{volumeeffect}. As a consequence of the reduced fermion coupling, production through gluon-fusion diagrams are favoured. This leads to a difference in the polarization of weak bosons produced in $G^*$ and $G^*_{bulk}$: a transverse polarization is favoured in the former case, a longitudinal polarization in the latter. 
Bulk graviton models have been increasingly considered in the searches using weak boson final states ($WW$, $ZZ$), once the decays to photon or lepton pairs ruled out their accessible parameter space. 

A variant of the RS model which also predicts the existence of a spin-2 graviton particle observable in diboson decays is the one of Arkani-Hamed, Dimopoulos, and Dvali \label{s:add} (ADD)~\cite{ADD1,ADD2}. This model predicts the existence of N flat extra-dimensions, which are compactified on a torus of radius $R$ much larger than the inverse of the fundamental Planck scale in (4+N)-dimensions, $1/m_D$, where $m_D$ is given by \par

\begin{equation}
{\overline{M^2}}_{Pl} = M_D^{N+2} R^N .
\end{equation}

\noindent
This implies a mass splitting between the KK excitations $1/R$ for each extra dimension. If R is large enough, as
required to explain the hierarchy of scales, the spectrum of graviton states predicted by the theory is almost continuous.
The experimental signature is thus not one of independently reconstructable narrow resonances, but rather an enhancement 
in the cross section at high invariant mass. The spectrum diverges unless an ultraviolet cut-off scale $M_S$ is imposed. We include tests of this model in this review as technically the KK excitations are still resonances that decay to SM boson pairs, although they do not produce a narrow peak in the mass distribution.

\subsection {\it Models with an Extended Higgs Sector}


\noindent
The particle discovered in 2012 by ATLAS and CMS~\cite{higgsatlas,higgscms}, and further characterized through measurements of mass~\cite{higgsmass}, spin~\cite{higgsspin}, production modes~\cite{higgsproductionmodes}, and couplings to fermions and bosons~\cite{higgscouplings} exploiting Run 1 and Run 2 data, is by now universally recognized as {\em the} Higgs boson. Indeed, the verification of all SM predictions for its properties has left little doubt on its identity. Irrespective of the above, the investigation of the Higgs sector is by no means finished business, as a whole class of models that extend the SM by including more complex scalar doublets predict more physical states, and allow one of them to be identical to the particle predicted by the minimal SM. So, {\em e.g.}, this is the case in the so-called ``decoupling limit''~\cite{decouplingsusy,decoupling2hdm}. More in general, the existence of a richer spectrum of Higgs-like particles is an open experimental and theoretical issue. The obvious, and minimal, extension of the SM Higgs sector is constituted by so-called two-Higgs doublet models~\cite{2HDM}, that introduce a second complex doublet of scalars in the Lagrangian and predict the existence of five new scalars. Like the Higgs boson of the SM, these result from the eight degrees of freedom of the two doublets, after spontaneous symmetry breaking uses three of them to provide $W$ and $Z$ bosons with longitudinal--polarization degrees of freedom. 

The most general form of the 2HDM scalar potential is described by 14 parameters, however some simplifying assumptions are usually made to exclude CP non-conservation and spontaneous breaking effects, as well as the absence in the potential of odd quartic terms in the doublets.  The potential then contains the following terms:\par

\begin{equation}
\begin{multlined}
m_{11}^2 \Phi_1^\dagger  \Phi_1 + m_{22}^2 \Phi_2^\dagger \Phi_2 - m_12^2 ( \Phi_1^\dagger \Phi_2 + \Phi_2^\dagger \Phi_1) + \frac{\lambda_1}{2} ( \Phi_1^\dagger  \Phi_1 )^2 + \frac{\lambda_2}{2} ( \Phi_2^\dagger  \Phi_2 )^2 \\
+ \lambda_3 \Phi_1^\dagger \Phi_1 \Phi_2^\dagger \Phi_2 + \lambda_4 \Phi_1^\dagger \Phi_2 \Phi_2^\dagger \Phi_1 + \frac{\lambda_5}{2}\left(  (\Phi_1^\dagger \Phi_2)^2 +  (\Phi_2^\dagger \Phi_1)^2 \right).
\end{multlined}
\end{equation}

\noindent
At its minimum, the fields have the form\par
\begin{equation}
\langle \Phi_1 \rangle_0 = \binom {0}{v_1/\sqrt{2}}, \, \langle \Phi_2 \rangle_0 = \binom {0}{v_2/\sqrt{2}} .
\label{eq:vev}
\end{equation}

\noindent
After spontaneous breaking, the Lagrangian will contain terms describing a light $h$ and a heavy $H$ neutral scalar, a neutral pseudoscalar $A$, and two charged scalars $H^{\pm}$. The two parameters which influence the most the phenomenology of these five particles are the angles $\alpha$ and $\beta$, which diagonalize respectively the neutral scalar squared-mass matrix and the charged and pseudoscalar squared-mass matrices. The angle $\beta$ also satisfies the relation $\tan \beta = v_2/v_1$, where $v_1$ and $v_2$ are the vacuum-expectation values of the two Higgs doublets, as specified in Eq.~\ref{eq:vev} above.

The phenomenology of the five Higgs particles varies quite significantly as one moves in the space of model parameters.  The experimental investigation of 2DHM scenarios therefore benefits from the focusing on well-specified benchmarks. Within 2HDMs, one may generally distinguish a few specific situations, which address differently the problem of avoiding tree-level FCNC potentially arising from the non-flavour-diagonal nature of the Yukawa couplings. Glashow and Weinberg showed~\cite{glashowweinbergfcncabsence} how, to ensure the absence of FCNC, it is sufficient that all fermions of a given gauge representation acquire mass through couplings to only one Higgs doublet. This leads to four main possibilities for 2HDMs. In the first one, called ``Type-I 2HDM'', all fermions couple to only one Higgs doublet, while  in the second, ``Type-II 2HDM'', up- and down-type quarks couple to different doublets. The precise definition of the other two types of models is not univocal in the literature, and they are not frequently used as benchmarks for LHC searches so we omit their definition here. The class of Type-II 2HDMs includes the well-known ``Minimal Supersymmetric Standard Model'' \label{s:mssm} (MSSM)~\cite{mssm1,mssm2,mssm3,mssm4,mssm5,mssm6}, along with a number of variants that allow for additional  freedom to escape tight direct and indirect experimental bounds. 
The MSSM is a valid setup which allows to define the general features of the Higgs sector in 2DHM models by using five main parameters: $\tan {\beta}$, the Higgsino mass parameter $\mu$, the mass of the $A$ boson $M_A$, and the common masses of scalars and gauginos at the GUT scale, $M_{0}$ and $M_{1/2}$. 

The Higgs sector in generic Type-II 2HDM is less constrained than in the MSSM, as the parameter $\alpha$ in the former is not constrained by the value of $\tan \beta$ and by the scalar and pseudoscalar masses as in the latter; in addition, the charged and pseudoscalar masses are not necessarily very similar in generic Type-II models, allowing for a richer spectrum of decays.  A few searches consider the model called hMSSM~\cite{hmssm1,hmssm2}\label{s:hmssm}, where the ``h'' stands for {\em habemus}; there, the light CP-even Higgs boson mass is fixed to 125 GeV throughout the space of model parameters. This is possible with a high SUSY-breaking scale, by making assumptions on the structure of the Higgs sector in terms of radiative corrections and couplings.


For the sake of understanding what are the searchable signatures of 2HDM involving boson pairs, we first consider the lighter neutral scalar, $h$, which we assume here to be the one already discovered at a mass of 125 GeV. Although all observed decay modes have shown to be in agreement with SM predictions, there is room for studies of decay modes that would allow the 2HDM and the SM to be told apart. One example is the $h \to ZA$ decay, which may occur for very light pseudoscalar masses, or via an off-shell $Z$ boson.  That channel has not received much attention in the literature, mainly because in the MSSM its rate is strongly suppressed, but this is not the case in generic 2HDM, where the value of $\cos^2 (\beta -\alpha)$, to which the decay rate is proportional, is not constrained to be small.

For the heavy neutral Higgs $H$ the coupling to bosons is proportional to $\cos (\alpha -\beta)$, with the same behaviour discussed for the light Higgs $h$ once one shifts $\alpha$ by $\pi/2$. The decay to $WW$ pairs is significant, although in Type-II models it may be suppressed for high values of $\tan \beta$; the possibility of a decay $H \to hh$ must be taken into account, as well as $H \to AA$; the unknown value of self-coupling parameters make these possibilities unconstrained in the considered models. The decay of the $H$ to a $ZA$ pair is also possibly dominant for certain range of parameters. 

Within SUSY theories one must also consider another particle which can decay to Higgs boson pairs, along with other SM bosons. This is the bound state of a stop quark and its antiparticle, called stoponium, $\eta_{t \bar{t}}$~\cite{stoponium}. If the two-body decay of stop quarks to flavour-preserving states is disallowed, the stop may be able to hadronize, creating an $\eta$-like state of quantum numbers $J^{PC}=0^{++}$ and width in the tens of MeV range. Such a particle would decay primarily into a pair of SM fermions or bosons, or into a pair of neutralinos. Finally, also models that predict the existence of a hidden sector of particles mixing with the observed Higgs boson~\cite{hiddensec_hh,hiddensec_higgsium} must be cited as a possible source of resonant Higgs boson pairs. As far as production of an $A$ boson is concerned, there are no couplings to vector boson pairs to consider. However, the possibility of decays $A \to Zh$ must be studied, if the mass of the $A$ boson is below the top pair threshold of 350 GeV, above which the $A \to t \bar{t}$ decay dominates. The $Zh$ decay rate is proportional to $\cos^2 (\alpha -\beta)$. In summary, there is considerable freedom in 2HDM models to allow for diboson signatures of the neutral Higgs bosons to be relevant and accessible in LHC searches. Experimental results on some of them are presented in Sec.~\ref{s:higgsX}.

Another model worth mentioning is the Next-to-Minimal Supersymmetric Standard Model (NMSSM) \cite{nmssm1,nmssm2,nmssm3,nmssm4}, \label{s:nmssm} which includes in the Higgs sector an additional singlet. The NMSSM reduces the fine tuning required in generic MSSM theories~\cite{finetuningmssm} and solves the so-called $\mu$ problem~\cite{muproblem}; in addition, it may provide a contribution to electroweak baryogenesis~\cite{nmssmbaryo2}. The spectrum of particles predicted in the NMSSM includes three CP-even neutrals  denoted by $h_1$, $h_2$, and $h_3$; two CP-odd neutrals $a_1$ and $a_2$; and a pair of charged states. One of the lighter two $h$ states can be the 125 GeV scalar. In general, models that predict a very light pseudoscalar $a$ (with $m_a \sim 10$ GeV or less) entail the possibility for the decay $h \to aa$, which has been studied especially in connection with the $a \to \tau \tau$ signature, which is the favored decay mode; the decay to muon pairs offers a cleaner experimental signature and is also promising, if the $a$ couples weakly to SM particles~\cite{nmssmpheno1, nmssmpheno2}.

\subsection {\it Technicolour Theories and Composite Models}

\noindent
A dynamical explanation of electroweak symmetry breaking is offered by the theory introduced in the late seventies by Weinberg and Susskind, now called technicolour~\cite{smshortcomings4,weinbergtechnicolor}. Technicolour predicts the existence of a number of new particles, called technifermions; these may generate mass terms for $W$ and $Z$ bosons by their binding energy, as well as give rise to techni-hadrons, bound states of a new strong interaction. In analogy with quantum chromodynamics \label{s:qcd} (QCD), technihadrons with quantum numbers $I^G(J^{PC})=1^-(0^{-+}),1^+(1^{--}), 1^-(1^{++})$ are labeled respectively as $\pi_{T}$, $\rho_{T}$, $a_{T}$. The introduced interaction is invariant under a $SU(N_{T})_{T}$ gauge group; for $N_{T}=3$ one has a new QCD-like interaction at a higher energy scale. Technicolour theories mimicking QCD have been strongly constrained by electroweak precision fits, in particular by the measurement of the S and T parameters~\cite{PeskinTakeuci}.
Low-scale technicolour theories (LSTC)~\cite{lowscaletech1,lowscaletech2} \label{s:lstc} offer an escape to electroweak as well as FCNC constraints by introducing a ``walking gauge coupling'' and calling for a large number of new fermion doublets. In this scheme the $\omega_{T}$, $\rho_{T}$, and $a_{T}$ masses are lower than twice the mass of the lowest-mass $\pi_{T}$, and thus only decay by electroweak interactions, mostly to photons and gauge bosons, with narrow widths. The $\rho_{T}$ and $a_{T}$ are nearly degenerate~\cite{lowscaletech2}; the experimental signature of those two particles is therefore unable to distinguish them; they would appear together as a single bump in the $WZ$ mass distribution in fully reconstructed final states, or as a broader feature in modes including neutrinos. However, the branching fraction to $WZ$ pairs is smaller than 10\% if the $\rho_{T}$ mass is smaller than twice the $\pi_{T}$ mass, when the decay $\rho_{T} \to \pi_{T}W$ is the most frequent~\cite{lowscaletech1,hilltechnicolor}. 

The spontaneous breaking of electroweak symmetry has also been combined with TeV-scale strong dynamics by a number of theories that extend the SM. The new strong sector might result from a new interaction~\cite{smshortcomings1,weinbergtechnicolor,smshortcomings4} or from composite Higgs bosons~\cite{kaplangeorgi,continowarpedcomposite,giudicestrongcoupling}. In all cases new heavy particles are called for, which usually decay to pairs of SM vector bosons. 
The specific class of models called Little Higgs~\cite{ref8LH,ref1inrizzo2007a,ref8LHc,ref9LH} foresees that the Higgs boson arises as a pseudo-Nambu-Goldstone boson in a spontaneously broken global symmetry; this idea was first brought forth by Georgi and Kaplan~\cite{comphiggs1,comphiggs2} in their ``composite Higgs'' model, and was later resurrected to solve its fine-tuning problems~\cite{ref1inrizzo2007b}. The name ``little Higgs'' is due to the fact that in these models the Higgs boson is much lighter than all other pseudo-Goldstone bosons in the theory. These theories imply the existence of new heavy bosons with the same quantum numbers of the electroweak ones of the SM, $W^{\pm}_H$, $Z_H$ and $A_H$, as well as a heavy charge +2/3 singlet (usually called a ``top partner'') and a scalar triplet~\cite{ref1inrizzo2007b,ref12LH}; these particles are the minimum extension of the SM that allows to cancel the leading quadratic divergence in the Higgs mass at one-loop level, eliminating the problem of fine tuning between widely different energy scales. A new discrete symmetry $T$ is required~\cite{ref13LH,ref14LH} in order to bring the mass of the new particles below the TeV; SM particles and the heavy singlet are even under T, while the new bosons are T-odd.  A sub-class called ``littlest Higgs'' models also include the existence of ``mirror fermions'', one for each SM fermion, which have negative T-parity~\cite{ref15LH}. For what concerns resonant diboson signatures of relevance to this review, mention must be made of the scalar triplet $\Delta^{--},\Delta^{-}, \Delta^0$ of Georgi and Machacek~\cite{LHphenom1,LHphenom2}; the doubly charged scalar, which can be produced by vector boson fusion mechanisms at the LHC, is a smoking gun of this model, but its most promising signature involves decays to same-sign lepton pairs, while the branching fraction to $W$ pairs is typically below 10\%~\cite{LHphenom3,LHphenom4}. The Higgs triplet model~\cite{LHphenom1} has been studied by ATLAS (see {\em infra}, Sec.~\ref{s:wz}). In that framework a parameter $s_H^2$ governs the phenomenology through the definition of the fraction of the squared masses of gauge bosons generated by the new Higgs triplet, when $1-s_H^2$ is the fraction attributed to the SM Higgs doublet. The cross section and width of the new Higgs particles result proportional to the value of $s_H$.



\subsection {\it Models Predicting Resonances Decaying to Gluon Pairs }

Besides the aforementioned possibility of the decay to gluon pairs of a RS graviton, two other models must be cited which predict the existence of resonances with a significant branching fraction of the decay $X \to gg$, and which can therefore be investigated by searches in the high-end tails of the observed dijet mass distributions by ATLAS and CMS. The possibility of resonant decays to gluon pairs arises in models predicting string resonances~\cite{stringres1,stringres2,stringres3,stringres4}. Fundamental strings may have a mass at the TeV scale in models with large extra dimensions; this may allow the production of Regge excitations of quarks and gluons. One assumes that the theory is weakly coupled, such that reliance can be made on perturbation theory to compute scattering amplitudes; this also implies that the energy scale of strong gravity effects is higher, leaving the lowest Regge excitations unaffected by their complex phenomenology. The cross section and natural width of the Regge excitations are independent on the details of the compactification; the cross section may however depend on the colour factors and spin of the excited states~\cite{string1,stringres1}. 

Gluon-gluon resonances also arise in 
colour-octet scalar models~\cite{S8}, which predict a coupling of a colour-octet scalar field with gluons through a Lagrangian term of the form\par

\begin{equation}
{\cal{L}} = \frac{k g_v}{\Lambda} d^{abc} S_8^a {\cal{G}}^b_{\mu \nu} {\cal{G}}^{c,\mu \nu},
\end{equation}

\noindent
where $g_s$ is the strong coupling constant, $k$ is the coupling of the scalar field, the constant $\Lambda$ determines the energy scale of the interaction, $d^{abc}$ are the structure constants of SU(3), $S_8$ describes the scalar field, and the $\cal{G}$ tensor describes the gluon fields. In the commonly used benchmark in searches for colour octet scalars, the interaction scale is set to coincide with the mass of the observable resonance, $\Lambda=M$, and an equality with the QCD coupling is set by assuming $k=1$~\cite{harriskousouris}. The width of the resonant scalar state is given by\par

\begin{equation}
\Gamma = \frac{5}{6} \alpha_s k^2 \frac{M^3}{\Lambda_2}.
\end{equation}

\noindent
$\Gamma$ can become not negligible for large resonance masses, when the decay gives rise to a broad distribution in the mass of jet pairs, including long low-mass tails due to the convolution of the Breit-Wigner shape with the falling Parton Distribution Functions (PDF) \label{s:pdf} of the initial-state gluons. This results in complications in the signal extraction procedures with respect to the more common searches for narrow structures.


\clearpage
\section {The LHC and the ATLAS and CMS Experiments}
\label{s:experiments}

\noindent


\subsection {\it The LHC During Run 1 and Run 2}
\label{s:LHC}

\noindent
The LHC collider is hosted in the 26.7-km tunnel used until 2001 by the LEP-II collider, at a depth of approximately 100 meters under the border between Switzerland and France, near the city of Geneva. With a design beam energy of 7 TeV and a proton-proton instantaneous luminosity of $10^{34} cm^{-2}s^{-1}$, the LHC was conceived to push our investigation of subnuclear physics at the high-energy frontier, as well as to provide insight in B physics and in the physics of heavy ion collisions. To perform those tasks, the accelerator is endowed with four main  experiments built inside underground caverns where beams are brought to intersect: the all-purpose ATLAS and CMS detectors, plus a $B$-physics-oriented detector studying highly-asymmetric collisions, LHCb, and a detector optimized for the study of heavy ion collisions, ALICE. Originally targeting in particular the discovery of the Higgs boson, a goal achieved in 2012, the LHC is now expected to bring in a deeper knowledge of the SM, in particular in the electroweak and Higgs sector. Yet it is hoped to also enable the discovery of new physics beyond the SM, in the form of new high-mass particles and interactions.

The bending of the beams  in the synchrotron ring is produced by 1232 14-m-long superconducting magnets, cooled by superfluid helium at the temperature of 1.9K to provide in their interior a dipole field of up to 8.3 Tesla. Dipoles contain the two beams in a single magnetic structure. A lattice of additional superconducting magnets producing quadrupole and higher-order fields provides strong focusing and  squeezing of the transverse dimensions of the beams in coincidence with the interaction regions. The LHC receives proton beams from the SpS accelerator at 450 GeV and accelerates them in 28 minutes to the final collision energy (up to 4 TeV during Run 1, and up to 6.5 during the ongoing Run 2) with two independent sets of radiofrequency cavities operating at 400 MHz. Protons in the beams are bunched into up to 2808 equally-spaced packets, providing a nominal interaction rate of 40 MHz in the core of the ATLAS and CMS detectors. At the design instantaneous luminosity of $10^{34} cm^{-2} s^{-1}$ each bunch crossing results on average in about 25 inelastic proton-proton collisions.While most of these are low-energy interactions that do not yield interesting physics processes, they make the reconstruction of physics objects produced in the hard interactions of interest considerably more difficult. In addition, particles such as neutrons produced in a bunch crossing may also contribute to detector signals in subsequent ones. These effects are collectively addressed as ``pile-up''. A careful accounting of pile-up contributions to the measurements performed by the detectors is required in order to retain the detectors performance in terms of resolution and calibration. 

After a few pilot runs at the end of 2009, when the collision energy (up to 2.36 TeV) and the small integrated luminosity only allowed for a first {\em pot-pourri} of demonstrative performance results by the experiments, the LHC started stable operations with 3.5 TeV proton beams in March 2010, producing 7-TeV collisions in the core of ATLAS and CMS at a rate of 20 MHz. Due to the complex tuning procedures of the accelerator complex, the integrated luminosity acquired during 2010 only amounted to $45 pb^{-1}$ in ATLAS and $43 pb^{-1}$ in CMS; the corresponding datasets were mainly used for a few pilot measurements and calibrations. In 2011 the larger number of bunches and bunch densities allowed the accelerator to deliver over $6fb^{-1}$ of integrated luminosity to the experiments; a total of $5.1fb^{-1}$ were acquired by ATLAS and $5.5fb^{-1}$ by CMS; of these, respectively $4.7fb^{-1}$ and $5.1fb^{-1}$ were declared good for physics analysis. Those 7-TeV datasets allowed to produce, among a host of other physics measurements, a number of searches for resonances in diboson final states. Although the results of those searches have all been outdated, they are anyway concisely reported on along with the more recent results in Sec.~\ref{s:results}. 

The year 2012 concluded the Run 1 phase of LHC running with beam energies upgraded to 4 TeV. The stable LHC operations and higher instantaneous luminosity (with peak values reaching $ 7.7 \times 10^{33} cm^{-2} s^{-1}$) resulted in  $21.3fb^{-1}$ ($21.8fb^{-1}$) of integrated luminosity acquired by ATLAS (CMS) at a center-of-mass energy of 8 TeV, with respectively $20.3fb^{-1}$ ($19.7fb^{-1}$) of it usable for most physics analyses. This allowed for the first observation of the Higgs boson~\cite{higgsatlas,higgscms}, jointly announced by the ATLAS and CMS collaborations in July 2012 on the basis of the analysis of the 2011 datasets combined with those resulting from the first $5 fb^{-1}$ of 2012 collisions. A wealth of searches and measurements were also published. During the forthcoming years the two experiments produced results of resonance searches based on the full Run 1 datasets; some of those results still stand as the most sensitive ones for a few selected final states. All results based on the full Run 1 statistics are duly reported in Sec.~\ref{s:results}.

Following two years of shutdown during which the accelerator complex was subjected to significant upgrades, in 2015 CERN resumed colliding operations with the new LHC Run 2, for the first time reaching the proton-proton center-of-mass energy of 13 TeV on June $5^{th}$ 2015, when the accelerator operated with a bunch crossing time of $25 ns$. The 2015 datasets corresponded to $3.9fb^{-1}$ ($3.8fb^{-1}$) of integrated luminosity acquired by ATLAS (CMS), of which $3.2fb^{-1}$ (resp. $2.3fb^{-1}$) were used for most physics analyses by the experiments. The 2016 operations allowed the LHC to reach the record instantaneous luminosity of $1.4 \times 10^{34} cm^{-2} s^{-1}$, and added to those datasets a further $35.6fb^{-1}$ (resp. $37.8fb^{-1}$) of acquired integrated luminosity, $32.9fb^{-1}$ (resp. $35.9fb^{-1}$) of which usable for most physics analyses. The recently ended 2017 operations further added 51.3 $fb^{-1}$ of delivered luminosity to ATLAS and CMS. The experimental collaborations published a number of  results from diboson final states based on Run 2 data collected in 2015 and 2016; several preliminary results based on 2017 data have also been produced, and the process will continue for the next few years. Nonetheless, we feel that this is as good a time as any to summarize the status of our reach for new resonances, as 2018 datasets will hardly produce qualitative changes to the current situation.

\subsection{\it The ATLAS Detector}
\label{s:ATLAS}

\noindent
ATLAS (A Toroidal Lhc ApparatuS) \label{s:atlas} is one of the two general-purpose detectors constructed to study collisions of protons and heavy ions produced by the LHC. The apparatus measures 46 meters in length and 25 meters in diameter, for a total weight of  7000 metric tons. It is forward-backward symmetric and has almost $4 \pi$ coverage in solid angle to particles outgoing from its center. ATLAS features an inner tracking system constituted by layers of silicon pixel and strip detectors and a straw-tube transition radiation tracker. The tracking system is  surrounded by an axial 2-Tesla magnetic field produced by a thin, 2.3m diameter superconducting solenoid. Around the solenoid are placed electromagnetic and hadronic calorimeters, in turn surrounded by a muon spectrometer inside a system of toroid magnets.  For a complete description of the ATLAS detector see~\cite{atlas1,atlas2}.

The inner tracking systems~\cite{atlastracking} overall provide coverage for particles emitted with pseudorapidities $|\eta|<2.5$~\footnote{ Both ATLAS and CMS rely on a coordinate system with the z axis pointing along the clockwise-circulating proton beam at the interaction point, the y axis pointing upwards, and the x axis pointing toward the center of the ring. Two angles determine the diretion of an outgoing particle: the azimuthal angle $\phi=\cos^{-1} (x/r_{xy})$, with $r_{xy}=(x^2+y^2)^{1/2}$, and the pseudorapidity $\eta = - \log \tan (\theta /2)$, with $\theta=\cos^{-1}(z/r)$ and $r=(x^2+y^2+z^2)^{1/2}$.}; the straw tubes, extending their coverage to $|\eta|<2.0$, add measurement capability useful for electron identification. The calorimeter system is divided in several subsystems. The measurement of electrons and photons is provided by an electromagnetic sampling calorimeter with lead sheets alternating with liquid argon as active medium, divided in three layers and with coverage in the region $|\eta|<3.2$, segmented longitudinally in shower depth. Outside of the lead/argon calorimeter, a steel/scintillator calorimeter provides hadronic coverage up to $|\eta|<1.7$~\cite{atlasecal}. Extended coverage is provided with copper and liquid argon as active medium by end-cap sections covering the range $1.5 < |\eta|<3.2$ for hadrons, and copper or tungsten and liquid argon sampling for both electromagnetic and hadronic showers in the forward region up to $|\eta|<4.9$~\cite{atlashcal}. Muons are detected and measured in the range $|\eta|<2.7$ by three large air-core  superconducting toroidal magnet systems, instrumented with drift tubes and cathode strip chambers, with triggering capabilities offered by additional resistive plate chambers and thin gas chambers  in the restricted range $|\eta|<2.4$~\cite{atlasmuon}. 

The ATLAS detector was partly upgraded during the 2013-14 shutdown. The most significant addition was a fourth layer of pixels, called ``insertable B-layer'',  placed at a radius of 33mm from the beam axis. The added layer significantly enhanced the detector capability to identify $b$-quark-originated hadronic jets, with improvements  in the impact parameter resolution of up to 40\% for low transverse-momentum charged particles.  Furthermore, the muon coverage in the endcap regions was completed, and new luminosity and beam monitors were added. The infrastructure was also improved by additional muon shielding and by a new beam pipe. 

Both in Run 1 and Run 2 ATLAS relied on a two-level trigger system to identify events of interest and send the relative digitized information to permanent storage. The first-level trigger is a hardware system that uses calorimeter and muon detector information to reduce the event rate to 100 kHz (75kHz in Run 1). This is followed by a software-based high-level trigger, which reduces the accepted event rate to an average of 1 kHz. The trigger system also withstood several upgrades between Run 1 and Run 2, most notably by the addition of a hardware fast tracker, operating at the level-1 output rate of 100 kHz. For a detailed description of the trigger system of the ATLAS detector see~\cite{atlastrigger}.

\subsection{\it The CMS Detector}
\label{s:CMS}

\noindent
Measuring 22 meters in length by 15 in diameter, CMS \label{s:cms} (a Compact Muon Solenoid) is smaller than its direct competitor, but at 14,000 metric tons it weighs twice as much. CMS also relies on silicon pixels and strip tracking in the region close to the beam, but unlike ATLAS it contains electromagnetic and hadronic calorimeters within a solenoid of 6 meter diameter, which provides a 3.8-Tesla axial field in its interior. Outside the solenoid, four layers of muon stations alternate with iron which allows for the return flux of the magnetic field. A general overview of the CMS detector can be found in~\cite{cms}. 

The all-silicon tracking system of CMS is composed of 1440 pixel modules and 15148 strip modules arranged in a total of 12 layers and covering the pseudorapidity region $|\eta|<2.5$.  Thanks to the strong axial field, the transverse momentum resolution is of $1.5\%$ and the transverse impact parameter resolution ranges from 25 to 90 $\mu m$  for non-isolated hadrons of momenta up to 10 GeV and $|\eta|<1.4$~\cite{cmstracker}. The calorimeter system, divided in electromagnetic (ECAL) \label{s:ecal} and hadronic (HCAL) \label{s:hcal} sections, is composed of a barrel and two endcap sections covering the pseudorapidity range $|\eta|<3.0$, and is complemented by forward units that extend the coverage to $|\eta|<5.0$. The ECAL uses lead-tungstate crystals of 25.8 radiation lengths 
to provide a resolution of better than 2\% for electrons of transverse energy of 45 GeV in the barrel region ($|\eta|<0.8$), and from 2\% to 5\% at higher pseudorapidity~\cite{cmsecal}. The HCAL is composed of brass and scintillator layers, with 5.8 interaction lengths ($\lambda_I$) at zero pseudorapidity, rising to 10.6 $\lambda_I$ at $|\eta|=1.3$; the ECAL in front of it provides additional $1.1 \lambda_I$ of material.
Muon stations embedded in the steel flux-return yoke outside the solenoid are equipped with gas detectors using three technologies: drift tubes, cathode strip chambers, and resistive-plate chambers. They provide coverage in the region $|\eta|<2.4$.  For muons whose trajectory is measured both in the inner tracker and in the muon stations, the transverse momentum resolution ranges between 1\% and 5\% for transverse momenta up to 1 TeV~\cite{cmsmuon}.

The online collection of events interesting for physics analysis is controlled by a two-level trigger system. The first level is composed by custom hardware processors that use fast calorimeter and muon information to select events at a maximum rate of 100 kHz. The latter are pipelined into the High-Level Trigger (HLT) \label{s:hlt} system, which is composed of processors operating a speed-optimized reconstruction of the full event characteristics in $4 \mu s$, allowing for an accurate selection of the most interesting events, with an output rate to permanent storage of up to 1 kHz. An extension of the triggering capabilities of CMS comes from the recently developed technique called {\em data scouting}~\cite{datascouting}. This consists in the storing of a limited amount of information from each event in a dedicated stream. To be meaningful, the information must include an online treatment of the calibration and cleanup procedures usually applied in offline analysis. The technique allowed {\em e.g.} the extraction of limits on the cross section of dijet resonances that extend to lower mass values than those reachable by analyses based on ordinary jet triggers~\cite{dijetscouting}. 

During the 2013-2014 shutdown the CMS detector withstood several minor yet important upgrades; we mention here the most notable four. The tracker was equipped to operate at low temperatures to mitigate the effects of radiation on its performance. The central beam pipe was replaced to prepare for the installation of a new pixel tracker. A fourth measuring station was added to each muon endcap, complemented by the installation of 125-ton composite shielding at each end of the detector. A pixel luminosity telescope was installed at each side of the central detector around the beam pipe. A further upgrade was carried out after the end of 2016 data taking to install a fourth layer of silicon pixel sensors in close proximity to the beam. The added layer significantly improves the performance of CMS tracking, especially in the measurement of track impact parameter and consequently the $b$-tagging capabilities of the system.

\section {Physics Objects Reconstruction, Calibration, and Measurement}
\label{s:objects}


The ATLAS and CMS experiments fully rely on their redundant tracking and calorimetric systems to identify and precisely estimate the four-momentum of all produced particles that cross their detection elements. The reconstruction flow allows for a considerable interplay between the various steps -- {\em e.g.}, a primary event vertex must be located before the measurement of neutral energy deposits can be turned into precise four-momentum estimates, as well as prior to any attempt at identifying secondary decay vertices from long-lived hadrons. A separate discussion of the identification and measurement of each of the ``high-level objects'' through which different final states can be categorized is nonetheless the easiest way to describe the procedures. 
Once individual particles such as electrons, muons, and $\tau$ lepton candidates, photons and isolated hadrons, as well as collective phenomena like hadronic jets, $b$-tags, and missing transverse energy are measured, their four-momenta must be corrected to account for known biases of various nature. These range from effects dependent of running conditions ({\em e.g.} the instantaneous luminosity at the moment when the event was collected) to shifts related to measurement procedures, detector non-linearities, or specific properties of the measured objects. Dedicated working groups within the experiments monitor the functioning of the sub-detector elements, and continuously update calibration constants and procedures in order to optimize the resolution of energy and momentum measurements throughout data taking. Below we briefly mention these procedures when relevant, for each considered high-level object.



\subsection{\it Electrons and Photons}
\label{s:photons}

In ATLAS, electrons and photons are identified starting from clusters of energy in the electromagnetic calorimeter. Clusters lacking a spatial match to the extrapolation of a reconstructed track or a conversion vertex identified in the inner detector constitute unconverted photon candidates, while ones associated with an identified conversion vertex are classified as converted photons; when a matching track exists, which is consistent with belonging to an electron, the cluster becomes an electron candidate. Electrons with transverse momentum~\footnote{ Transverse momentum, denoted as $p_T$, is the component of a particle momentum in the plane orthogonal to the $z$ coordinate (which coincides with the beams direction at the detector nominal center). Transverse energy, denoted as $E_T$, is also frequently used to identify the product of a particle's energy times the sine of the angle of its momentum vector with respect to the $z$ axis. } $p_T>7$ GeV are identified in the pseudorapidity range $|\eta|<2.5$. Their clusters must match spatially to the extrapolation of the trajectories of charged tracks measured in the inner detector~\cite{electronsAtlas}. A set of criteria based on the observed characteristics of the cluster and the track are devised to reduce the background from jets with a leading track and high electromagnetic component. Analyses targeting the decay of $W$ and $Z$ bosons usually also require the electron candidates to have transverse energy $E_T$ above  25 or 30 GeV; the associated track is required to be consistent with the identified primary vertex using impact parameter significance criteria in the transverse plane as well as the spatial difference between the track extrapolation and the primary vertex along the beam axis. The isolation of electrons from other objects is usually imposed by requiring that the sum of calorimeter energy not attributed to the electron shower within a cone of radius $R=0.3$ in $(\eta , \phi)$ space be smaller than some threshold (typically 4 to 6 GeV)~\cite{electronAtlasefficiency2012}. Photon candidates are reconstructed from fixed-size clusters of cells in the electromagnetic (EM) \label{s:em} calorimeter, using the cluster shape as a selection criterion~\cite{photonefficiencyAtlas}. Photons of relevance in searches for decays of massive resonances are usually required to be located inside the precision region of the EM calorimeter ($|\eta|< 1.37$ or $1.52 < |\eta|< 2.37$, excluding the transition region between barrel and endcap calorimeters)~\cite{photonidAtlas}. To increase the purity of photon candidates, ATLAS searches require isolation criteria based on both calorimeter and tracking information.  Contributions from pile-up and underlying event to the calorimeter isolation are subtracted on an event-by-event basis~\cite{photonIsoAtlas1,jetareacacciari,photonIsoAtlas2}.
Energy scale corrections are applied to the measured electron energy in order that data and Monte Carlo (MC) \label{s:mc} match in the reconstructed  peak of well-identified $Z \to ee$ events~\cite{electronscaleAtlas}.  A multivariate regression algorithm~\cite{AtlasEMcalibMVA} was also developed to improve the calibration of the estimate of electron and photon energy, based on Run 1 data.  The uncertainty in the energy scale of energetic photons varies in the range 0.5-2\% as a function of pseudorapidity.


CMS also starts the identification procedure of electron and photon candidates from clusters in the electromagnetic calorimeter~\cite{cms_ele}. First, groups of clusters are used to form superclusters. In the barrel region these are formed from five-cristal strips in pseudorapidity, centered around the seed crystal. The extension in azimuth accounts for the effects of the magnetic field on electrons resulting from photons starting their shower before the ECAL. In the endcap calorimeters the clusters are formed from 5x5 matrices of crystals. 
Electron candidates are reconstructed from superclusters with a matched track. The transverse momentum of the electron is estimated from a fit to the track trajectory that employs a Gaussian-sum filter algorithm~\cite{gaussiansumCMS}. The algorithm accounts for the multiple bremsstrahlung of soft photons by the electron crossing the layers of silicon sensors in the tracker. 
Electron selections differ based on the specific needs of the different analyses; use is commonly made of the profile of the energy deposit in the calorimeter, as well as the fraction of energy deposited in the hadronic calorimeter behind the electron cluster. To increase the purity of the selected candidates, the number of missed hits in the pixel tracker and other track quality criteria are also used. Electrons from vector boson decays are required to be isolated from other energy deposits or track activity. The variable $I_{rel}$ is computed as the sum of transverse momenta of Particle Flow (PF)~\cite{pf1cms,pf2cms,pf3cms,pf4cms} \label{s:pf} candidates (both charged and neutral) within a radius of $0.3$ around the electron direction, divided by the electron $p_T$ and corrected on an event-by-event basis for pile-up effects~\cite{neutralsub1}. 
For photons a variable called $R_9$, computed as the fraction of cluster energy contained in the innermost 3x3 matrix of crystals, is sensitive to the degradation of the $E_T$ resolution caused by the effects of the magnetic field when the showers are initiated in the tracker. To compute their four-momenta with precision, photon candidates are associated to the primary vertex having the largest sum of $p_T^2$ of associated charged tracks. The identification of photons is also based on a boosted decision tree algorithm~\cite{cms_photonmva1,tmva}, which receives as inputs the shower shape and isolation, as well as other observables that affect the shower characteristics due to pile-up and other effects; a veto of electrons that does not remove conversions is also applied~\cite{cms_csev}. The photon isolation is computed in a cone of radius $R=0.3$ in $(\eta, \phi)$ space around the candidate. The about 50\% of photon candidates that convert before entering the calorimeter are identified by their vertex if the resulting tracks pass at least three tracking layers.

The energy scale of the CMS electromagnetic calorimeter is calibrated to account for several detector effects~\cite{emcalibCMS}, and allow accurate $E_T$ estimates of electrons and photons. A number of observable characteristics of the cluster are used to extract corrections that depend on $E_T$, $\eta$, $R_9$, and on the extension of the cluster in azimuth: these include the point where the cluster intercepts the interface between crystals, the location of the shower with respect to the local lateral granularity, and the fraction of energy of the shower starting  before the calorimeter. A fine-tuning of the calibration developed for $h \to \gamma \gamma$ searches, where photon energy measurements are crucial, is then operated by reconstructing as photons the electron showers from $Z \to ee$ decays, when no tracking information is used except for the vertex position in the measurement of the mass of the electron pairs. The comparison of the mass distribution observed in real data to that obtained in simulated events allows to account for slight underestimations of the energy resolution. 
The uncertainty in the energy scale of photons after the above calibrations amounts to 0.25\% for photons of $E_T>20$ GeV.


\subsection{\it Muons}

ATLAS reconstructs muons within the $|\eta|<2.7$ acceptance of its muon system. Muon candidates are obtained from one of three main reconstruction methods~\cite{ATLASmuon,ATLASmuon13}: a combined fit, segment tagging, or calorimeter tagging. The combined muon reconstruction starts with independent tracks fits in the inner detector and in the muon stations. A combination is performed by an outside-in fit that starts with the track in the muon stations and extrapolates it inward to the inner detector track matching it. During the procedure, hits in the muon stations can be added or removed to improve the fit quality. When a muon station candidate is absent (either because the muon has insufficient momentum to cross more than one muon layer, or because it crosses regions of the muon stations with reduced acceptance)  segment-tagged muons can still be constructed by tracks in the inner detector that extrapolate to at least one local track segment in the drift tubes or cathode strip chambers. To further extend acceptance, calorimeter-tagged muons can be reconstructed from inner tracks that match to a calorimetric deposit compatible with the passage of a minimum ionizing particle. Muon candidates employed in most physics analyses must then pass several quality requirements, including {\em e.g.} the number of missed hits in the inner tracker and the value of the fit $\chi^2$,  to suppress backgrounds from pion and kaon decays and to ensure a good momentum measurement; depending on their quality, they are classified as loose, medium, or tight muons. The misidentification probability is monitored using a unbiased sample of  $K^0$ decays to charged pion pairs in real data collected with calorimeter-based triggers, while the identification efficiency is measured with the tag-and-probe method~\cite{ATLASmuon13} in $Z \to \mu \mu$ and $J /\psi \to \mu \mu$ decays. To select muons coming from the decay of $W$ and $Z$ bosons, isolation criteria are added to the list of requirements. ATLAS analyses use a track-based isolation variable obtained by constructing the sum of transverse momenta of tracks with $p_T>1$ GeV collected within a cone of muon-$p_T$-dependent radius, or a calorimeter-based isolation constructed with the energy of topological clusters~\cite{topologicalclustersATLAS} in a cone of radius $R=0.2$ around the muon candidate, after subtracting the expected muon and pile-up contributions. The momentum scale of muons is corrected with a parametrization determined by a likelihood fit to $Z$ and $J/ \psi$ dimuon candidates; after the calibration residual scale uncertainties are estimated to range between 0.05\% and 0.1\% for $Z$-decay muons. The relative momentum resolution ranges between 1.7\% and 2.9\%, mostly depending on muon pseudorapidity.

In CMS, most high-$p_T$ analyses consider muon candidates reconstructed in the pseudorapidity range $|\eta|<2.4$ by a combined fit performed outside-in with the Kalman filter technique~\cite{kalman}. The fit considers both tracker and muon detector hits; the latter improve the momentum resolution of muons with $p_T>200$ GeV~\cite{cmscosmicmu,cmsdetperform}. These ``global muons'' must have at least one hit in the pixel tracker, six in the strip layers, and two segments reconstructed in the muon detector planes. In addition, global muon candidates must fulfill other selection criteria, including impact parameter and momentum measurement requirements~\cite{cmsmuons7}. ``Tracker muons'', that only require a track matched to a single segment in the muon system, extend the acceptance to momenta below 5 GeV. To be considered in offline analysis, muons must typically fulfill several additional requirements based on the quality of the track fit and the matching of muon chambers segments to the trajectory measured in the silicon tracker, as well as on the impact parameter of the track. For muons coming from the decay of heavy objects, isolation requirements are also applied. The typical requirement is that the sum of $p_T$ of all tracks in a cone of $R=0.3$ around the muon candidate does not exceed 10\% of the muon $p_T$.  The identification efficiency is measured like in ATLAS using the tag-and-probe technique~\cite{CMStagandprobe} on clean samples of $Z \to \mu \mu$ and $J /\psi \to \mu \mu$ decays. The fake rate corresponding to the chosen muon selection criteria is measured for pions, kaons, and protons using large samples of $K^0 \to \pi \pi$, $\Lambda \to \pi p$, and $\phi \to KK$ decays. For tight identification criteria fake rates are usually below 0.1\%~\cite{cmsmuons7}. Muon scale calibration parameters are extracted with a likelihood technique that models the biases as a function of individual muon kinematic characteristics, using as a reference the $Z \to \mu \mu$ and $J/\psi \to \mu \mu$ mass distributions. The momentum scale after calibration is known to better than 0.2\% accuracy. The relative momentum resolution of calibrated muons, integrated in $\phi$ and $\eta$, ranges from 1.8\% to 2.3\% for $Z$-decay muons.

\subsection{\it $\tau$ Leptons}

The lifetime of $\tau$ leptons is of $2.9\times 10^{-13}s$, therefore their experimental signature in hadron collisions depends on its decay products. $\tau$ leptons exhibit a quite varied set of possible decays: the ones to electron plus neutrinos and muon plus neutrinos occur about 35\% of the time overall. The rest of the decays involve charged and neutral hadrons plus a $\tau$ neutrino; these are typically divided into the ``one-prong'' and ``three-prong'' categories depending on the number of produced charged hadrons. The identification of the all-leptonic decays involves the detection of electrons and muons, and may or may not include information about the unbalancing of the transverse momentum in the detector. The identification of hadronic $\tau$ lepton decays is more complex; the ATLAS and CMS experiments have devised refined techniques to improve their detection efficiency, given the importance of the $\tau$-lepton signature for a number of new physics searches as well as for Higgs boson measurements. 

The identification of hadronic $\tau$ decays in ATLAS~\cite{atlastau1} starts by defining candidates from anti-$k_T$ jets with a $R=0.4$ radius parameter. The visible energy of hadrons from $\tau$ decay is reconstructed from three-dimensional clusters of calorimeter cells in a narrow $R=0.2$ cone around the original jet axis. Tracks are associated to the calorimetric deposits if they pass quality criteria based on impact parameter and number of hits, and have $p_T>1$ GeV. The reconstruction of the individual charged and neutral hadrons in $\tau$ decays helps in the calculation of the visible energy from $\tau$ decay and in the classification of the decay mode. The higher resolution of the inner tracker improves the energy measurement with respect to estimates exclusively based on calorimeter deposits. A final $\tau$-specific calibration, obtained from MC, corrects for effects such as energy lost out of the narrow cone, the underlying event, and pile-up contributions.
A boosted decision tree is trained to discriminate $\tau$ candidates from hadronic jets using the calorimeter cluster shape and the track multiplicity. A veto is also applied to a likelihood value designed to flag electron candidates. 
A recently developed ``$\tau$ particle flow'' algorithm~\cite{atlastau2} classifies the $\tau$ candidate decay into one of the five main dceay modes (one-prong, one-prong with one $\pi^0$, one-prong with multiple $\pi^0$, three-prong, three-prong plus neutral pions) according to the identified charged and neutral hadrons it gave rise to. The resulting calibration of the $\tau$ energy, measured in $Z \to \tau \tau$ samples, achieves a resolution of 16\%.

The identification and measurement of $\tau$ lepton candidates in CMS is based on the PF algorithm. The original seed of a $\tau$ candidate is an hadronic jet, wherein neutral pion candidates are reconstructed. Charged tracks are combined to fully identify the decay mode of the hadronic $\tau$ decay system, and to compute its four-momentum~\cite{taumeascms}. The classification of the decay is operated with a boosted decision tree classifier~\cite{taumvacms}, which employs the isolation of the hadronic $\tau$ candidate from other detected particles, along with impact parameter information to discriminate real $\tau$ leptons from faking quark and gluon jets; additional discrimination against electrons and muons is provided by other variables sensitive to those particle species. The BDT-based  identification achieves an identification efficiency of 50\%, with  fake rates of less than 1\%. 
The energy scale of hadronically-decaying $\tau$ leptons after calibrations is known with a precision of 3\%~\cite{taumvacms}; the relative energy resolution ranges from 15\% to 20\%, with a dependence on $\tau$-lepton $p_T$.

\subsection{\it Jets}
\label{s:jets}

\noindent
In ATLAS jets are usually reconstructed starting from energy deposits in the calorimeter using the anti-$k_T$ algorithm with the radius parameter set to $R=0.4$. Four-momenta are obtained from the energy of clusters, assuming they are massless. For jets with a transverse momentum below 50 GeV a requirement is applied on the output of a multi-variate discriminant~\cite{jvt} to reject jets largely contributed by pile-up, based on tracking and vertex information. Jets withstand corrections that address the pile-up contributions and the calorimetric response~\cite{atlasjets1,atlasjets2,jetareacacciari}.  The calibration of the energy response depends on jet $p_T$ and $\eta$ and is performed with factors extracted from MC simulation and with the combination of a number of {\em in situ} techniques~\cite{jetcalibinsitu}, which exploit the $p_T$ balance between a jet and a reference object (photon, $Z$ boson, jets system) recoiling against it. A further calibration designed to reduce flavour-dependent response is also derived~\cite{globalsequentialcalib}. An uncertainty in the jet energy scale of less than 1\% was estimated for $|\eta|<1.2$ and jets of $55<p_T<500$ GeV in Run 1~\cite{atlasjes15}; the value was confirmed for jets of $100<p_T<500$ GeV using Run 2 data~\cite{atlasjes17}. In the study on Run 1 (Run 2) data the uncertainty was estimated  in the central region to be of 3\% (4.5\%) at low-$p_T$, where pile-up corrections have a larger impact. For jets in the region $|\eta|>0.8$, dijet balancing techniques allowed to estimate a jet energy scale uncertainty below 2\% for $p_T>80$ GeV. 


Jet reconstruction in most CMS analyses uses as input the momenta of particle candidates obtained from the PF algorithm. PF reconstructs and identifies each individual particle with a combination of information from the various detector elements. This is possible thanks to the excellent granularity of the CMS electromagnetic calorimeter, and to the strong axial field produced by the solenoidal magnet inside the tracking and central calorimeter. PF candidates belong to five distinct categories obtained from their PF constituents: electrons, photons, muons, charged and neutral hadrons. The four-momenta of all identified PF candidates are clustered by the anti-$k_T$ algorithm~\cite{antikt1,antikt2} or the Cambridge-Aachen (CA) \label{s:ca} recombination algorithm~\cite{CA1,CA2} as implemented in the FastJet package~\cite{fastjet1}. The $k_T$ algorithm~\cite{kt1,kt2} is used to recluster wide jets in the calculation of the variable called ``N-subjettiness'' used to identify boosted objects (see {\em infra}, Sec.~\ref{s:boostjets}). Charged hadrons that do not originate from the event vertex produced by  the hard subprocess of interest are subtracted from the list of clustered particles; a further mitigation of pile-up effects is operated by an average neutral-energy density subtraction~\cite{jetareacacciari}. Jet energies are corrected as a function of transverse momentum and pseudorapidity using MC simulations and data in dijet, multijet, $\gamma$+ jet, and leptonic Z+jet events~\cite{energycorrcms1,energycorrcms2}; noise-originated jets, as well as ones not originated from the studied hard interaction, are eliminated by specific jet identification criteria~\cite{jetnoiseremovalcms}. For jets of 100 GeV (1 TeV) the energy resolution is of about 10\% (5\%)~\cite{cmsjesjer}. The jet energy scale of CMS jets after all calibrations is known with an accuracy of less than 3\% overall, and better than 1\% for jets of $p_T>30$ GeV in the barrel region $|\eta|<1.3$~\cite{cmsjesjer}.

\subsection {\it Boosted Objects}
\label{s:boostjets}

The hadronic decay of heavy SM objects --$W$ and $Z$ bosons, Higgs bosons, and top quarks-- produces pairs (or triplets in top decay) of energetic quarks that fragment independently, giving rise to experimentally distinguishable hadronic jets, except when hard initial- or final-state gluon radiation obfuscates the picture.
If the decaying particle is highly boosted, however, the final state quarks are emitted with small spatial separation, and the streams of hadrons resulting from quark fragmentation partly overlap in $(\eta,\phi)$ space. In such conditions it proves advantageous to first reconstruct the event with a clustering algorithm using a large radius parameter, and successively examine the substructure of the resulting jets in detail. The field of ``boosted objects tagging'' (or ``V-tagging'' when $W$ and $Z$ bosons are the explicit target) has boomed in the last decade, with the development of a number of techniques that allow the identification of vector bosons and top quarks within high-$p_T$ wide jets, allowing to overcome the otherwise enormous QCD backgrounds in the identification of hadronic decays of heavy objects.
 
The ATLAS experiment operates an identification of boosted top quarks, $W$, $Z$, and Higgs bosons starting from wide jets; $R=1.0$ anti-$k_T$ jets have been used in many analyses. A local cluster-weighting algorithm~\cite{atlasloccluw} is used to calibrate the energy of their constituents and a reclustering is then performed by the $k_T$ algorithm with a $R=0.2$ parameter; such sub-jets are removed from the list if they contain less than 5\% of the large-R jet. The four-momentum of the large-R jet is then recomputed using the resulting sub-jets momenta. Finally, jet energy and mass can be calibrated to match the particle-level originating jet using MC simulation information~\cite{atlaslargerjetcalib, atlastrimmedjets}.
Another approach combines topological clusters reconstructed in the calorimeter and undergoing a local calibration~\cite{atlastopoclus,atlaslocalcalib} into jets, using the CA algorithm with a $R=1.2$ parameter. The iterative combination allows to identify the two last sub-jets as possible candidates of decay of bosons ($W$, $Z$, $h$). A cleanup~\cite{softdrop,atlasgrooming} can then be applied to reduce noise and pile-up contributions, and avoid the consequent degradation in resolution.

In CMS, wide jets reconstructed with the anti-$k_T$ or CA algorithm are first subjected to a grooming technique~\cite{grooming} which re-clusters the constituents into ``sub-jets'' after removing the product of soft QCD radiation not directly related to the hard interaction. A number of techniques exist for this task. A jet pruning algorithm~\cite{ellissubstructure,ellissubstructure2} used by many analyses reclusters jets with the CA algorithm, iteratively discarding their soft components during the recombination procedure, identified by angular and transverse momentum requirements. Finally the pruned mass of the jet, as computed from the four-momenta of the components that have survived the pruning phase, is used to identify decay candidates of heavy objects. The technique significantly reduces the jet mass in single-quark- and gluon-originated jets~\cite{cmssubstructure}.  Recently further advanced techniques for handling pile-up subtraction have been devised by the experiment. The algorithm called PUPPI~\cite{puppi} (from ``Pile-Up Per Particle Identification'') combines tracking information with the event pile-up properties into a four-momentum-rescaling weight assigned to charged and neutral candidates. The weight accounts for the likelihood that the particles originate from pile-up. Another algorithm, called ``soft-drop''~\cite{softdrop1,softdrop2} can be used with jets whose momenta have been recomputed with PUPPI, to remove contributions from soft radiation. Corrections are finally applied to remove jet $p_T$ dependences in the measured jet mass~\cite{corrsoftdrop}. The resolution in the jet mass after these corrections is typically of 10\%. 

A further variable which proves effective in separating QCD jets from ones due to the decay of heavy objects in both ATLAS and CMS analyses is the N-subjettiness~\cite{nsubjettiness,nsubjettiness2,subjettiness2}. The iterative $k_T$ algorithm~\cite{kt1,kt2} is used to re-cluster the original jet, successively recombining individual components  until N sub-jets remain. One then defines the N-subjettiness $\tau_N$ as\par

\begin{equation}
\tau_N = \frac{1}{d_0} \sum_k p_{T,k} \, min (\Delta R_{1,k}, \Delta R_{2,k}, ..., \Delta R_{N,k})
\end{equation}

\noindent
In the above formula $d_0$ is a normalization factor that accounts for the original jet radius, and the sum runs on all constituents; $\Delta R_{n,k}$ is the angular distance between the axis of the n-th subjet and the constituent. $\tau_N$ is larger for jets compatible with being constituted by N components. A frequently used derived quantity is the ratio $\tau_{21}=\tau_2 / \tau_1$, which optimally distinguishes jets originated by hadronic decays into quark pairs from single quark or gluon hadronizations; other combinations of the $\tau_j$ variables are also used.


\subsection{\it B Tagging}
\label{s:btagging}

The identification of heavy quarks as originators of hadronic jets is a very important tool in the search for new physics. $b$-quark-originated jets arise in over 99\% of top quark decays, and are also the most frequent final state of Higgs boson decays, as the $h \to b \bar{b}$ branching fraction is about 58\%~\cite{HSMBR}; in addition, $b$-tagging can increase the purity of hadronic decays of the $Z$ boson, because of the relative rarity of $b$-quark production in QCD backgrounds. Theoretical prejudices pointing at the top quark as a possible door to physics beyond the SM, together with the importance of Higgs and weak bosons as possible final states of new higher-mass states, make $b$-quark jets a crucial signature at the LHC. For those reasons, the ATLAS and CMS collaborations devoted extreme care in the construction of high-resolution tracking detectors capable of measuring the impact parameter of charged particle tracks with resolutions of the order of $10\mu m$, with the main goal of enabling a high-efficiency identification of $b$-quark-originated jets. This exploits the ${\cal{O}} (ps)$ lifetime of $B$ hadrons, which results in secondary decay vertices originating particles whose back-propagated tracks do not intersect the primary vertex position. The upgrades operated to the pixel detectors of the two experiments mentioned in Sec.~\ref{s:experiments}, as well as the future ones designed for the high-luminosity phase of the LHC,  had indeed the goal of improving, or at least securing, the performance of the experiments' near-vertex tracking capabilities in the increasingly complex environment of LHC collisions.

The large amount of information that needs to be considered in the identification of $b$-quark jets makes the task a quite complex one. The strongest indicium of a heavy flavour decay is the presence of a secondary vertex reconstructed from charged tracks in the jet, displaced from the primary vertex and not compatible with being the result of two-body decays of neutral hadrons such as $K_s$ or $\Lambda$, or photon conversions. Useful information is also contained in the composition of the jet and its fragmentation properties, as well as the presence of electron or muon candidates of small transverse momentum (``soft leptons''). Over the course of the past few years the experiments have therefore devised increasingly refined multi-variate techniques to produce effective $b$-tagging algorithms, improving their efficiency for a given fake rate along the way. 

In ATLAS, $b$-tagging is usually performed by an algorithm called MV2~\cite{atlasbmva}, which uses a boosted decision tree classification implemented in the TMVA package~\cite{tmva}. MV2 receives as input the output of a number of independent algorithms. Among them are SV~\cite{atlasSV}, which directly reconstructs secondary vertices inside the jet; the IP2D and IP3D algorithms~\cite{atlasbmva}, which use the properties of tracks, constructing log-likelihood-ratio discriminants to tell apart heavy-flavour-originated jets from light-quark or gluon ones; and JetFitter~\cite{jetfitter}, which exploits the topological structure of weak decays of heavy hadrons, attempting a full reconstruction of the decay chain inside the jet with a Kalman filter technique~\cite{kalman}. 

CMS also developed a number of methods to $b$-tag hadronic jets~\cite{cmsbtagging,cmva1,cmva2}. Of relevance to this review are the most performant ones, which have been applied to searches involving heavy bosons decays to $b$-quark pairs. The first one, called CSV (for ``Combined Secondary Vertex''), \label{s:csv} combines the information of identified secondary vertices in the jets with other individual and collective properties of tracks with large impact parameter, such that even when no secondary vertex can be directly fit, the information can still be effectively used to discriminate heavy-quark jets from ones originated from light-quark or gluons. The combination is performed using a likelihood ratio in an earlier version, and with an artificial neural network in the version used more recently. 
A second algorithm, called CMVA (for ``Combined Multi-Variate Algorithm'')~\cite{cmva1,cmva2}, \label{s:cmva} further combines two different versions of the CSV discriminant output with information coming from jet-probability tagging (which uses the probability of tracks being originated from the primary vertex) and from two algorithms searching for soft electrons and muons inside jets. The combination is performed with a boosted decision tree based on the scikit-learn package~\cite{scikitlearn}.

The performance of the $b$-tagging algorithms is assessed with similar techniques by the two experiments. MC simulations of $b$-enriched final states such as top-pair decays are used along with real top-enriched and multi-jet events in data. Correction factors depending on jet $p_T$ are applied to scale the MC-measured efficiencies to what is observed in real data~\cite{atlasbtagsf}. The experiments set the working point of their algorithms by specifying reference values of the rate of $b$-tagging light-quark and gluon jets. In CMS the ``loose'' operating point corresponds to a 10\% rate of spurious tags, and is employed in searches where signal efficiency is more important than background rejection; a ``medium'' working point is defined by a 1\% rate on light-quark and gluon jets; and a ``tight'' working point corresponds to 0.1\% rate of spurious tags. To those values correspond typical $b$-tagging efficiencies of $\sim 85\%$, $\sim 70\%$ and $\sim 50\%$, respectively; these numbers are weakly dependent on $p_T$ over a wide range of jet momenta, and typically decrease for very large-$p_T$ jets. In ATLAS the working points of the MV2 algorithm in Run 2 correspond to efficiencies of 85\%, 77\%, 70\% and 60\% for light-jet mistagging rates of 3\%, 0.8\%, 0.3\%, and 0.08\% respectively. These numbers are higher than the Run 1 ones also thanks to the fourth silicon pixel layer constituting part of the ATLAS upgrades for Run 2~\cite{atlasbmva}. $B$-tagging is also crucial in the identification of vector bosons ($Z$ and $H$) and top quarks producing a single wide jet of high $p_T$. Accordingly, both ATLAS and CMS have developed $b$-tagging algorithms suitable to be applied to narrow jets within the core of wide jets~\cite{atlashbbtagger,cmsboostbtagger}.

\subsection{\it Missing Transverse Energy}

Energetic neutrinos emitted from the decay of $W$ and $Z$ bosons are an important signature in diboson physics. The measurement of their longitudinal momentum is prevented at the LHC because of the unknown boost of the reaction center-of-mass in the laboratory frame and of the incomplete hermeticity of the detectors close to the beam axis, but their transverse momentum can be inferred by momentum conservation. This allows the reconstruction of transverse $W$ boson mass in $W \to l \nu$ decays, as well as the detection of a signal of $Z \to \nu \nu$ decays in many topologies. 

The missing transverse momentum in ATLAS is calculated from all reconstructed jets and leptons as the negative of their vectorial sum in the transverse plane~\cite{atlasmet,atlasmet15}. Calorimeter objects in the calculation are calibrated to the electromagnetic energy scale. Muons in the pseudorapidity range $|\eta|<2.7$ are considered separately and corrected for by accounting for the energy deposit they left in the calorimeter, when they are isolated. For non-isolated muons the momentum determined in the muon spectrometer, after the calorimeter energy loss, is added to the calculated missing transverse energy vector. Corrections are also applied to account for soft tracks associated to the primary event vertex. After all calibrations, the resolution in the missing transverse energy is well parametrized by the form $\sigma = k \sqrt{\sum E_T}$ with $k=0.4-0.5$ GeV$^{-0.5}$, where $\sum E_T$ is the sum of transverse energies of detected objects in the event.

Most CMS analyses use a calculation of the missing transverse momentum based on PF objects~\cite{cmsmet}, defined as the negative of the sum of all PF momenta in the transverse plane. A less used definition, called ``calo-MET''~\cite{cmscalomet},  considers all calorimetric deposits and a correction for identified muon candidates. The resolution of missing transverse energy is improved by correcting all identified jets to match the energy of particle-level jets. Pile-up contributions add little unbalancing to the transverse energy flow as they very seldom produce energetic neutrinos; however the fluctuation of neutral components originated from pile-up, combined with calorimeter non-linearities in the response to neutral and charged hadrons, requires a correction that is operated using the vectorial sum of all particles associated to pile-up vertices. Residual azimuthal dependences of the missing transverse energy due to imperfect detector alignment, $\phi$-dependent calibrations, and other small effects is also corrected for. A more recent calculation of the missing transverse energy in CMS, developed to address high-luminosity running conditions~\cite{cmsmet}, employs a multivariate algorithm to reconstruct the missing transverse energy vector.  The algorithm performs a regression with boosted decision trees trained with simulated $Z \to \mu \mu$ decays and $\gamma$ + jet events, to improve the measurement of the missing $E_T$ in high-pile-up conditions. The regression targets the hadronic recoil computed from PF inputs in two successive regression steps. Its application can recover the degradation of PF-based missing transverse energy resolution in high-pile-up conditions. 


\section {Analysis Techniques}
\label{s:exptechniques}

In order to perform data analysis at the LHC in the search for new phenomena, and to improve the experimental sensitivity to processes of interest, a number of techniques have been devised and successively incrementally refined over the course of the past two decades by the ATLAS and CMS collaborations.  In some cases the two collaborations have converged on agreed-upon methodologies, especially for what concerns statistical techniques and in view of a coherent assessment of systematic uncertainties, opening the way to simpler combinations of results. In other cases, and especially for what concerns the treatment of backgrounds and the estimate of selection efficiencies, they have followed independent paths, nonetheless often coming to very similar conclusions on the most effective techniques.

 The broadness of the topic makes it impossible to cover in this review the many different methods, so here we only provide a brief overview of a sample set of methods, choosing among those are most frequently employed to improve the purity of data selections, to estimate the resulting efficiency on signal and background processes, to model the rate and differential distribution of background processes in the selected datasets, to assess the presence of the searched signals using the framework of hypothesis testing and interval estimation, and to assess the impact of sources of systematic uncertainty in the measurements. The reader interested in more detail on any of the specific algorithms is advised to start from the provided references to articles discussing analysis results, and find therein explicit links to the employed tools.

\subsection {\it Selection Efficiency Measurements}

Both CMS~\cite{CMStagandprobe} and ATLAS~\cite{ATLAStagandprobe} employ the well-known ``tag and probe'' technique to estimate the efficiency of lepton identification requirements. These usually, but not exclusively, rely on the large datasets of $Z \to ee $ and $Z \to \mu \mu$ decays acquired with single and double lepton triggers by the experiments. By applying relatively loose selection criteria the purity of these samples is already quite high; they can thus be effectively used to infer the characteristics of real electrons and muons of high momentum. Usually one of the two $Z$-decay products is subjected to tight trigger, identification, and isolation requirements: this is the ``tag'' lepton, whose characteristics ensure the high purity of the overall  selection. To the other lepton candidate, called ``probe'', only a restricted set of requirements are imposed, such that its other characteristics can be studied. The same selection is applied on data and a homologous MC simulation, and the ratio between the efficiency of further requirements on the additional characteristics of the proble lepton is computed in real and simulated $Z$ decays:\par
\begin{equation}
R = \epsilon_{data} / \epsilon_{MC}.
\end{equation}
\noindent
The ratio $R$ and its associated statistical and systematic uncertainties can then be used to correct the signal efficiency of selections applied to cross section measurements and searches that employ electrons and muons of the studied characteristics of the probe.

The tag-and-probe technique is also used to aid in the determination of the jet energy scale and in the assessment of the resulting systematic uncertainty~\cite{tagprobejetscaleCMS,tagprobejetscaleATLAS}. Events with jet pairs, events with a photon recoiling against a jet, and $Z \to ll$ + jet events can all be used to extract information on the calorimeter response to jets as a function of their features -- most frequently, their pseudorapidity -- when the other object is measured in a fiducial region of well-known and already calibrated response. The photon plus jet and $Z$ plus jet events are especially useful for this, given the high energy resolution achievable on the four-momentum of the electroweak boson. 


\subsection {\it Modeling of Signal and Background Processes}
\label{s:modeling}

The study of signal processes of interest in new physics searches generally require MC generators that may incorporate the theory predictions for the corresponding physics models, producing simulated signal events that can be studied to optimize the data selection, evaluate the collection efficiency, and ultimately provide an estimate of the signal that results from the analysis chain. A variety of software tools are available to simulate the resonances considered in this review. Madgraph~\cite{madgraph} and Pythia~\cite{pythia} are among the most commonly used, but different studies employ a variety of other tools, such as Herwig++~\cite{herwig}, Jhugen~\cite{jhugen1,jhugen2}, Sherpa~\cite{sherpa}, and Powheg-box~\cite{powheg-box}; other packages are mentioned in connection to the description of the searches in Sec.~\ref{s:results}. 

All generators require a model of the parton fluxes in the colliding protons. A number of different PDF sets have been proposed to model the initial parton fluxes of proton-proton collisions~\cite{pdfsets,ct14,mmht14,nnpdf30}. A set of guidelines~\cite{pdf4lhc10,butterworthpdf} are commonly followed to estimate PDF-related uncertainties in LHC analyses. They rely on comparisons between a choice of different sets (CT14~\cite{ct14}, MMHT14~\cite{mmht14}, NNPDF3.0~\cite{nnpdf30}) and on specific procedures to account for the variation that different models produce on the studied observables.
The generation output is passed through a complete simulation of the detector response provided by Geant4~\cite{geant4}, or faster, lighter simulations that employ a parametrization of the interaction of final state particles with the detection elements~\cite{fastsim}.

A virtually ubiquitous theme of searches for new phenomena in hadron collider data is the correct modeling of the rate and detailed features of all physics processes that may contribute to the selected samples. In many situations a good model of some of the background processes is offered by control samples of experimental data, or by other data-driven techniques that {\em e.g.} precisely estimate the misidentification rates of the employed algorithms. These techniques are the preferred option for the study of instrumental backgrounds arising when one or more of the final state particle signals are spurious: the detailed modeling by computer simulations of the reconstructed properties of the physics objects can be challenging, hence reliance on parametrizations of the fake rate in real data is warranted in these cases. An example is the signature of an energetic photon, which can be faked by a neutral pion or by an hadronic jet; a study of the fake rate as a function of the observed properties of the identified photon cluster can be performed in independent datasets and provide a satisfactory  model of the effect. Another example is the signature of a $b$-tag in a jet, which is subjected to a contamination from light quarks and gluons; experimental techniques that assess the number of $b$-tags due to those contaminations in a given collected dataset usually rely on parametrizations of the fake rate as a function of the jet properties. 

In particular cases, physics backgrounds from QCD processes may also often prove impossible to precisely model with simulated data, due to the very large cross section of the reactions to be considered and the consequently unmanageable demands posed on computing time. The LHC experiments in those cases obtain an estimate of background shapes and normalization using custom-made methods that employ signal sidebands or other data-driven information, in a variety of ways that depend on the specific features of the final states under examination. In many of the searches for narrow resonances described in this review, however, the modeling of backgrounds with either MC simulations or data-driven techniques would not be precise enough, due to systematic uncertainties of various nature. In those cases reliance is  made on the empirical parametrization of the studied distributions with smooth functions of several parameters. A rigorous {\em a priori} definition of the classes of functions used for the parametrization, as well as a prescription for the recipe selecting the proper number of parameters of the function, is then essential to constrain experimental biases. 

An often used technique for model selection in LHC searches is the Fisher F-test~\cite{ftest}, which compares the goodness of fit of models employing an increasing number of parameters. The models under test must be {\em nested}: $f_1(x;p_1,...,p_n)$ is said to be nested in $f_2(x; p_1,...,p_{n+1})$ if there exists a value of $p_{n+1}$ (usually $p_{n+1}=0$) such that $f_1(x)=f_2(x)$ for all $x$ and any set $p_1,...,p_n$. The F-test is performed by first defining a type-I error rate $\alpha$, typically 5\% or 10\%. Then binned data are fit to the two models in turn, from whose results one computes\par

\begin{equation}
F^* = \frac{\frac{RSS_1-RSS_2}{p_2-p_1}}{\frac{RSS_2}{n-p_2-1}}
\end{equation}

\noindent
where $p_1$ and $p_2$ are the number of parameters of the two functions, $n$ is the number of data points being fit, and $RSS_1$ and $RSS_2$ are the residual sums of squares or the $\chi^2$ values resulting from the two fits. The value $F^*$ computed as above  is then compared to the distribution of possible values arising if the additional degrees of freedom of the larger model do not capture any feature of the data better than the smaller model, except what is expected from the added freedom to accommodate statistical fluctuations in the data. If that is the case, F should distribute as Fisher distribution  with the corresponding number of degrees of freedom, $F(p_2-p_1;n-p_2)$; a tail integral from $F^*$ to infinity of the distribution results in the p-value of the ``null hypothesis'' that the added degrees of freedom reduce the fit $\chi^2$ of the larger model by what is expected from random fluctuations. The comparison of that p-value with $\alpha$ allows the selection of one of the two models. The F-test should be iterated by considering larger and larger models until the null hypothesis is accepted, whence the smaller of the two models tested in the last iteration is selected as the correct one. Searches in the invariant mass spectrum of jet pairs are a good example of the above technique; they traditionally use~\cite{harriskousouris} power-law functions that parametrize the steeply falling spectrum using a ``QCD inspired'' ansatz, such as :\par

\begin{equation}
\frac{d\sigma}{dm} = \frac{p_0}{m^{p_1}} (1-m/\sqrt{s}+p_3 m^2/s)^{p_4} .
\end{equation}

\noindent
Other examples of recent diboson resonance searches that employ functional parametrizations of the studied spectra are given in Sec.~\ref{s:results}. 

It is important to note that since the value of F only refers to a comparison of models and is not a measure of goodness-of-fit, the choice of the model with the F-test does not guarantee that the fit will have desirable properties: the procedure might fail if there is not enough resilience in the initial choice of the family of models under test, so the experiments usually consider enough families of functions to ascertain that the chosen model captures well enough the features of the data, and also require the p-value corresponding to the fit $\chi^2$ for the chosen model to be above a pre-defined threshold. While in typical cases the presence of a narrow signal component in a wide spectrum of fitted data would not affect significantly the conclusions of the model selection procedure, a careful study of the resulting coverage properties of the confidence intervals on the strength of the searched for signal is in all cases required, using pseudo-experiments with varying signal contaminations. The searches for a Higgs boson signal and the subsequent measurements performed in the $\gamma \gamma$ final state by ATLAS~\cite{hggatlas1,hggatlas2} and CMS~\cite{hggcms1,hggcms2} have set a standard for the application of these techniques to resonance searches. 

In alternative to the above methods, the ``Bumphunter'' algorithm first developed by CDF~\cite{bumphunter1,bumphunter2} is used in some ATLAS analyses to scan binned mass distributions in the data, in search for deviations from a fitted background estimate. The algorithm considers every possible set of contiguous bins, of width ranging from two bins to half of the range of the full distribution, and quantifies the statistical significance of localized excesses. 

In all other situations, backgrounds are routinely modeled with MC simulations. A number of software tools for the generation of simulated physics processes are available to experimentalists, and different choices are made by ATLAS and CMS for their different analyses, depending on the process to be described. Next-to-leading-order calculations are by now the standard to model top quark pair-production, where ATLAS employs Powheg-box~\cite{powheg-box} or MC\&NLO~\cite{mcnlo} while CMS relies instead on Madgraph 5~\cite{madgraph}. For electroweak backgrounds like $W+$ jets or $Z+$ jets production ATLAS uses Alpgen~\cite{alpgen} or Sherpa~\cite{sherpa}, while CMS usually employs Madgraph 5; the parton showering and fragmentation of final state partons is implemented by interfacing the output of the generators with Pythia~\cite{pythia}. When the use of matrix-element generators or parton showering models may result in double counting of the partonic configurations, recipes for parton-jet matching are employed~\cite{mlm}. For non-resonant diboson production a number of possible choices exist, including Powheg~\cite{powheg-box}, Sherpa~\cite{sherpa}, Alpgen~\cite{alpgen}, Pythia~\cite{pythia}, or Herwig~\cite{herwig}.  Rare multi-object backgrounds, like top pair production in association with vector bosons, are usually modeled with Madgraph 5. 

\subsection {\it Multivariate Analysis Techniques}

\noindent
Multivariate analysis (MVA) \label{s:mva} techniques have known a booming expansion over the past thirty years, fostered by fast increase in the availability of cheap computing power. The implementation of the most advanced methods has however lagged behind in particle physics, despite the extremely good match that high-energy physics (HEP) \label{s:hep} problems offer to the classical use cases of those methods. So, for instance, while already in 1992 neural networks (NN) \label{s:nn} were tested in tough classification problems such as quark-gluon discrimination at the Tevatron~\cite{NNCDF}, the diffusion of NN for particle searches and measurements is only becoming significant now. The reason for the slowness of large particle physics experiments in adopting MVA techniques was the general preference of sound, well-tested and old-fashioned analysis methods which, while suboptimal, were not perceived as dangerous as MVAs, which could instead be thought to potentially introduce ``unknown unknowns'' in the extraction of the wanted results. Of particular concern of this conservative groupthink was  the fact that the algorithms lend themselves to be used as black boxes, with little or no human intervention; the decision boundaries of MVA algorithms used in classification problems are typically quite hard, if not close to impossible, to fully decrypt and make sense of. The fact that decision trees (see {\em infra}) are on the easier side and neural networks on the harder side along this metric is a reason why the latter have been the latest to be adopted and trusted in HEP applications. 

One generally divides problems that lend themselves to be solved with MVA techniques into classification and regression ones. In the first case the goal is to construct a function of the data features that takes on different, discrete values for recognizable different classes. In the second case, the output of the function must instead predict a continuous variable. Broadly speaking, in searches for new physics processes the crucial issue is the classification of events based on their observed features and kinematics, as any final state can be contributed by a number of different background sources in addition to the searched signal. There exist a number of classification tools that allow for the separation of data in disjunct classes based on their features:  the most widely known are nearest neighbours, support vector machines, decision trees, random forests, and neural networks. These methods belong to the category of statistical learning methods called ``supervised learning'' ones, as before being applied to classify the data under study they rely on a training stage where they learn to distinguish the features of the various event classes. In HEP analysis boosted decision trees (BDT) \label{s:bdt} were the first MVA algorithm to know a rapid diffusion, starting a few years after the turn of the century. While a decision tree is a simple structure where a succession of one-dimensional boundaries on each available feature are defined to best separate two classes of events, BDTs are large collections of such structures randomly created from training data, used collectively to improve the classification accuracy through ``boosting'' techniques designed to optimize a differentiable {\em loss function}.
 The most used implementation of BDTs in HEP was for a long time the one available in the TMVA package~\cite{tmva}, but recently LHC experiments have started to publish results based on other implementations~\cite{scikitlearn,xgboost} which may offer better performance and flexibility. Other methods are also offered within the TMVA package, and have slowly also found their way to scientific publications of physics results; in the last few years neural networks, particularly in their ``deep'' form (DNN, for ``deep neural network'') \label{s:dnn} have started to be used for LHC searches. Frequently used implementations are ones based the Keras library~\cite{keras} or the package called TensorFlow~\cite{tensorflow}. Neural networks are built out of several layers of nodes. In a first input layer the network receives the value of each of the features that define the object under evaluation. Each additional layer contains nodes that combine the inputs using custom activator functions; the nodes output the result to selected nodes of the next layer. A final output layer produces, in the simplest two-class implementation, a binary output. Deep networks have tens, or even hundreds of internal ``hidden'' layers.  Some examples of the use of DNN architectures for LHC results are mentioned in Sec.~\ref{s:results}.


\subsection {\it Statistical Techniques}

\noindent
Over the course of the past two decades the statistical techniques employed by HEP experiments to extract information from their datasets have withstood a considerable evolution  which, at least in part, has been fostered by the availability of larger computing power than  previously used. In fact, crucial problems commonly present in the statistical interpretation of HEP data, such as the accurate incorporation of systematic uncertainties into the calculation of  confidence intervals, constitute quite CPU-intensive tasks. 

\subsubsection {\it Hypothesis Testing, Confidence Intervals, and Signal Extraction}

\noindent
The search for the Higgs boson, first at the Tevatron and then at the LHC, has in particular attracted a renewed interest of the experimental collaborations toward the construction of a carefully designed test statistic suitable for the hypothesis tests needed to probe the existence of that long-sought particle. The need to combine many partly independent searches, each affected by uncertainties related to channel-specific features, as well as by uncertainties partly or fully correlated with those of other channels, has led to intensive studies by an inter-experimental working group~\cite{hcwg}, and to the production of custom routines implementing the agreed-upon analysis procedures~\cite{roostats}. In synergy with those activities, some significant breakthroughs have been achieved in practical statistical tools. One of them enabled the use of approximate formulae for confidence intervals, significantly easing the task of scanning large-dimension parameter spaces in new physics searches~\cite{asimov}. A second study provided a simple method to estimate the significance of a departure of data from the ``null hypothesis'' --the one according to which data are sampled from a background-only model-- in the context of ``simple versus composite'' hypothesis tests, {\em i.e.} in the presence of a nuisance parameter defined only in the ``alternative hypothesis'' --the one defined by a signal-plus-background model~\cite{lee}; the classical example of this setup is the search for a new particle in a mass distribution, where the unknown particle mass plays the role of the nuisance parameter. HEP physicists have dubbed the inflation of the estimated signal significance as ``Look-Elsewhere Effect'' (LEE) \label{s:lee}. The method devised in~\cite{lee} enables a non-CPU-intensive evaluation of the ``global significance'' of an observed effect, accounting for the multiplicity of ways by which the signal may appear in the search. One should however note that only the local significance of a signal search may be well-defined, given the arbitrariness intrinsic in the definition of the range of variability allowed to the nuisance parameter describing the composite alternative hypothesis~\cite{5sigma}.  

The majority of searches for new resonances discussed in this review employ the formalism and test statistic originally designed for the Higgs boson search and discovery at the LHC~\cite{higgs2011}; we find it useful to summarize it below. The starting point is to associate each nuisance parameter affecting the measurement with a prior distribution $\rho(\theta|\tilde{\theta})$, which is assumed to originate from a suitable ``hyper-prior'' $\pi_\theta(\theta)$ modified by the result of some real or imaginary measurement $p(\tilde{\theta}|\theta)$:\par

\begin{equation}
\rho(\theta|\tilde{\theta}) = p(\tilde{\theta}|\theta ) \pi_\theta(\theta).
\end{equation} 

\noindent
One may then easily model the probability density functions $\rho$ as {\em e.g.}, Gaussian, Log-normal, or Poisson distributions without losing the ability to assume flat hyper-priors in the nuisance parameters. The formalism thus allows a purely frequentist treatment of systematic uncertainties, and their study with frequentist tools.

Once a Likelihood function $L(X|\mu, \theta)$ is computed for some data $X$ to study a signal of strength $\mu$ (usually a pure number multiplying the theory-predicted signal rate, such that theory predictions correspond to $\mu=1$), incorporating the required nuisance parameters by multiplication by $p(\tilde{\theta}|\theta)$ factors, the following {\em profile likelihood} test statistic can be defined:\par

\begin{equation}
\tilde{q}_\mu = -2 \log \frac{L(data|\mu, \hat{\theta_\mu})}{L(data|\hat{\mu},\hat{\theta})}. 
\label{eq:qtilde}
\end{equation}

\noindent
In this formula, $\hat{\mu}$  and $\hat{\theta}$ are the parameter values that produce the global maximum of the likelihood function, and $\hat{\mu}$ is constrained to vary in $[0,\mu]$. The value $\hat{\theta_\mu}$ corresponds to the maximum likelihood conditional to setting the signal strength to $\mu$. Given data $X$, use of the test statistic $\tilde{q_\mu}$ allows to extract the p-value $p_b$ under the background-only hypothesis $\mu=0$, as well as the one corresponding to any value of $\mu$ in the signal+background hypothesis, $p_\mu$, respectively as tail integrals of the $\tilde{q_0}$ and $\tilde{q_\mu}$ distributions from the observed values to infinity. With the p-values one may finally construct the variable~\cite{read,junk} \par

\begin{equation}
CL_s(\mu) = \frac{p_\mu}{1-p_b}  .
\end{equation}

\noindent
Given data $X$ and the pre-defined Type-I error rate $\alpha$, the critical region in the range of $\mu$ ({\em i.e.}, the region excluded at Confidence Level (CL) \label{s:cl} $1-\alpha$), corresponds to values where  $CL_s(\mu) < \alpha$. At the LHC it is customary to set $\alpha=0.05$, and to quote 95\% confidence-level limits; the $CL_s$ criterion described above should however be recognized to over-cover by construction~\cite{statpdg}. Its merit is to prevent situations when statistical fluctuations allow an insensitive test to provide stringent exclusion limits. 

In case one observes a very significant signal, {\em i.e.} data $X$ for which $p_{\mu=0}$ is very small~\footnote{When using $p_0$ for significance calculations, one should derive the value of $\tilde{q_\mu}$ in Eq.~\ref{eq:qtilde} by allowing for any value of $\hat{\mu} \ge0$.}, it may result impractical to  quantify the local significance by brute-force calculation of the tail integral of the background-only distribution of the test statistic. One may then use the result of~\cite{asimov}, which have shown how a very good estimate is provided simply by\par

\begin{equation}
p = \frac{1}{2} \left[ 1 - erf \left( \sqrt{q_0^{obs}/2} \right) \right] .
\end{equation}

\noindent
Besides the $CL_s$ criterion cited above, other techniques for deriving confidence intervals have been used in LHC publications; among them, it is necessary to mention Bayesian techniques~\cite{statpdg}, as well as the frequentist ``Unified'' approach of Feldman and Cousins~\cite{feldmancousins}, which allows to solve the inconsistent treatment of other methods when confidence intervals are derived for parameters defined in a subset of the real axis. Despite being more principled in the common case of relevance to this review, when upper limits are set on the (positive-defined) cross section for a unknown new process, the Feldman-Cousins method has not found application there due to the unwillingness of the experiments to quote confidence intervals on the production rate of new physics phenomena which do not include the null value when they do not wish to claim an observed-level significance.

\subsubsection{\it Treatment of Systematic Uncertainties}

\noindent
The treatment of systematic uncertainties in new particle searches by ATLAS and CMS is similar, in the sense that the two experiments usually employ the same analysis techniques to search for signals and set upper limits on signal strength parameters. In fact, those techniques have been largely developed through a common effort, as is the case of the very important methodology and corresponding algorithms originally used for the Higgs boson searches and measurements~\cite{hcwg} but now also commonly employed for Beyond-the-Standard Model (BSM) \label{s:bsm} searches, or tweaked to reflect an agreed-upon consensus and to establish useful standards in inter-experimental working groups. The benefit of this similarity is a much easier combination of the experimental results; {\em e.g.} the consistent treatment of common sources of uncertainty allows for a correct treatment of correlations. Combination procedures have however limited application in the case of new resonance searches, as the collaborations have less interest in combining null results than to collaborate in producing a more credible and robust signal when the chance arises to do so. Unfortunately, the latter situation has not arisen so far, except of course in the case of the Higgs boson discovery and the measurement of its parameters~\cite{higgsmass,higgsspin}. Within each collaboration the combination of search results of different final states of the same production process is more common, and there the common breakdown of the various sources of systematic uncertainty into the independent contributions eases significantly the task of a proper statistical treatment. It is important to note, though, that systematic uncertainties in new physics searches are often sub-dominant with respect to statistical ones; combined with the accelerating speed at which the LHC has been producing data in the past eight years, this reduces the importance and the impact of a careful assessment of all sources of systematics.

The construction of a complete likelihood function which incorporates the effect of a variation of nuisance parameters is the typical {\em modus operandi} of statistical inference at the LHC. Effects that impact the normalization of background processes are usually modeled by Log-normal or Gamma priors, while systematic sources that produce smearings due to the combined effect of many small shifts are of course modeled by Gaussian priors. In frequentist treatments such as the one designed for the Higgs search already mentioned {\em supra}, systematic effects are assessed by expressing the related nuisance parameters as a function of the parameters of interest, evaluating the so-called ``profile likelihood''. This is typically done through the generation of a large number of pseudo-experiments; one important detail is that in the generation of the pseudo-experiments the value of all the nuisance parameters should be varied within their assumed probability density functions, rather than kept fixed. In Bayesian calculations, the treatment of nuisance parameters is operated with a marginalization, by integrating over their PDF~\cite{conway,statpdg}. 

A problem of considerable importance in the search of new particle signals is the impact on statistical inference of the imperfect knowledge of the probability density function generating the background component of the data. When functional forms are used to parametrize the background shape as discussed {\em supra} (Sec.~\ref{s:modeling}), a common procedure consists in fitting various different functions to the data, determining the variation in the parameter of interest that results from different plausible fits. However, the arbitrariness intrinsic in the choice of the functional forms considered in these procedure is difficult to handle. A recent technique to assess in a consistent way the systematic uncertainty in such cases has been proposed in~\cite{wardle}. The idea is to associate to the choice of background model a discrete nuisance parameter, which can be profiled in the likelihood maximization similarly to continuous nuisances. In this case one does not rely on any single well-determined model for the interpretation of the background shape, but rather uses all models together. The technique has been successfully used by CMS measurements of Higgs boson properties based on the $h \to \gamma \gamma$ decay mode~\cite{hggwardle}. An open issue in the application of the discrete profiling method is the choice of penalty term that should be added to the likelihoods employing functions that use a larger number of parameters. This can be operated using the Akaike information criterion~\cite{akaike}, or by using the p-value of the resulting fits.

All searches for new resonances are affected by a systematic uncertainty coming from the imperfect knowledge of the integrated luminosity to which the analyzed datasets correspond. For Run 1, the luminosity uncertainty amounts to 2.8\% in ATLAS~\cite{atlaslumR17,atlaslumR18} and to 2.6\% in CMS~\cite{cmslumR1}; Run 2 searches described in this review are affected by respectively 3.2\%~\cite{atlaslumR18} and 2.3 to 2.5\%~\cite{cmslum2015syst,cmslum2016syst} systematics in the two experiments. In searches that involve final state including hadronic jets, the jet energy scale is usually one of the dominant sources of uncertainty; in dijet mass spectra, in particular, the QCD background falls very steeply with increasing dijet mass, so even a small uncertainty in the energy scale may turn into a large rate uncertainty. In some cases the knowledge of the jet energy resolution also plays a significant role, usually through a modification of the shape of the signal distribution. 

When jets are $b$-tagged or tagged to contain boosted objects, the identification efficiency is one major source of uncertainty in the signal yield. When the final states are purely leptonic other uncertainties may instead dominate, such as the lepton identification and trigger efficiency, as well as uncertainties related to the knowledge of the momentum scale (for muons) or energy scale (for electrons). In all cases the presence of an explicit jet veto in the data selection may affect significantly the precision of signal efficiency, as was recognized recently after a discrepancy in $WW$ production cross section with SM predictions was back-traced to imprecise estimates of soft QCD effects~\cite{wwxsanomaly}. Other theoretical uncertainties  affecting the signal rate include the knowledge of the PDF in the relevant kinematic range of production of the resonance, as well as systematics related to the modeling of initial and final state radiation, and the factorization and renormalization scale.  Sources of uncertainty affecting background estimates produced by MC simulation include the knowledge of the theoretical cross sections, initial and final state radiation modeling, choice of scales and the choice of generator used for the modeling, as well as the choice of the PDF model. When backgrounds are estimated with data-driven techniques many of the above sources do not apply, and the typical leading uncertainty is then related to the finite statistics of the employed control regions (in case those are used for the extraction of fake rates or shape extrapolations), or the above mentioned choice of the functional form used to model the background shape.


\section {Analysis Results}
\label{s:results}

\noindent
In this section results of searches for resonances in diboson final states will be discussed by categorizing them into four classes: resonances decaying to $W$ and $Z$ bosons (Sec.~\ref{s:VV}), resonances decaying to final states including photons (Sec.~\ref{s:gammaX}), resonances decaying to final states including Higgs bosons (Sec.~\ref{s:higgsX}), and resonances decaying to gluon pairs (Sec.~\ref{s:gg}). We introduce this broad range of experimental studies below with a brief overview of the status of our knowledge on diboson resonances acquired by hadronic machines before the turning on of the LHC.
 
\subsection {\it Early Searches }

\subsubsection {\it Diboson Studies at the $Sp\bar{p}S$}
\label{s:spps}

\noindent
The dawn of hadron collider searches for resonances decaying to pairs of elementary bosons can arguably be set to the year 1983, when the UA1 experiment at CERN~\cite{UA1} first reported the observation of the SM electroweak bosons~\cite{rubbia,Wdisc1,Wdisc2,Zdisc1,Zdisc2}. In the following few years of running, the $S p \bar p S$ collider provided UA1 and UA2~\cite{UA2} with proton-antiproton collisions at a center-of-mass energy of 630 GeV. Signatures of pairs of weak bosons ($WW$ and $WZ$) were sought for in final states with multiple leptons or two leptons and two hadronic jets.  Indeed, in 1987 the UA1 experiment reported observing two events containing the signature of a high-transverse-momentum $W$ boson decaying to lepton-neutrino pair accompanied by two recoiling hadronic jets from a luminosity of 0.7 $pb^{-1}$ collected between 1982 and 1985 at center-of-mass energies of 546 and 630 GeV~\cite{ua1wjj}. The unavailability of reliable calculations of QCD radiation processes in those years allowed for speculations on the possible origin of those events from the decay of a heavier resonance; it later became clear that better simulations of $W+$ jets processes needed to be used to interpret the data, and that the occurrence of energetic gluon radiation was not as rare as until then thought in electroweak production processes.

The reference model for heavy $W^\pm$ and $V^0$ resonances used by UA1 and UA2 in their searches assumed that the couplings of those particles were a carbon copy of the ones of SM bosons. Using that working hypothesis the experiments set 90\% lower mass limits  at 220 GeV for a heavy $W^\pm$ and at 173 
 GeV for a heavy $V^0$ particle~\cite{ua2wprime,ua1wprime,ua1wprime0}. 
UA1 and UA2 also studied their jet final states in search of new phenomena, and produced qualitative limits on the existence of resonances decaying to jet pairs. Although the gluon-gluon final state was never specifically considered in those analyses, we mention them here for completeness~\cite{ua1jj0,ua1jj1,ua1jj2,ua2jj1,ua2jj2}.

\subsubsection {\it Tevatron Results Summary}
\label{s:tev}

\noindent
The Tevatron collider operated from October 1985 to September 2011 at the US Fermi National Accelerator Laboratory, colliding protons and antiprotons at center-of-mass energies of 1.6 (in 1985, with negligible collected data), then 1.8 (from 1987 to 1996, with a total of $\sim 120 pb^{-1}$ delivered to the experiments), and finally 1.96 TeV (from 2001 onwards, with a total delivered luminosity of $\sim 11 fb^{-1}$).  The CDF and DZERO experiments mined the resulting datasets across four decades, producing a number of searches for diboson-decaying resonances in all potentially sensitive final states. For space limitations we omit to give those results the attention they would deserve, and only offer in Table~\ref{t:tevatron} a sample list of the most recent Tevatron Run 2 results obtained by CDF and DZERO on resonances predicted by models foreseeing their decay to diboson final states.

\begin{table}[h!]
\begin{center}
\caption {\em Summary of recent Tevatron search limits (at 95\% CL) for new resonances in diboson final states. The $\dagger$ symbol indicates that the mass limit includes information from other decay modes. Limits on gravitons are reported for $\tilde{k}=0.1$. 
The reported DZERO limits on the technirho mass refer to $m(\pi_{T})=100$ GeV; the CDF ones to a large range of $m(\pi_{T})$ values. The DZERO searches for $Z\gamma$ final states and other ones for which no mass limits are quoted set limits on the cross section as a function of resonance mass; the CDF $H^{\pm}$ search set limits on the production cross section in the $(m_H^{\pm},m_H^0)$ plane. } 

\begin{tabular}{|l|c|c|c|c|c|c|c|}
\hline
Model             & Boson  & Final   & Exclusion   & Exp. & Int. Lum.      & Year & Ref. \\
or particle    &     pair    & state   & (GeV) &         & ($fb^{-1}$) &        & \\
\hline
$\rho_{T}$  & $W \pi_{T}$ & $l +bj $ & $180-250$ & CDF & 1.9 & 2010&~\cite{ref241toback} \\
\hline
RS graviton  & $ZZ$ & $ll q \bar{q}$,$lll'l'$ & $<491$ & CDF &2.5-2.9 & 2011 &~\cite{cdfzz2011} \\
\hline
RS graviton     & $WW$ &$l \nu + j$  & $ <607$      & CDF & 2.9 & 2010 &~\cite{ref216toback} \\
 $Z'$           &$WW$ & $l \nu +j$ &   $247-544$   & CDF & 2.9 & 2010 &~\cite{ref216toback} \\
$W'$           & $WZ$ & $l \nu +j$ &  $285-516$      & CDF & 2.9 & 2010 &~\cite{ref216toback} \\
\hline
Fermiophobic Higgs & $\gamma \gamma$ & - & $<106$ & CDF & 3.0 & 2009 &~\cite{cdffermiophobic} \\
\hline
RS graviton & $\gamma \gamma$ & - & $<1111^{\dagger}$ & CDF & 5.4-5.7 & 2011 &~\cite{cdfrsgraviton2011}\\
\hline
 RS graviton      & $ZZ$ & $ll+X$    & $300-1000$      & CDF & 6.0 & 2012 &~\cite{cdfzz} \\
\hline
2HDM $H^{\pm}$ & $W H$ & $l \nu b \bar{b}$ & -  & CDF & 8.7 & 2012 &~\cite{2hdmcdf}\\ 
\hline
Chromophilic $Z'$ & $gg$ & - & - & CDF & 8.7 & 2012 & ~\cite{cdfzprimegg}\\
\hline
$X$   & $Z \gamma$ & $ll \gamma$ & - & DZERO & 0.3 & 2006 &~\cite{dzerozgamma1}\\
\hline
$\rho_{T}/\omega_{T}$ & $W \pi_{T}$& $e+j$ & $190-210^{*}$ & DZERO & 0.39 & 2007 &~\cite{ref240toback}\\
\hline
$X$   & $Z \gamma$ & $ll \gamma$ & - & DZERO & 1.1 & 2008 &~\cite{dzerozgamma2}\\
\hline
RS graviton & $\gamma \gamma$ & - & - & DZERO & 1.05 & 2010 &~\cite{d0rsgraviton2009} \\ 
\hline
$W'$                & $WZ$ & $lll$      & $188-520$ & DZERO & 4.1 & 2010 &~\cite{ref218toback} \\
\hline
 SSM $W'$      & $WZ$ & $l+jets$ & $<690$ & DZERO & 5.4 & 2010 &~\cite{dzeroWWWZ}\\
RS graviton      & $ZZ$ & $l+jets$  & $<754$ & DZERO & 5.4 & 2010 &~\cite{dzeroWWWZ}\\
\hline
RS graviton & $\gamma \gamma$ & - & $560-1050$ & DZERO & 5.4 & 2010 &~\cite{d0rsgraviton2010}\\
\hline
Fermiophobic Higgs & $\gamma \gamma$ & - & $<112.9$ & DZERO & 8.2 & 2011 &~\cite{d0fermiophobic} \\
\hline
\end{tabular}
\end{center}
\label{t:tevatron}
\end{table}

\noindent
In general, all Run 2 analyses by CDF and DZERO reported results in good agreement with SM predictions, with the exception of a localized excess of four $ZZ \to lll'l'$ events clustering at the mass of 325 GeV, observed by CDF in a dataset corresponding to  $6 fb^{-1}$ of integrated luminosity at $1.96$ TeV~\cite{cdfzz}. Accounting for the 4-body mass resolution, the local significance of the observation was estimated at $3.4\sigma$, but the combination of other observed leptonic decay channels of $ZZ$ production did not confirm the effect.

\subsection{\it LHC Searches for Resonances Decaying to Weak Boson Pairs}
\label{s:VV}

\noindent
Final states including pairs of weak vector bosons ($WW$, $WZ$, $ZZ$) have been a favourite target of the ATLAS and CMS collaboration since their start of investigation of multi-TeV collisions in 2010. In addition to the higher energy than the Tevatron collider, the LHC experiments soon reaped benefits from the newly developed sub-jet vector-boson tagging techniques in high-$p_T$ wide jets, which added a new dimension to the quest. Due to the large branching fraction of $W$ and $Z$ bosons to jet pairs, the addition of mixed topologies (e.g. one hadronic and one leptonic decay) or fully-hadronic final states significantly enhanced the experimental sensitivity to new high-mass particles. In this section we review the published results by grouping them by the bosonic final state they considered.

\subsubsection{\it Searches for $WW$ Resonances}
\label{s:ww}

The $WW$ final state has been considered by ATLAS and CMS searches in all main decay signatures of $W$ bosons except the $\tau \nu$ one, which however still contributes to the $e \nu$ and $\mu \nu$ signatures through leptonic $\tau$ decays, as it does in other LHC searches of leptonic final states of diboson-decaying resonances reported in this article. Table~\ref{t:WW} provides a summary of the searches and the relevant references. $W$ boson pairs can be produced by gravitons predicted by the Randall-Sundrum model; ATLAS and CMS have considered both the bulk graviton model and the original version of the Randall-Sundrum theory, with choices of the $\tilde{k}$ parameter varying from 0.1 to 1.0.  
Additional heavy Higgs bosons predicted by 2HDM extensions of the SM have also been the target of these studies. Fermiophobic Higgs bosons have an enhanced branching fraction to the $WW$ final state; they have also been considered in some publications. Some works also considered heavy scalar singlets and triplets, and generic models with heavy narrow resonances.

\begin{table}[h!]
\begin{center}
\caption{\em Summary of ATLAS and CMS searches that considered the $WW$ decay of heavy particles predicted by new physics theories. Many of the publications study multiple final states of vector boson pairs together; only theories accessible with the $WW$ signature are listed in the  second-to-rightmost column of this table. The year quoted in the fourth column corresponds to the one of publication of the arXiv preprint.}

\begin{tabular}{|l|c|c|c|c|c|c|}
\hline
Expt. & CM energy & Int. Lum. & Year & Decay modes & Considered models & Ref. \\
                   & (TeV)        & ($fb^{-1}$) &    &                     &                               &                \\
\hline
ATLAS         & 7 & 4.7 & 2012 & $l \nu l \nu$ & RS $G^*$, $G^*_{bulk}$ &~\cite{wwatlas1} \\ 
\hline
ATLAS         & 7 & 4.7 & 2013 & $l \nu q \bar{q}'$ & RS $G^*$, $G^*_{bulk}$ &~\cite{wwatlas2} \\ 
\hline
ATLAS & 8 & 20.3 & 2015 & $l \nu q \bar{q}'$ & RS $G^*_{bulk}$ &~\cite{wwatlas3} \\ 
\hline
ATLAS & 8 & 20.3 & 2015 & $q \bar{q}' q \bar{q}'$ & RS $G^*_{bulk}$ &~\cite{wwatlas4} \\ 
\hline
ATLAS & 8 & 20.3 & 2015 & $l \nu q \bar{q}'$, $l \nu l \nu$ & Heavy neutral Higgs &~\cite{wwatlas5}\\ 
\hline
ATLAS & 8 & 20.3 & 2015 & Combination & RS $G^*_{bulk}$ &~\cite{wwatlas6}\\ 
\hline
ATLAS & 13 & 3.2 & 2016 & $q \bar{q}' l \nu$, $q \bar{q}' q \bar{q}'$ & Scalar singlets, HVT bosons, $G^*_{bulk}$ &~\cite{wwatlas7} \\ 
\hline
CMS & 7 & 5 & 2012 & Combination & Fermiophobic Higgs bosons&~\cite{wwcms1}\\  
\hline
CMS & 7 & 5 & 2012 & $q \bar{q}' q \bar{q}'$ & RS $G^*$ &~\cite{wwcms2} \\ 
\hline
CMS & 8 & 19.7 & 2014 & $q \bar{q}' q \bar{q}'$ & RS $G^*$ and $G^*_{bulk}$ &~\cite{wwcms3} \\ 
\hline
CMS & 8 & 19.7 & 2014 & Combination & RS $G^*$ and $G^*_{bulk}$ &~\cite{wwcms4}\\ 
\hline
CMS & 13 & 2.7 & 2016 & $l \nu q \bar{q}'$, $q \bar{q}' q \bar{q}'$ & RS $G^*_{bulk}$, HVT bosons &~\cite{wwcms5}\\ 
\hline 
CMS & 8+13 & 19.7+2.7 & 2017 & Combination & HVT Singlets and triplets, RS $G^*_{bulk}$ &~\cite{wwcms6}\\
\hline
\end{tabular}
\end{center}
\label{t:WW}
\end{table}

\begin{figure}[h!]
\begin{center}
\begin{minipage}{0.57\linewidth}
\includegraphics[scale=0.4]{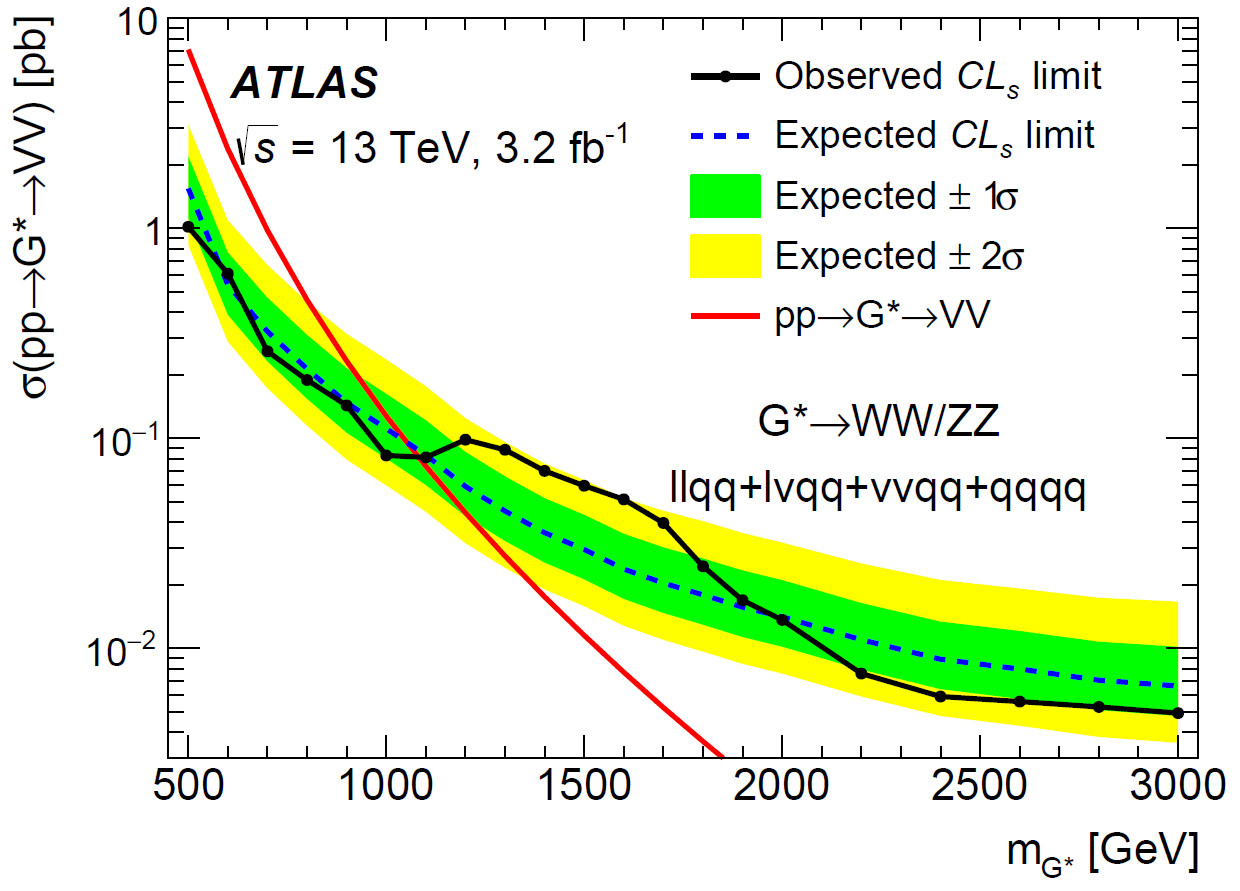}
\end{minipage}
\begin{minipage}{0.42\linewidth}
\includegraphics[scale=0.4]{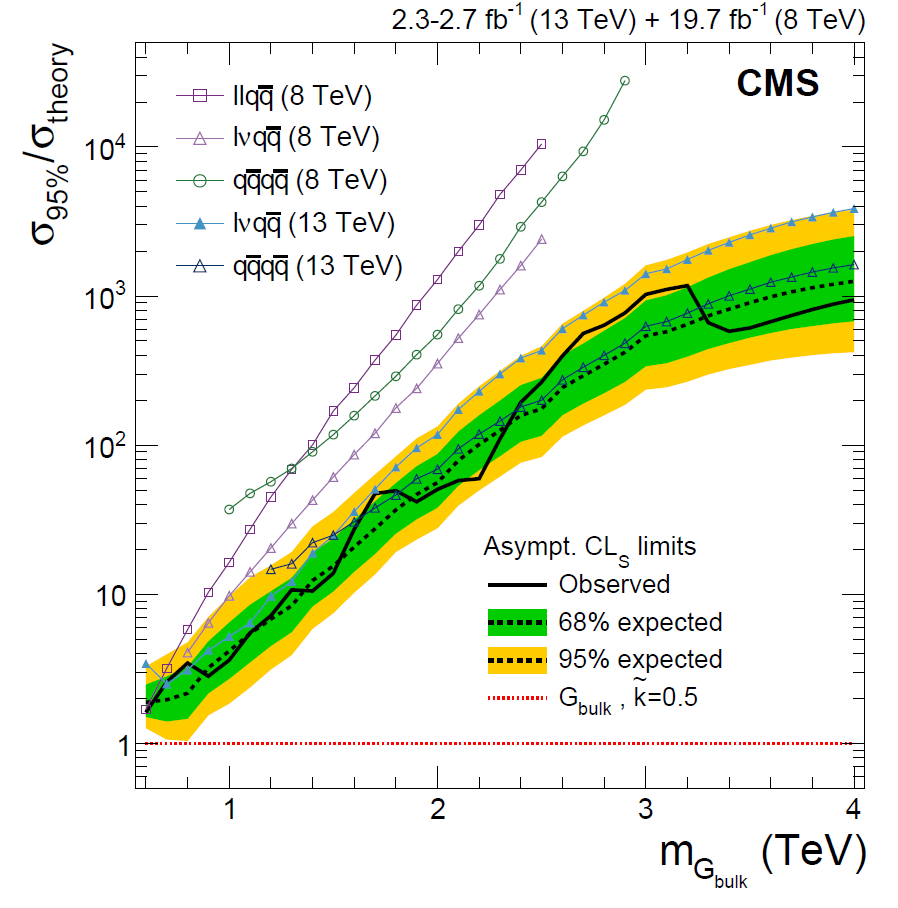}
\end{minipage}
\caption{\em 95\% CL upper limits on the cross section of production of bulk gravitons obtained by the most stringent searches: the ATLAS result reported in~\cite{wwatlas7} (left), and the CMS result reported in~\cite{wwcms6}.}
\label{f:wwgbulk}
\end{center}
\end{figure}

The ATLAS collaboration set 95\% CL limits on the mass of the RS graviton $G^*$ of~\cite{RandallSundrum1} with the $WW$ decay mode for $\tilde{k}=0.1$. Using 7-TeV collisions data the lower limit was set at 1.23 TeV~\cite{wwatlas1} by studying the $l \nu l' \nu$ final state, and to 0.94 TeV~\cite{wwatlas2} by searching for $l \nu q \bar{q}'$ decays.  
CMS produced a search for RS gravitons along with several other resonances in~\cite{wwcms2}, by considering the all-hadronic final state of vector boson pairs; they could not place mass limits due to the high-mass region investigated in the study, which considered sub-jet tagging of high-$p_T$ wide jets. A limit at 1.2 TeV was later reported in~\cite{wwcms3} by considering again the all-hadronic final state in the full statistics of 8-TeV data collected in 2012.

On the bulk graviton $G^*_{bulk}$ for  $\tilde{k}=1.0$, lower mass limits were set by ATLAS at 0.84 TeV~\cite{wwatlas1}, at 0.71 TeV in~\cite{wwatlas2}, 0.76 TeV in~\cite{wwatlas3}. No lower mass limits were quoted by the 2015 ATLAS study~\cite{wwatlas4}, which targeted $G^*_{bulk}$ resonances of very high mass using high-$p_T$ wide jets. The study of~\cite{wwatlas6} reported a limit at 0.81 TeV by combining the $WW$ and the $ZZ$ signatures. Finally, the latest published ATLAS study~\cite{wwatlas7} extended the mass limit to $m_{G^*_{bulk}}>1.1$ TeV (see Fig.~\ref{f:wwgbulk}, left), which is the highest lower limit to date in the considered model.
CMS considered the parameter values  $\tilde{k}=0.2$ and 0.5 for its searches reported in~\cite{wwcms3,wwcms4,wwcms5,wwcms6}. Its tightest limit on the bulk graviton cross section, reported in~\cite{wwcms6}, is shown in Fig.~\ref{f:wwgbulk} (right).
It must be mentioned that the theoretical calculation of the cross section for production of the $G^*_{bulk}$ graviton was revised in~\cite{gbulkrevised1,gbulkrevised2}. The limits obtained in early searches such as those of~\cite{wwatlas1, wwatlas2, zzcms1} cannot therefore be compared directly with those of more recent results which employed the improved calculation.

The HVT model was considered by one ATLAS analysis~\cite{wwatlas7} and two CMS analyses~\cite{wwcms5,wwcms6} studying the $WW$ final state. ATLAS considered both model A and model B with narrow widths for the heavy bosons. The lower limit at 95\% C.L on the mass of the resonance for the latter was set at 2.6 TeV by combining $WZ$ and $WW$ final states. In the first analysis targeting heavy-vector triplet bosons~\cite{wwcms5}, CMS set 95\% CL limits on model A and B by considering singlet and triplet resonances, also combining the results of the $WW$ channel with those of the search in the $WZ$ final state. In the singlet case $W'$ bosons were excluded below 2.0 or 2.2 TeV  and $Z'$ bosons with masses below 1.6 and 1.7 TeV for models A and B, respectively, and in the triplet hypothesis, spin-1 resonances were excluded with masses below 2.3 and 2.4 TeV, respectively. The second CMS study~\cite{wwcms6} slightly tightened the above bounds to 2.3 TeV for a singlet $W'$, 2.2 and 2.3 TeV for a singlet $Z'$ in model A and model B respectively, and 2.4 TeV for heavy triplets in both models, again by combining different channels and including signatures other than the $WW$ one.

Information on other models could also be obtained by some of the LHC searches performed in the $WW$ final state. Heavy Higgs bosons were considered in the ATLAS search described in~\cite{wwatlas5}. That study considered three different scenarios: heavy neutrals with a narrow natural width, ones with an intermediate width $\Gamma_H$ in the range $0.2<\Gamma_H/\Gamma_{H_{SM}}<0.8$, and ones with a lineshape predicted by a model called ``complex pole scheme'' \label{s:cps} (CPS)~\cite{cps1,cps2,cps3}, and width corresponding to the one predicted for the SM Higgs. The CPS produces a precise determination of the lineshape of Higgs bosons; for masses below 400 GeV the difference with a Breit-Wigner distribution is however negligible. 95\% CL upper limits on the production cross section of a heavy Higgs state multiplied by its branching fraction to $WW$ pairs were separately obtained for the gluon-gluon fusion production and the vector-boson fusion production hypotheses, for Higgs masses ranging from 300 to 1500 GeV. 
In the ATLAS search on Run 2 data already mentioned above~\cite{wwatlas7}, a heavy CP-even scalar singlet model was also considered~\cite{franceschini}. Scalar resonances were excluded for masses below 2.65 TeV.  In the CMS search reported in~\cite{wwcms1}, a fermiophobic Higgs boson was considered. In fermiophobic Higgs models, the Higgs boson decays to vector bosons are strongly enhanced, making the searches more sensitive. CMS combined the result of searches in the $WW$, $ZZ$, and $\gamma \gamma$ final states, including a total of 32 independent sub-channels. The mass region between 110 and 300 GeV was investigated, and an exclusion of the mass range 110-194 GeV obtained at 95\% CL.

As one would expect from statistical considerations alone, some of the above searches resulted in localized excesses in the studied mass distributions. The statistical significance of the largest signals was assessed at the level of two standard deviations after accounting for the trials factors resulting from the multiplicity of independent locations where fluctuations could arise in the considered distributions. Most of the excesses were inconsistent across different channels and did not raise particular interest, but one excess found in the invariant mass of dijet-tagged wide-jet pairs by ATLAS~\cite{wwatlas4} was considered more seriously by the theoretical physics community, with suggested explanations ranging from bulk gravitons~\cite{wwatlas4}, to composite particles~\cite{atlasjjbump2a,atlasjjbump2b,atlasjjbump2c,atlasjjbump2d,atlasjjbump2e,atlasjjbump2f}, to $W'$ or $Z'$ bosons~\cite{atlasjjbump3a,atlasjjbump3b,atlasjjbump3c,atlasjjbump3d,atlasjjbump3e,atlasjjbump3f,atlasjjbump3g,atlasjjbump3h,atlasjjbump3i,atlasjjbump3j,atlasjjbump3k,atlasjjbump3l}. All three selections of $WW$, $WZ$, and $ZZ$ event candidates showed hints of a signal in the 2 TeV mass region, with local significances evaluated respectively at the level of $2.6\sigma$, $3.4 \sigma$, and $2.9 \sigma$. About 20\% of the events included in the three selections were common to all, and the global significance of the combined dataset was estimated at $2.5 \sigma$. The observed effect did not increase when the dijet-tagged search was combined with results of analysis of the leptonic decay modes in~\cite{wwatlas6}. CMS saw a very mild excess of events at slightly lower mass (1.9 TeV) in a search in the same all-hadronic final state~\cite{wwcms3}, but the effect (a local significance of about $1.5 \sigma$) was found most prominent in the ``single-V-tagged'' selection targeting excited quark decays $q^* \to qV$. Eventually, later searches did not confirm the effect, which was finally assessed as a statistical fluctuation.

\begin{figure}[h!]
\begin{center}
\begin{minipage}{0.49\linewidth}
\includegraphics[scale=0.45]{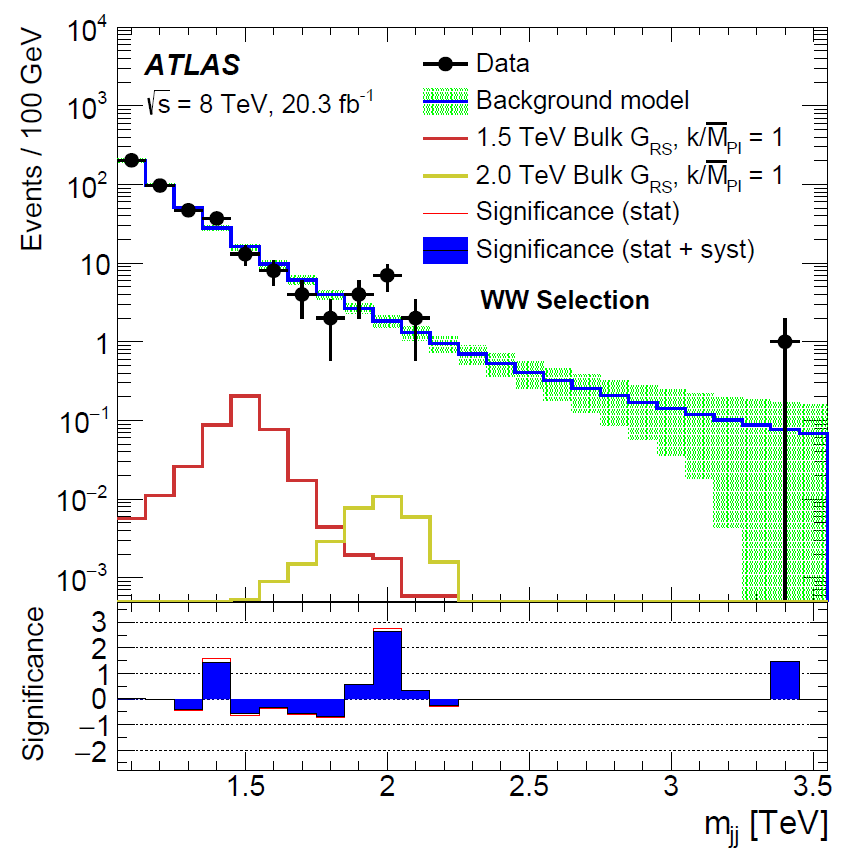}
\end{minipage}
\begin{minipage}{0.49\linewidth}
\includegraphics[scale=0.46]{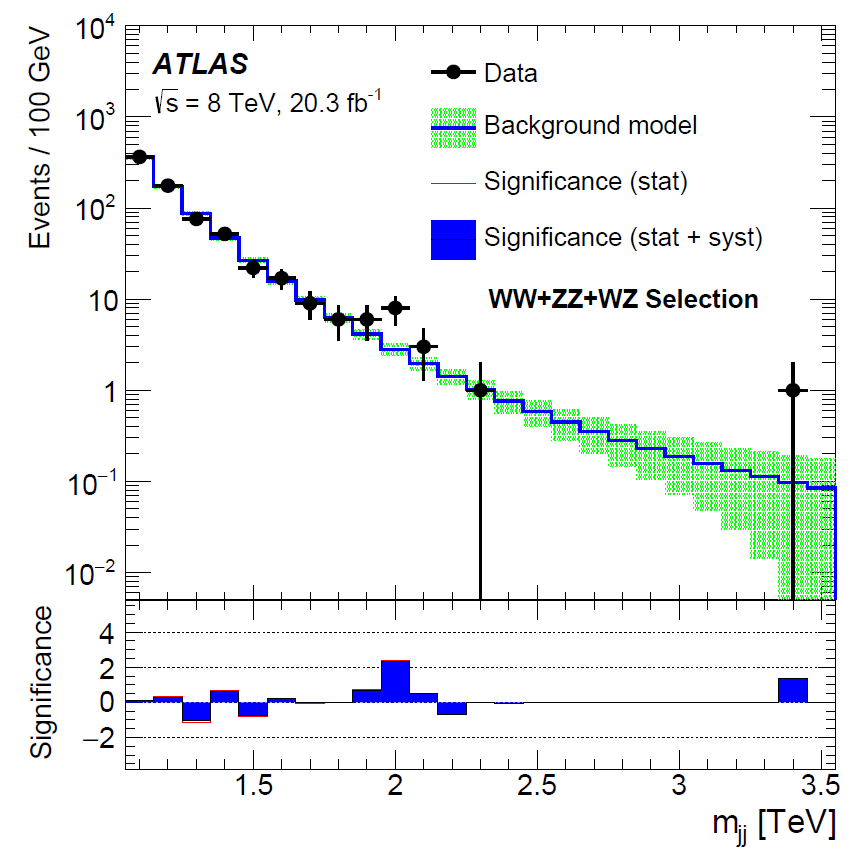}
\end{minipage}
\caption{\em Left: reconstructed mass distribution of $WW$ event candidates in the ATLAS search reported in~\cite{wwatlas4}. The data (points with vertical uncertainty bars) exhibit a small excess over background predictions (curve) in the 2 TeV region. Simulated signals of RS bulk gravitons are shown for comparison for $G^*$ masses of 1.5 and 2.0 TeV; the lower panel shows the background-subtracted distribution of the data, in units improperly labeled as ``significance'' which are derived by a division of the residuals by their statistical (full histogram) or statistical plus systematic uncertainty (empty histogram). Right: combination of data in the $WW$, $WZ$ and $ZZ$ event selections in the same search. }
\label{f:atlasjjbump}
\end{center}
\end{figure}

\subsubsection{\it Searches for $WZ$ Resonances}
\label{s:wz}

\noindent
The methodology of searches for $WZ$ decay candidates is very similar, when not coincident, to that of $WW$ searches reported above or $ZZ$ searches dealt with in Section~\ref{s:zz} below; the considered final state categories also have usually a significant overlap (in leptonic decays) or are identical (in boosted dijet topologies). In fact, many of the results mentioned here were reported by ATLAS and CMS in the same publications cited in the previous section. For the benefit of easier consultation we however consider those publications anew below. In Table~\ref{t:WZ} we list the articles that considered signatures of $WZ$ decaying resonances, and list the targeted models.

\begin{table}[h!]
\begin{center}
\caption {\em Summary of ATLAS and CMS searches that considered the $WZ$ final state of decays of heavy particles predicted by new physics theories. Many of the publications study multiple final states of vector boson pairs together; only theories probed by the $WZ$ signatures are listed in the second-to-rightmost column of this table. The year quoted in the fourth column corresponds to the one of publication of the arXiv preprint. }
\begin{tabular}{|l|c|c|c|c|c|c|}
\hline
Expt. & CM energy & Int. Lum. & Year & Decay modes & Considered models & Ref. \\
                   & (TeV)        & ($fb^{-1}$) &    &                     &                               &                \\
\hline
ATLAS & 7 & 1.02 & 2012 & $l \nu l' l'$ & EGM $W'$, LSTC $\rho_T$ &~\cite{wzatlas1} \\ 
\hline
ATLAS & 7 & 4.7 & 2013 & $l \nu q \bar{q}$ & EGM $W'$ &~\cite{wwatlas2} \\ 
\hline
ATLAS & 8 & 20.3 & 2014 & $l \nu l' l'$ & EGM $W'$, HVT &~\cite{wzatlas2} \\ 
\hline
ATLAS & 8 & 20.3 & 2014 & $l l q \bar{q}'$ & EGM $W'$ &~\cite{wzatlas3} \\ 
\hline
ATLAS & 8& 20.3 & 2015 & $l l q \bar{q}'$ & $H^{\pm}$, HTM &~\cite{wzatlas4} \\ 
\hline
ATLAS & 8 & 20.3 & 2015 & $l \nu q \bar{q}$ & EGM $W'$ &~\cite{wwatlas3} \\ 
\hline
ATLAS & 8 & 20.3 & 2015 & $q \bar{q} q \bar{q}'$& EGM $W'$ &~\cite{wwatlas4} \\ 
\hline
ATLAS & 8 & 20.3 & 2015 & Combination & EGM $W'$ &~\cite{wwatlas6} \\ 
\hline
ATLAS & 13 & 3.2 & 2016 & Combination & HVT bosons &~\cite{wwatlas7} \\ 
\hline
CMS    & 7 & 5.0 & 2012 & $l \nu l' l'$ & SSM $W'$, $\rho_T$ &~\cite{wzcms1}  \\ 
\hline
CMS    & 7 & 5.0 & 2012 & $l l q \bar{q}'$, $\nu \nu q \bar{q}'$ &  SSM $W'$  &~\cite{wzcms2}  \\ 
\hline
CMS    & 7 & 5.0 & 2012 & $q \bar{q} q \bar{q}'$& $W'$ &~\cite{wwcms2}  \\ 
\hline
CMS    & 8 & 19.7 & 2014 & $q \bar{q} q \bar{q}'$ & $W'$ &~\cite{wwcms3}  \\ 
\hline
CMS    & 8 & 19.7 & 2014 & Combination & model-independent &~\cite{wwcms4}  \\ 
\hline
CMS    & 8 & 19.5 & 2014 & $l \nu l' l'$& EGM $W'$ &~\cite{wzcms3}  \\ 
\hline
CMS    & 13 & 2.7 & 2016 & Combination & HVT and model-independent &~\cite{wwcms5}  \\ 
\hline
CMS    & 13 & 15.2 & 2017 & $l \nu l' l'$ & $H^{\pm}$ &~\cite{wzcms4}  \\ 
\hline
CMS    & 8+13 & 19.7+2.7 & 2017 & Combination & HVT &~\cite{wwcms6} \\ 
\hline
\end{tabular}
\end{center}
\label{t:WZ}
\end{table}

\begin{figure}[h!]
\begin{center}
\begin{minipage}{0.49\linewidth}
\includegraphics[scale=0.34]{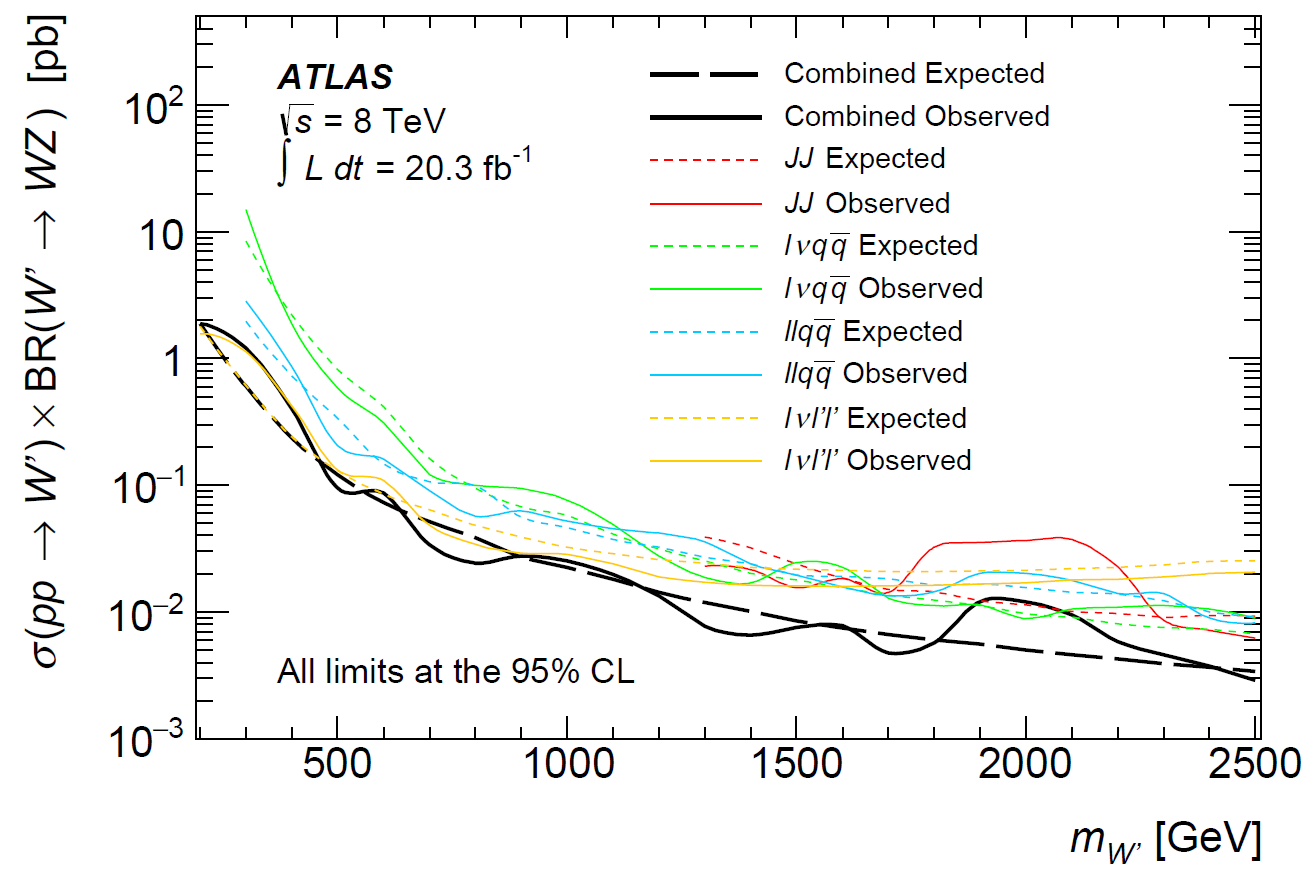}
\end{minipage}
\begin{minipage}{0.49\linewidth}
\includegraphics[scale=0.35]{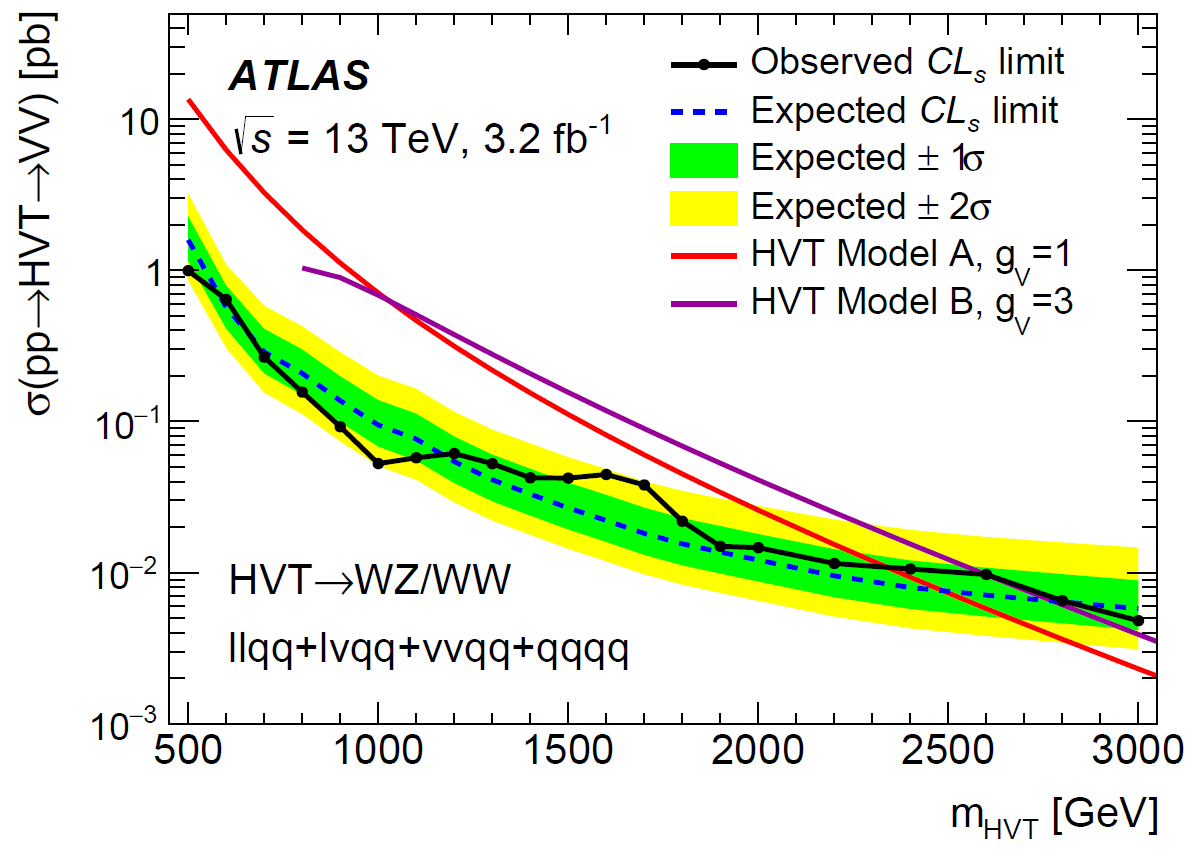}
\end{minipage}
\caption{\em Left: 95\% CL upper limits on the production cross section of a $W'$ boson times its branching fraction to $WZ$ pairs obtained by ATLAS analyses in different final states, and their combination~\cite{wwatlas6}. Right: 95\% CL upper limits obtained by ATLAS on the heavy vector boson production cross section as predicted in the HVT models A and B~\cite{wwatlas7}.}
\label{f:egmatlas}
\end{center}
\end{figure}

As is evident in Table~\ref{t:WZ}, the extended gauge model was considered as a benchmark by most ATLAS searches for resonances decaying to $WZ$ pairs, and often also by CMS. ATLAS excluded at 95\% CL $W'$ bosons with masses below 0.76 TeV in an early search in the leptonic final state~\cite{wzatlas1}, then below 0.95 TeV in a mixed final state with $W \to l \nu$ decays and non-boosted jets~\cite{wwatlas2}, and below 1.52 TeV in an 8-TeV search again using the leptonic signature~\cite{wzatlas2}. A 1.59 TeV limit was soon therafter obtained in the $l l q \bar{q}'$ final state in the same dataset~\cite{wzatlas3}, using both resolved and merged jets from $W$ decay; the same methodology, but employing $W \to l \nu$ decays and $Z \to q \bar{q}$ ones, obtained a slightly lower exclusion at 1.49 TeV~\cite{wwatlas3}. In the fully-hadronic search~\cite{wwatlas4}, as mentioned {\em supra} (Sec.~\ref{s:ww}) ATLAS saw a $3.4\sigma$ excess of events in the 2-TeV region of their $WZ$ selection (see Fig.~\ref{f:atlasjjbump}), and excluded $W'$ bosons in the 1.3-1.5 TeV range. The strongest limit came from the combination of search channels reported in~\cite{wwatlas6}, which excluded EGM $W'$ bosons below 1.81 TeV (see Fig.~\ref{f:egmatlas}, left). The CMS experiment considered explicitly the EGM model in only one search in 8-TeV data~\cite{wzcms3}, which used a fully leptonic final state and obtained a 95\% CL lower limit at 1.55 TeV on $W'$ bosons. In the other cases the SSM was considered, with lower mass limits set at 1.14 TeV in a fully leptonic search in 7-TeV data~\cite{wzcms1}, and at 0.94 TeV in a semileptonic search with $W$ bosons decaying into a wide jet~\cite{wzcms2}. The search reported in~\cite{wwcms2} targeted high $W'$ masses with V-tagged dijet events, and obtained tight cross section limits but did not explicitly exclude a mass range for the resonance; the same final state was considered in data collected at 8 TeV, when a lower mass limit for a heavy $W'$ boson was set at 1.7 TeV~\cite{wwcms3}. 

\begin{figure}[h!]
\begin{center}
\begin{minipage}{0.49\linewidth}
\includegraphics[scale=0.45]{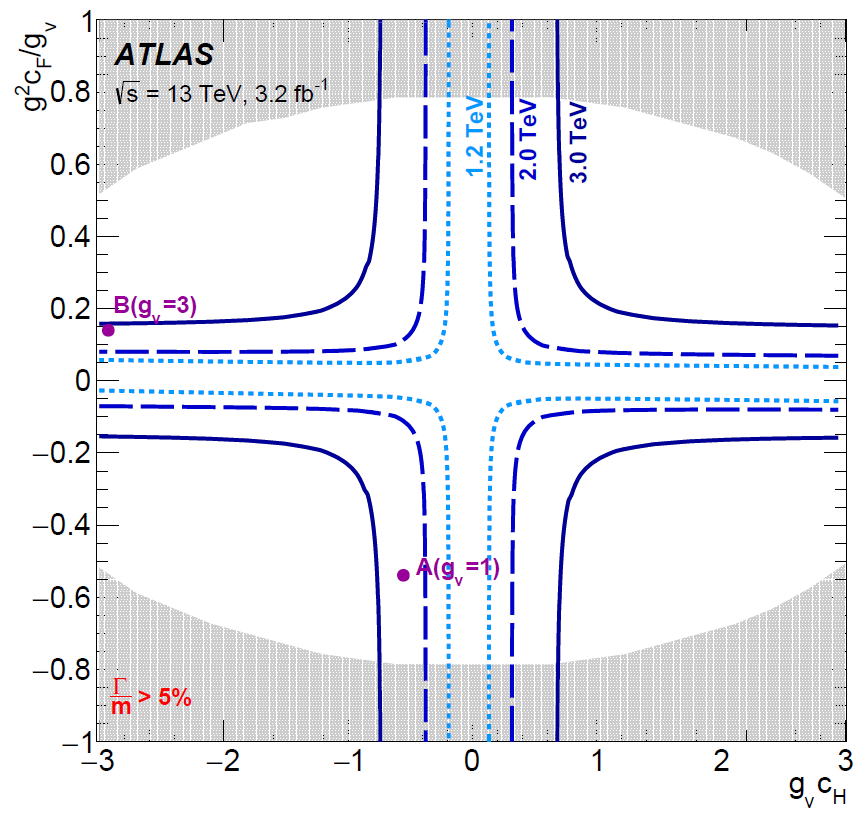} 
\end{minipage}
\begin{minipage}{0.49\linewidth}
\includegraphics[scale=0.45]{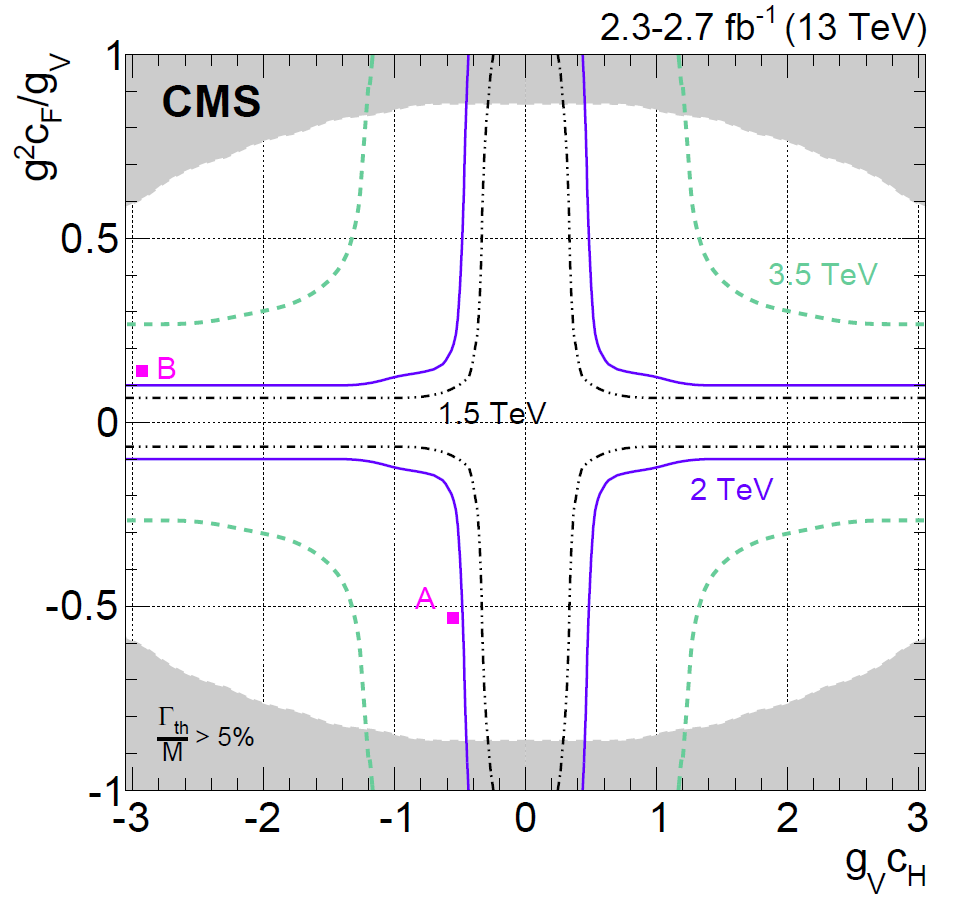} 
\end{minipage}
\caption{\em 95\% CL exclusion contours in the HVT parameter space obtained by the ATLAS analysis reported in~\cite{wwatlas7} for resonances of mass 1.2 TeV, 2 TeV and 3 TeV. Points of parameter space in the hatched area correspond to widths of the resonance above 5\% of its mass, which were not directly tested by the search. Right: the corresponding result obtained by the CMS analysis~\cite{wwcms5}, which considered resonance masses of 1.5, 2.0, and 3.5 TeV. Again, the hatched area corresponds to $\Gamma_{W'}/m_{W'}$ values above 0.05, which would modify the signal shape and change the experimental results.}
\label{f:WZhvt}
\end{center}
\end{figure}

Several searches that targeted the $WZ$ signature also set limits on the HVT. ATLAS obtained 95\% CL limits at 1.49 TeV for model A and 1.56 TeV for model B using the fully leptonic final state~\cite{wzatlas2}; those limits got significantly extended to 2.35 and 2.6 TeV, respectively, by a combined search employing 13-TeV data collected in 2015~\cite{wwatlas7}.
CMS excluded $W'$ singlets in the HVT below masses of 2.0 (2.2) TeV for model A (B) in a search combining the fully-hadronic and the semileptonic final states in 2015 data~\cite{wwcms5}. A more recent combination using both 8- and 13-TeV datasets allowed to extend the limits to 2.3 (2.4) TeV, respectively~\cite{wwcms6}. Results of the ATLAS and CMS searches in 13-TeV data collected in 2015, in the parameter space of the HVT, are compared in Fig.~\ref{f:WZhvt}.

Other investigated models which could contribute to the signature of resonant $WZ$ pairs were the low-scale technicolour model of Eitchen and Lane~\cite{lowscaletech2}. This was considered {\em e.g.} by the already mentioned early ATLAS study of 7-TeV collisions~\cite{wzatlas1}, which set lower 95\% CL limits on a $\rho_T$ technimeson at 0.46-0.47 TeV, depending on the mass relation of the $\rho_T$ with the $a_T$ technimeson. CMS also reported searching for a $\rho_T$ technimeson in an early fully-leptonic search~\cite{wzcms1}, setting more stringent mass limits in the 0.17-0.69 TeV range under the assumption that $m_{\pi_T}=0.75 m_{\rho_T}-25 $GeV; the excluded range was of 0.18-0.94 TeV if $m_{\rho_T}<m_{\pi_T}+m_W$.

The ATLAS search reported in~\cite{wzatlas4} considered the Georgi-Machacek Higgs triplet model \label{s:htm} (HTM)~\cite{LHphenom1}, when a charged Higgs boson is produced by vector-boson fusion mechanisms. They studied 8-TeV collisions events with a $Z$ boson decaying to $ee$ or $\mu \mu$ pairs and the $W$ boson decaying into a wide jet with substructure, and reported exclusions of the mass range 0.24-0.7 TeV with the parameter $s_H=1$, using a cross section calculated in~\cite{bolzoni,loganzaro}. CMS also considered the Georgi-Machacek model in their search for vector-boson-fusion $H^{\pm}$ production~\cite{wzcms4}, with a final state containing the fully leptonic $WZ$ decay signature  and two VBF-tagging jets of large combined mass and pseudorapidity separation. They reported 95\% CL exclusion of regions of the plane defined by the $H^{\pm}$ mass and the $s_H$ parameter.

Finally, CMS extracted model-independent limits on the cross section of a $W'$ boson in two of their publications on the $WZ$ final state~\cite{wwcms4,wwcms5}, where they produced estimates of the reconstruction efficiency for the considered bosons, allowing for a simple interpretation of the upper limits on the cross section for generic models.

\subsubsection{\it Searches for $ZZ$ Resonances}
\label{s:zz}

\noindent
$ZZ$ resonances can be sought for in their decays to four charged leptons, two leptons and two neutrinos, a wide jet with substructure and two leptons (both charged or neutral), or to a fully-hadronic dijet topology using $Z$-tagging substructure in both jets. The final states with four charged leptons are the cleanest, with backgrounds mainly due to $Z$ pair production and other minor electroweak contributions; however they have a very small branching ratio overall, so they do not offer themselves to searches for rare processes at high invariant mass; that role is better covered by the final states including highly-boosted jets with substructure. ATLAS and CMS have pursued all the available signatures, combining results and obtaining stringent limits on resonances that may decay to $Z$ boson pairs. Table~\ref{t:ZZ} contains a summary of the relevant searches, the considered new physics models, and the references.

\begin{table}[h!]
\begin{center}
\caption {\em Summary of ATLAS and CMS searches that considered the $ZZ$ final state of decays of heavy particles predicted by new physics theories. Many of the publications study multiple final states of vector boson pairs together; only theories probed by the $ZZ$ signatures are listed in the second-to-rightmost column of this table. The year quoted in the fourth column corresponds to the one of publication of the arXiv preprint.}

\begin{tabular}{|l|c|c|c|c|c|c|}
\hline
Expt. & CM energy & Int. Lum. & Year & Decay modes & Considered models & Ref. \\
                   & (TeV)        & ($fb^{-1}$) &    &                     &                               &                \\
\hline
ATLAS & 7 &  1.02 & 2012 &  $ll q \bar{q}$, $ll l'l'$ & RS $G^*$ and model-independent &~\cite{zzatlas1} \\ 
\hline
ATLAS & 8 & 20.3 & 2014 & $ll q \bar{q}$ & RS $G^*_{bulk}$ &~\cite{wzatlas3} \\ 
\hline
ATLAS & 8 & 20.3 & 2015 & $q \bar{q} q' \bar{q}'$ & RS $G^*_{bulk}$ &~\cite{wwatlas4} \\ 
\hline
ATLAS & 8 & 20.3 & 2015 &  Combination & Heavy Higgs &~\cite{zzatlas2} \\ 
\hline
ATLAS & 8 & 20.3 & 2015 &  Combination & RS $G^*_{bulk}$ &~\cite{wwatlas6} \\ 
\hline
ATLAS & 13 & 3.2 & 2016 &  Combination & RS $G^*_{bulk}$ &~\cite{wwatlas7} \\ 
\hline
CMS & 7 &  5.1 & 2012 &  Combination & Fermiophobic Higgs &~\cite{wwcms1} \\ 
\hline
CMS & 7 &  4.9 & 2012 & $l l q \bar{q}$ & RS $G^*$ &~\cite{zzcms1} \\ 
\hline
CMS & 7 & 5.0  & 2012 & $l l q \bar{q}$, $\nu \nu q \bar{q}$ & RS $G^*$ &~\cite{wzcms2} \\ 
\hline
CMS & 7 &  4.9 & 2012 & $q \bar{q} q' \bar{q}'$ & RS $G^{*}$ &~\cite{wwcms2} \\ 
\hline
CMS & 8 & 19.7 & 2012 & $q \bar{q} q' \bar{q}'$ & RS $G^*_{bulk}$ &~\cite{wwcms4} \\ 
\hline
CMS & 13 & 2.7 & 2016 & $q \bar{q} q' \bar{q}'$ & RS $G^{*}_{bulk}$ and model-independent &~\cite{wwcms5} \\ 
\hline
CMS & 8-13 & 19.7+2.7 & 2017 & Combination & RS $G^{*}_{bulk}$ &~\cite{wwcms6} \\ 
\hline
\end{tabular}
\end{center}
\label{t:ZZ}
\end{table}

The reference for most LHC searches in the $ZZ$ final state is the Randall-Sundrum model with bulk gravitons. The 95\% CL exclusion limits obtained by ATLAS on the rate of $ZZ$ resonances, interpreted as upper bounds on the production cross section for $G^*_{bulk}$ states, were turned into constraints of the graviton mass. The range 325-845 GeV was excluded by the early analysis reported in~\cite{zzatlas1}; the bound $m_G^*<740$ (540) GeV was obtained by the search for $ZZ \to ll q \bar{q}$ candidates in the full 8-TeV dataset~\cite{wzatlas3} by considering $\tilde{k}=1.0$ (0.5), respectively. In the same data sample, the analysis of the fully-hadronic final state~\cite{wwatlas4} resulted in tight cross section bounds for masses above 1.3 TeV but no mass range exclusion. Finally, the already reported combination of multiple searches in 8-TeV data~\cite{wwatlas6} resulted in a lower limit at 810 GeV; a combination of results of searches in the 2015 dataset of 13-TeV collisions~\cite{wwatlas7} produced a 95\% CL exclusion of masses below 1.1 TeV, as already reported {\em supra} (Sec.~\ref{s:ww}). 

CMS sought for RS gravitons in six analyses; the first three, using data corresponding to an integrated luminosity of about 5 $fb^{-1}$ of 7-TeV collisions~\footnote{Small differences in the integrated luminosity used by these analyses are due to the varying requirements of sub-detector operativity for the different considered final states.}, considered the original RS model; the latter three considered the bulk graviton version. In the first study~\cite{zzcms1}, based on 7-TeV data, an exclusion at 95\% CL of gravitons with mass $m_{G^*}<945$ GeV ($<720$ GeV and in the range 760-850 GeV) was obtained by assuming $\tilde{k}=0.1$ (0.05), respectively. In the second one~\cite{wzcms2}, limits were presented for both a LO calculation and the inclusion of NLO effects~\cite{kumar1,kumar2}; the latter significantly increases the predicted cross section. For $\tilde{k}=0.05$ masses below 803 (879) GeV were excluded at NLO (LO). As already reported {\em supra} (Sec.~\ref{s:ww}), the fully-hadronic search of~\cite{wwcms2} could not exclude any mass range, although it could obtain stringent cross-section limits for very high graviton masses. For bulk gravitons, the 8-TeV and 13-TeV analyses reported in~\cite{wwcms4,wwcms5,wwcms6} were already mentioned in Sec.~\ref{s:ww}.  

The $ZZ$ signature was used by ATLAS to search for a heavy Higgs boson in~\cite{zzatlas2}. The results of searches in $ll q \bar{q}$, $\nu \nu q \bar{q}$, $ll l' l'$, $l l \nu \nu$ decay modes were used to set exclusion limits at 95\% CL separately for gluon-gluon and VBF production modes of the heavy scalar, in the narrow-width approximation. An interpretation in the space of model parameters of Type-I and Type-II 2HDM was also offered for $m_H=200$ GeV. CMS combined the $ZZ$ search for a fermiophobic Higgs boson in 7-TeV collisions data with the results of $WW$ and $\gamma \gamma$ searches, as reported above (Sec.~\ref{s:ww})~\cite{wwcms1}. 
Finally, both ATLAS~\cite{zzatlas1} and CMS~\cite{wwcms5} presented model-independent limits using the $ZZ$ signature. ATLAS used four-lepton events to set 95\% CL bounds on the fiducial cross section $\sigma(pp \to X \to XX)<0.92 pb$ for any source of events with a mass $m_{ZZ}$ above 300 GeV.


\subsection{\it LHC Searches for Resonances Decaying to Final States Including Photons ($W\gamma$, $Z\gamma$, $\gamma \gamma$)}
\label{s:gammaX}

\subsubsection{\it Searches for $W \gamma$ and $Z \gamma$ Resonances}

\noindent
The final state of a weak vector boson and a photon may arise in the decay of spin-1 technimesons, as well as decays of spin-0 or spin-2 heavy composite scalar particles originated by a new strong interaction. The experimental signature of these processes is often quite clean when leptonic final states of the vector boson are selected, as backgrounds are then mostly due to initial state radiation of hard photons accompanying the $W$ or $Z$ boson. However, for high-mass searches hadronic $W$ or $Z$ decay are also considered, as their larger branching fraction makes them profitable where the production cross section dies out. Jet substructure techniques have been used with success in this case to fight the large QCD backgrounds.

A study by the ATLAS collaboration in 4.6 $fb^{-1}$ of 7-TeV proton-proton collisions~\cite{vgammaatlas1} considered events with photon candidates and leptonic decays of the $W$ and $Z$ bosons to measure cross sections for $W \gamma$ and $Z \gamma$ production, limit anomalous gauge couplings, and search for technicolour resonances. For the new resonance search they based their signal model on the ``technicolour strawman model'' of~\cite{lanemrenna}, implemented in the Pythia package~\cite{pythia}. They considered neutral and charged technimesons, with four technicolours and a mass splitting between the new particles given by $m_{a_T}=1.1 m_{\rho_T}=1.1 m_{\omega_T}$ and $m_{\rho_T}-m_{\pi_T}=m_W$, in accordance to searches for low-scale technicolour~\cite{lowscaletech1,lowscaletech2} previously performed in $WZ$ and dilepton decay modes. With these choices for the relationship among the technihadron masses, all searched for resonances have widths narrower than the experimental resolution. The signals of $\omega_T \to Z \gamma$ and $a_T \to W \gamma$ were parametrized by a Crystal Ball function, with a Gaussian component added to the $\omega_T$ shape. The fits of observed mass distributions to SM backgrounds and the signal templates produced cross section limits which, once compared to the considered technicolour model, were turned into 95\% CL lower limits on the mass of $\omega_T$ at 494 GeV, and on the mass of  $a_T$ at 703 GeV.
The same final state was later considered for a search in the full 2012 dataset of 8-TeV collisions, corresponding to a luminosity of 20.3 $fb^{-1}$~\cite{vgammaatlas2}. The $a_T$ technihadron decay to $W \gamma$ was excluded at 95\% CL for masses below 960 GeV, and the $\omega_T$ was excluded for all masses in the range 200-890 GeV (except for the region 700-750 GeV), by searching for its $Z \gamma$ decay. Two benchmarks of the composite scalar model~\cite{compscalar1,compscalar2} were also studied with the $Z \gamma$ final state: a first one with couplings suggested by a ``naive dimensional analysis''~\cite{nda1} and a second with larger couplings to electroweak gauge bosons~\cite{nda2,nda3,compscalar2}. No exclusion was obtained in the first benchmark, while in the second one masses of the heavy composite scalar were excluded in most of the range below 1180 GeV.

\begin{figure}[h!]
\begin{center}
\begin{minipage}{0.49\linewidth}
\includegraphics[scale=0.35]{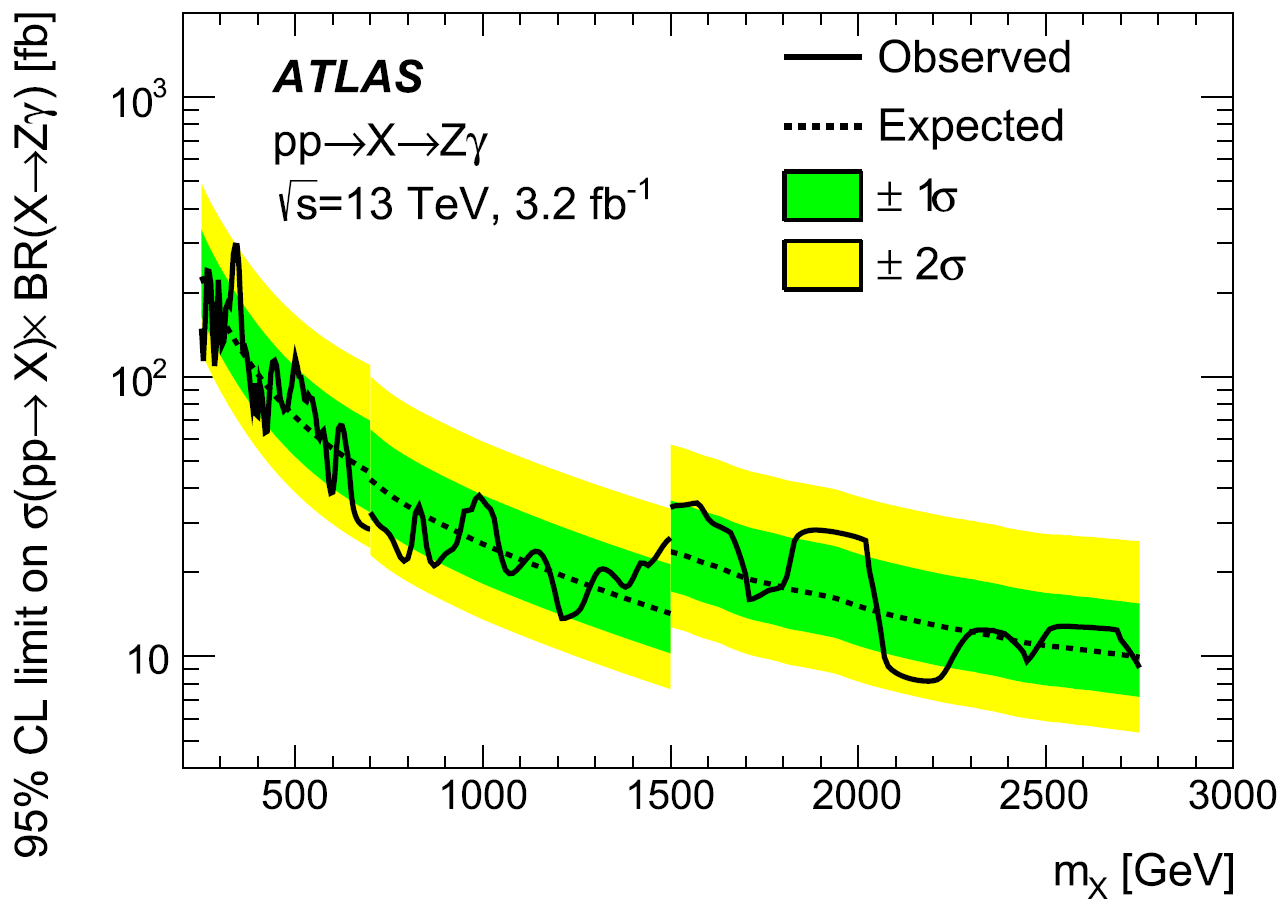}
\end{minipage}
\begin{minipage}{0.49\linewidth}
\includegraphics[scale=0.4]{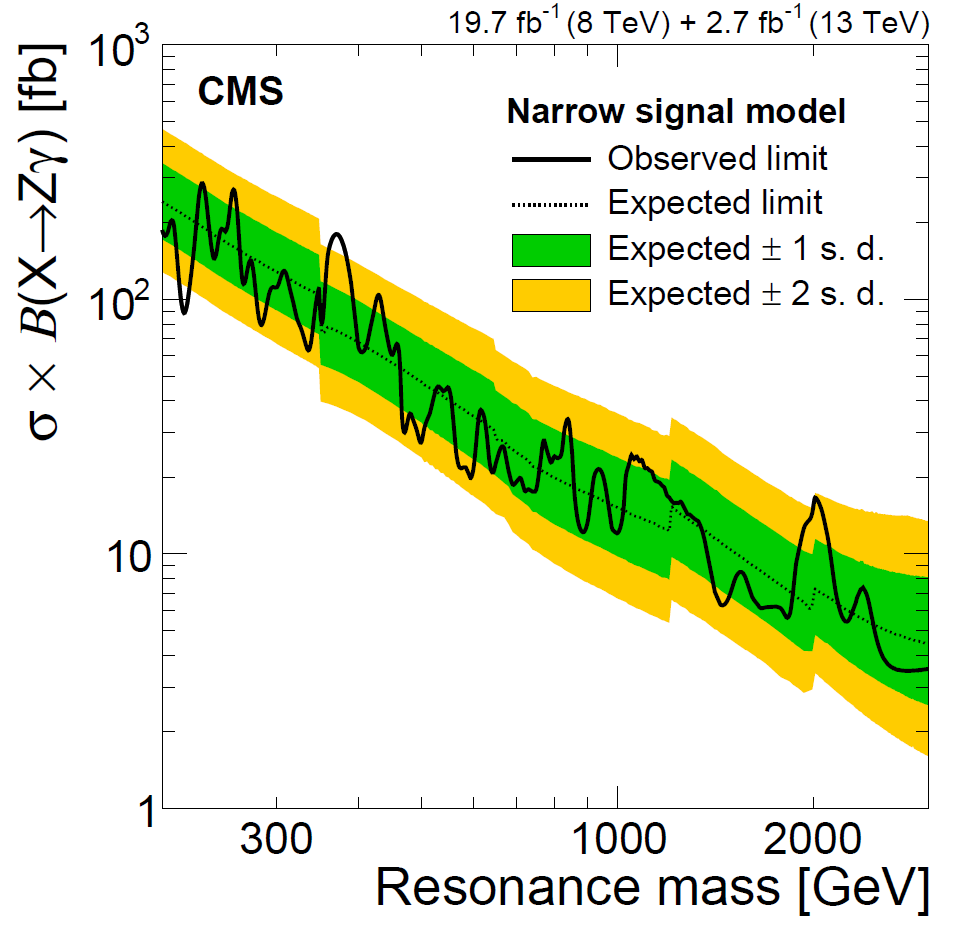}
\end{minipage}
\caption{\em 95\% CL upper limits on the cross section of production of a heavy resonance times its branching fraction to $Z \gamma$ pairs as a function of resonance mass $m_X$. Left: limits obtained by the ATLAS analysis reported in~\cite{zgammaatlas2}. Right: upper limits on resonant $Z \gamma$ production versus resonance mass, obtained by the CMS analysis reported in~\cite{cmszgamma2}, which combined results with those of a previous search~\cite{cmszgamma1}.}
\label{f:zgammaatlas2}
\end{center}
\end{figure}

In a more recent search, performed on 3.2 $fb^{-1}$ of integrated luminosity from the 13-TeV 2015 run~\cite{zgammaatlas2}, ATLAS considered only the $Z \gamma$ decay mode of heavy resonances, but sought for both $ee$ and $\mu \mu$ decays of the $Z$ as well as hadronic decays, employing jet substructure techniques. The data were selected to contain a photon with $p_T>250$ GeV and a $Z$ boson; the $Z \gamma$ invariant mass was required to be above 200 GeV in the leptonic final state, and above 640 GeV in the hadronic $Z$ case. The extracted upper limits at 95\% CL on resonances decaying to the $Z \gamma$ final state are shown in Fig.~\ref{f:zgammaatlas2}.
Finally, a search for $Z \gamma$ resonances was performed by ATLAS in a larger dataset corresponding to a luminosity of 36.1 $fb^{-1}$ collected at 13 TeV~\cite{zgammaatlas3}. The search considered decays of the $Z$ boson to electron-positron or muon pairs, and spanned the range 200-2500 GeV in the mass of the decaying resonance. No significant departure from the background model was observed, and 95\% CL limits were extracted on the cross section of the new particle times its branching fraction to a $Z \gamma$ pair. The limits varied between 88 $fb$ and 2.8 $fb$ for spin-0 resonance masses in the 250 -2400 GeV range in the hypothesis of gluon fusion production, and from 117 (94) $fb$ to 3.7 (2.3) $fb$ for a spin-2 resonance produced via gluon-gluon fusion (respectively quark-antiquark annihilation).

\noindent
CMS also searched for narrow $X \to Z \gamma$ resonances in two recent publications. The first one was based on the study of a total luminosity of 19.7 $fb^{-1}$ of 8-TeV collisions and 2.7 $fb^{-1}$ of 13-TeV collisions, acquired respectively in 2012 and 2015, and considered final states produced by $Z$ boson decays to electron-positron or muon pairs~\cite{cmszgamma1}.  By collecting events with photon candidates associated with a high-$p_T$ lepton pair allowed to select data mostly constituted by electroweak $Z$ boson production with hard photon radiation from the initial state, with some residual contamination from $Z$+jet production with misidentified photons. The background was collectively parametrized by the sum of three exponential functions. No significant signal was observed in the investigated spectra, and 95\% CL limits were extracted on the product of cross section times branching fraction of the heavy resonance to $Z \gamma$ in the range from 280 to 20 $fb$ for values of $m_X$ in the 200-2000 GeV range. The most recent result~\cite{cmszgamma2} was based on the analysis of the same integrated luminosity of the previously mentioned search, but considered decays of the $Z$ boson into both light-quark pairs and $b \bar{b}$ quark pairs, by using jet substructure techniques to identify the $Z$ decay products inside jets reconstructed with a $R=0.8$ cone.  The search investigated the mass range from 650 GeV to 3 TeV, using photon plus wide-jet events selected by requiring photons of $p_T>170 (180)$ GeV and jets with $p_T>170 (200)$ GeV for the analysis of 8-TeV (13-TeV) data. Two non-overlapping event categories were considered to be sensitive separately to generic hadronic $Z$ decays or ones to $b \bar{b}$ pairs. The $Z \to b \bar{b}$ selection required that at least one $b$-tag were found by the CSV algorithm above its medium-purity working point in one of  the narrow-cone jets identified within the wide jets, when the other jet also passed the loose working point of the CSV algorithm. The search did not find any excess over a smooth background parametrization in the two event classes. Limits on the  resonance cross section times its branching fraction into $Z \gamma$ final state were obtained by combining the results with those of the previous search for $Z \gamma$ using leptonic $Z$ decays~\cite{cmszgamma1}. The 95\% CL limits ranged from 300 $fb$ to 3.5 $fb$ for resonance masses in the interval 200-3000 GeV (see Fig.~\ref{f:zgammaatlas2}).

\subsubsection{\it Searches for $\gamma \gamma$ Resonances}

\noindent
In general, the Landau-Yang theorem prevents the decay of a spin-1 resonance to photon pairs~\cite{novtovv1,novtovv2}; that final state provides access to spin-0 states such as those predicted by extended Higgs sectors, or to spin-2 states like gravitons of extra dimensions theories. Higher-spin parents are also possible, although not considered fashionable by existing literature. The fact that the 125 GeV Higgs boson displays couplings resembling those predicted in the SM means that in the context of 2HDMs the favoured parameter space region is close to the alignment limit $\alpha=\beta$. In such a scenario, decays to photon pairs of the $A$ and $H$ boson are favoured.

Over the course of the past few years the CMS and ATLAS experiments produced several searches for $X \to \gamma \gamma$ decays in Run 1 and Run 2 data~\cite{cmsgg1,cmsgg2,cmsgg3,cmsgg4,atlasgg1,atlasgg2,atlasgg3,atlasgg4,atlasgg5}. While the 7-and 8-TeV searches produced null results in the investigated mass spectra, the first analysis of 2015 data by ATLAS reported a peaking excess of events at a mass of about 750 GeV, with a local statistical significance exceeding three standard deviations. In conjunction, the CMS experiment also reported a similar effect at the same diphoton mass, with a lower significance. The reported anomalies, jointly made public in an end-of-the-year seminar at CERN in December 2015, caused an explosion of theoretical interest in potential new physics models that could produce the observed effect. Over the course of the following six months more than 600 articles were submitted to the Cornell arXiv~\cite{600articles}. That enthusiasm was however soon dampened by the sobering results of searches that included the larger 2016 datasets, which basically killed the earlier signal, ascertaining its statistical nature of an upward fluctuation. A part of the large body of produced literature in that short time span is still very useful as it offers creative new mechanisms for new physics beyond the SM, and it provides a demonstration of how diboson signatures may be a versatile investigative tool.

The first ATLAS search for resonances decaying into photon pairs~\cite{atlasgg1} was produced on a dataset corresponding to a luminosity of 2.1 $fb^{-1}$ of 7-TeV collisions acquired in 2011. No hints of a signal were seen, and 95\% CL lower limits on the $M_S$ parameter of the ADD model were set in the range between 2.27 and 3.53 TeV, depending on the considered number of extra dimensions. An exclusion of masses below 0.79 and 1.85 TeV, again at 95\% CL, was also obtained for RS gravitons with the dimensionless parameter $\tilde{k}=0.01$ and $0.1$, respectively. The search was later repeated using the full 4.9 $fb^{-1}$ of integrated luminosity collected in 2011~\cite{atlasgg2}, when limits in the ADD scenarios were increased to the range 2.52-3.92 TeV and limits on RS gravitons to 1.00 (2.06) TeV for $\tilde{k}=0.01 (0.1)$. The next result was reported based on the full luminosity of $20.3 fb^{-1}$ of 8-TeV collisions acquired in 2012~\cite{atlasgg3}; that search could also not detect any significant departure from the background model in the searched range $409 < m_{\gamma \gamma}<3000$ GeV; 95\% CL lower limits on the mass of RS gravitons were set at 1.41 and 2.66 TeV for the two values of $\tilde{k}$ cited above. 

\begin{figure}[h!]
\begin{center}
\begin{minipage}{0.49\linewidth}
\includegraphics[scale=0.4]{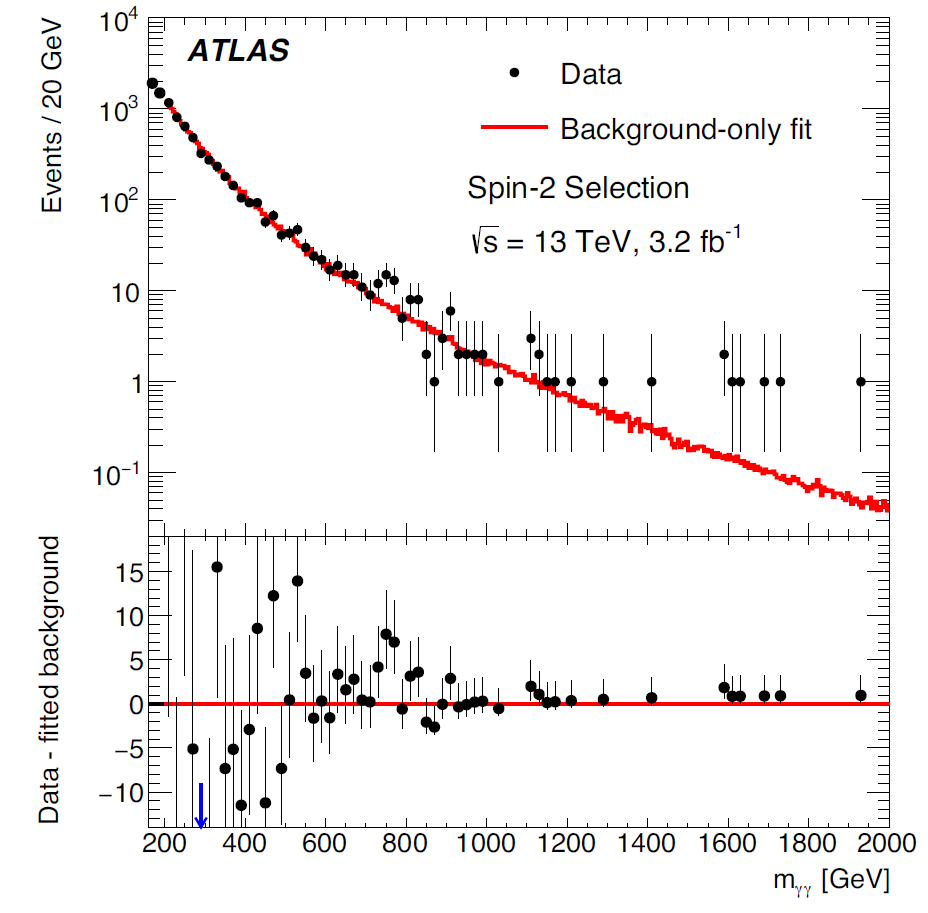}
\end{minipage}
\begin{minipage}{0.49\linewidth}
\includegraphics[scale=0.4]{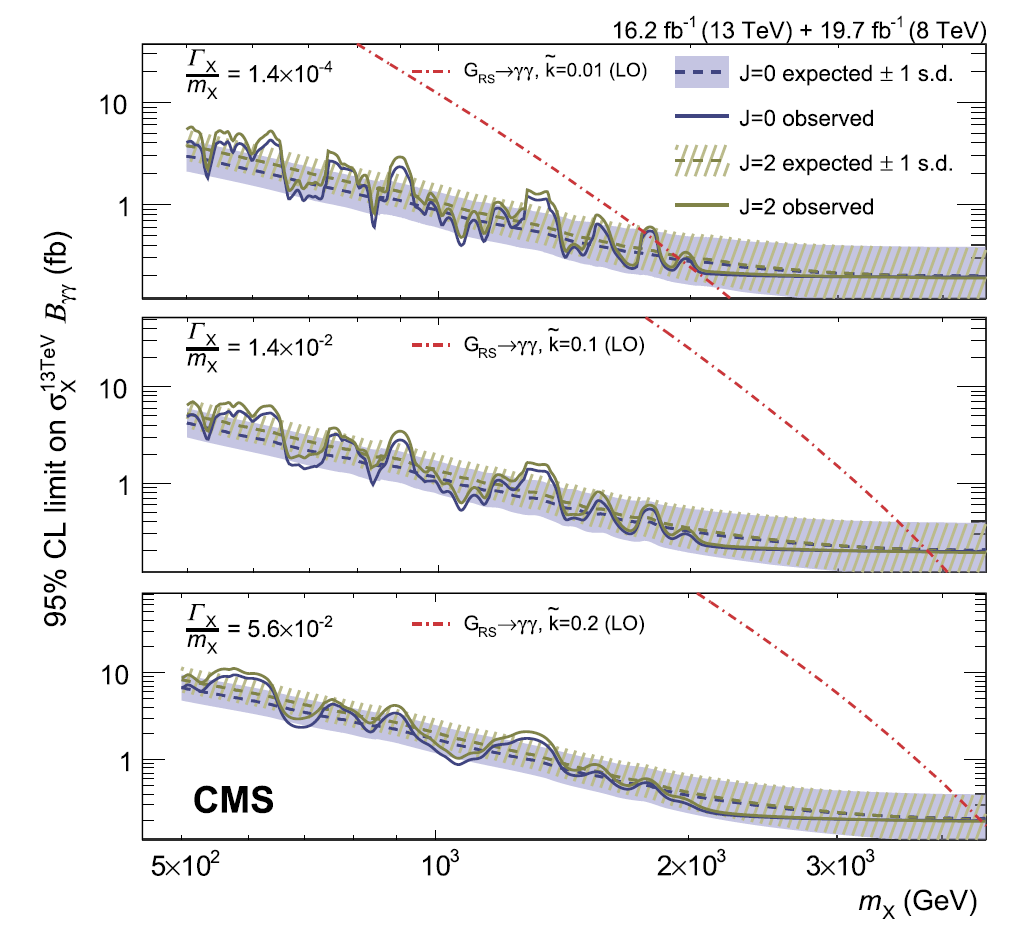}
\end{minipage}
\caption{\em Left: mass distribution of photon pairs obtained by ATLAS from the analysis of a luminosity of 3.2 $fb^{-1}$ of 2015 data collected at 13-TeV center-of-mass energy~\cite{atlasgg4}. The background is fit with a smooth functional form and is subtracted in the lower inset. Right: 95\% CL upper limits on the 13-TeV cross section of diphoton resonances times branching fraction to photon pairs produced by CMS using 8- and 13-TeV data~\cite{cmsgg4}, as a function of the resonance mass $m_X$. The grey and green curves show the exclusion limits for the scalar and RS graviton, respectively. From top to bottom, the limits refer to values  of the dimensionless coupling $\tilde{k}=0.01$, $\tilde{k}=0.1$, and $\tilde{k}=0.2$.
}
\label{f:ggresults}
\end{center}
\end{figure}

As mentioned already, the ATLAS search for spin-0 resonances and RS spin-2 graviton resonances found a significant structure in the mass distribution of selected photon pairs~\cite{atlasgg4} in the first 3.2 $fb^{-1}$ of 13-TeV collisions acquired in 2015. The signal had a broad Gaussian-like shape, of width of about 50 GeV, centered at a mass of 750 GeV. Two different methods for the prediction of the background shape, a first one based on a functional parametrization and a second one based on a MC simulation of photon pair production at next-to-leading order using Diphox~\cite{diphox} and a data-driven model of spurious backgrounds from photon+jet and dijet events, returned similar results. The local significance of the signal was estimated at $3.8\sigma$ or $3.9 \sigma$, depending on the tested signal hypothesis (spin-2 or spin-0, respectively); once corrected for the LEE the significance was estimated at $2.1 \sigma$. In the case of a narrow signal, the local significance estimates decreased to $3.3\sigma$ and $2.9 \sigma$ for the spin-2 and spin-0 hypothesis, respectively. The mass distribution of photon pairs reported by ATLAS in~\cite{atlasgg4} is shown in Fig.~\ref{f:ggresults} (left). As previously mentioned, preliminary results from the first half of 2016 data were sufficient to disprove the existence of a resonance at 750 GeV; ATLAS waited for the collection of the full 2016 dataset before producing a new publication~\cite{atlasgg5}. Due to its relevance we report it here although it is only available as a preprint at the time of writing. The reported study includes a total luminosity of 36.7 $fb^{-1}$ of 13-TeV collisions. An improved calibration procedure of photons brought the reanalysed 2015 data to show a smaller signal component in the mass range where the original effect had been located: significances for the two signal hypotheses (spin-2 and spin-0) were now estimated at $3.3\sigma$ and $3.2\sigma$, respectively. The combination of the 2015 and 2016 datasets produced results in better agreement with the background-only hypothesis, with global significances of the largest deviations well below the $1 \sigma$ level. The analysis allowed to exclude at 95\% CL gravitons with mass below 4.1 TeV in the RS model with $\tilde{k}=0.1$. 

The CMS collaboration produced four publications on searches for resonances decaying to photon pairs. In the first one~\cite{cmsgg1} they reported on the analysis of a luminosity of 2.2 $fb^{-1}$ of 7-TeV proton-proton collisions. The search considered both the RS graviton and the ADD model, setting 95\% CL limits in the range 0.86-1.84 TeV as a function of the normalized coupling strength $\tilde{k}$. For the ADD model the limits ranged in the 2.3-3.8 TeV. The second study considered the full integrated luminosity collected at 8 TeV in 2012~\cite{cmsgg2}, amounting to 19.7 $fb^{-1}$, and produced a search for $X \to \gamma \gamma$ resonances in the mass range 150-850 GeV, targeting both spin-0 and spin-2 resonances of natural widths in the 0.1 GeV to 0.1 $m_X$ range. Events were categorized into four classes of different purities depending on the pseudorapidity of the photons and the value of the variable $R_9$ (see Sec.~\ref{s:photons}), and the four searches were then combined. The data were found to be in good agreement with a smooth data-driven background model in all classes, and a 95\% CL upper limit on the signal cross section times resonance branching fraction to photon pairs was extracted as a function of the resonance mass. The results were also interpreted in the context of the 2HDM in the $(\tan \beta, cos(\beta-\alpha))$ plane.

The third study produced by CMS~\cite{cmsgg3} used a combination of data from 2012 ($19.7 fb^{-1}$ of 8-TeV collisions) and 2015 (3.3 $fb^{-1}$ of 13-TeV collisions). The 2015 data considered were both ones taken with the central solenoid at its nominal field of 3.8 Tesla, and ones taken when the magnet was de-energized for servicing; four classes of events were studied separately as in the previous publication, to improve the sensitivity of the search. A modest excess over background predictions was observed in the vicinity of 750 GeV, with a local (global) statistical significance of $3.4 \sigma$ ($1.6 \sigma$). The search also set 95\% upper limits on RS graviton production, from which lower limits at 95\% CL could be extracted for gravitons at masses of 1.6, 3.3, and 3.8 TeV for values of $\tilde{k}=0.01$, 0.1, and 0.2, respectively. Finally, in~\cite{cmsgg4} CMS produced a search in a data sample corresponding to 12.9 $fb^{-1}$ of 13-TeV collisions collected in 2016, and combined those results with the analysis of the datasets already reported earlier~\cite{cmsgg3}. The search did not find any significant signal in the investigated mass spectrum, and set limits at 95\% confidence level on the mass of spin-2 RS gravitons at 1.95 to 4.45 TeV depending on the values of the dimensionless coupling parameter $\tilde{k}$ in the range 0.01 - 0.2, as well as limits on the mass of scalar resonances produced by gluon-gluon fusion (see Fig.~\ref{f:ggresults}, right).

\subsection {\it LHC Searches for Resonances Decaying to Final States Including Higgs Bosons}
\label{s:higgsX}

\noindent
Following the conclusive observation of a Higgs boson in July 2012, the LHC experiments naturally started to include it as a taggable object in their searches for other massive states. The primary decay modes that benefited from this addition were ones including a $Wh$ or a $Zh$ pair; searches for heavy particles that could yield a pair of Higgs bosons also got underway, in conjunction with the start of a long-term plan targeting the measurement of the Higgs boson self coupling through the measurement of the non-resonant Higgs pair production cross section. In addition to processes that include Higgs bosons as a decay signal, new resonances lighter than $m_h$ were sought for, primarily in the decay $h \to Za$ or $h \to aa$. For completeness we include below a mention of these latter searches as an exception to our stated exclusive focus on SM bosons as decay objects.

\subsubsection{\it Searches for $Wh$ and $Zh$ Resonances}

\noindent
The first search reported by ATLAS for resonances decaying to the $Z h$ final state~\cite{ZHatlas1} focused on the CP-odd Higgs boson $A$ of generic 2HDM, reconstructed in three different final states: leptonic $Z$-boson decays with $h \to b \bar{b}$ and with $h \to \tau \tau$, and decays of the $Z$ to neutrinos in conjunction with the decay $h \to b \bar{b}$. The full 20.3 $fb^{-1}$ of 8-TeV collisions acquired in the 2012 run was considered for the search. For the decay mode including $\tau$ leptons, all three combinations of observable $\tau$ decays to hadrons or leptons were considered ($\tau_h \tau_h$, $\tau_h \tau_l$, and $\tau_l \tau_l$); the resonance mass was estimated by computing the combination $m_4 = m_{ll \tau \tau}-m_{ll}-m_{\tau \tau}+m_Z+m_h$, which increases the resolution by using known mass values for the $Z$ and $h$ bosons. In the $ll b \bar{b}$ category, two $b$-tagged jets were required, and additional criteria were imposed to reject the background from top pair production. In the $\nu \nu b \bar{b}$ search several kinematic criteria were imposed to reduce QCD backgrounds. No significant signal was observed in any of the search categories, and upper limits were set at 95\% CL on the product of $A$ boson cross section times the branching fraction of the $A \to Z h$ decay times the branching fraction of the Higgs boson to $b \bar b$ or $\tau \tau$. In the first case the limits ranged from 57 to 14 $fb$ as $m_A$ varied in the 220-1000 GeV, while in the second case they varied from 98 to 13 $fb$ for the same range of $A$ mass values.

In a parallel search performed on the same dataset as the one above, ATLAS~\cite{VHatlas1} searched for heavy resonances decaying to $Wh$ and $Zh$ bosons in the context of the HVT (see Sec.~\ref{s:theory}) and Minimal Walking Technicolour (MWT) \label{s:mwt} models. The MWT is a model with strongly-coupled dynamics developed in~\cite{mwt}, which includes a vector $R_1$ and an axial-vector $R_2$. The charged states can be sought for in their decay to $Wh$, while the neutral states of each triplet may produce a signal in $Z h$ final states. Lattice simulations would predict these new states to have masses in the 2 TeV range~\cite{mwtlattice1,mwtlattice3}. In the reported study Higgs boson candidates were reconstructed from $b$-quark jet pairs, while $W$ ($Z$) bosons were identified through leptonic decays respectively yielding $l \nu$ pairs ($l^+l^-$ or $\nu \nu$ pairs), with $l= e, \mu$. Events were required to contain two jets, either or both $b$-tagged by the MV2 algorithm~\cite{atlasbmva}; the invariant mass of the jet pair was required to be in the 105-145 GeV range. Further channel-specific selection cuts were applied in the three considered leptonic decay categories. All backgrounds were estimated with MC simulations of the relevant processes, except the one arising from QCD multijet production, estimated with data-driven methods. All search channels were found in agreement with background predictions in the four-body mass distributions (the transverse mass was used in the $\nu \nu b \bar{b}$ channel). The 95\% CL upper limits on the signal rates were converted into exclusion regions in the parameter space of the considered theoretical models. 

A mention should also be made here of a third search performed by ATLAS in the same 8-TeV dataset collected in 2012. That analysis considered the cascade decay of a heavy neutral scalar $H^0$ to a charged Higgs plus $W$ boson pair, with a subsequent decay of the charged Higgs state to a $Wh$ final state (assuming that $h$ was the 125 GeV scalar)~\cite{atlascascade}. One of the $W$ bosons was required to decay to an electron-neutrino or muon-neutrino pair, and the other to a pair of jets. Events were collected by electron and muon triggers and selected to contain a charged lepton, missing transverse energy above 30 (20) GeV for the $e \nu$ ($\mu \nu$) $W$ signature, and at least four anti-$k_T$ jets reconstructed with a radius parameter $R=0.4$, two of which $b$-tagged by the MV1 algorithm~\cite{atlasbmva}. After a reconstruction of the two-body decay candidates (the two $W$ bosons and the $h$ scalar), one $W$ was combined with the $h$ decay products to form an $H^{\pm}$ candidate; the higher-mass of the two possibilities was considered. Finally, the combined mass of the three bosons was used to construct the heavy $H^0$ mass. A boosted decision tree discriminant was employed to tell apart the signal from the dominant $t \bar{t}$ background. The observed yields of selected signal candidates were found in good agreement with SM background predictions, and 95\% CL upper limits on the signal cross sections were reported and interpreted in the context of the 2HDM. 

\begin{figure}[h!]
\begin{center}
\includegraphics[scale=0.7]{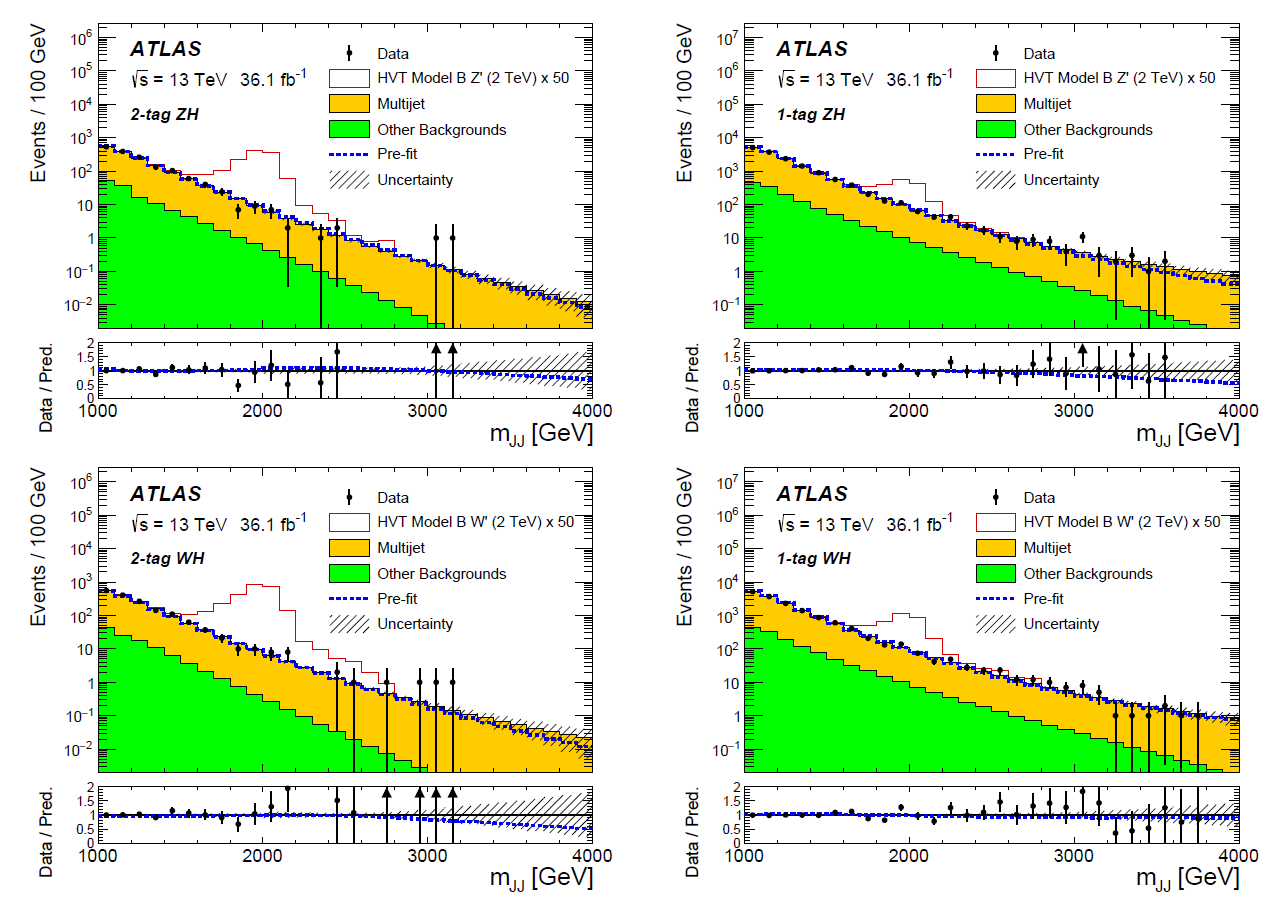}
\caption{\em Mass spectra of signal-region events collected by ATLAS in the all-hadronic search for resonances decaying to $Vh$ pairs, performed in a dataset corresponding to an integrated luminosity of 36.1 $fb^{-1}$ of 13-TeV collisions. The four panels correspond to events with two b-tags (left) or one b-tag (right) selected with the $Zh$ (top) or $Wh$ (bottom) signature. The dashed curves show the pre-fit background expectation; the signals expected from HVT model-B resonances of 2-TeV mass are shown multiplied by a factor 50 for comparison. The lower portions of the four panels show the ratios of observed data divided by background prediction; arrows there indicate off-scale points.}
\label{f:vh_atlas3}
\end{center}
\end{figure}

The search for $W'$ and $Z'$ resonances in $Wh$ and $Zh$ final states, with $h \to b \bar{b}$ and decays of weak bosons to electrons, muons, and neutrinos in the three categories defined by the number of charged leptons ($l^+ l^-$, $l \nu$, and $\nu \nu$) was repeated by ATLAS in the luminosity of 3.2 $fb^{-1}$ of 13-TeV collisions acquired in 2015~\cite{VHatlas2}. The channel with zero detected charged leptons had significant acceptance to $W' \to Wh$ decays due to undetected leptons from $W$ decays or contributions from $W \to \tau \nu$ decays, hence it was combined with the 1-lepton channel in the $W'$ search. The $h \to b \bar{b}$ decay was tagged by reconstructing a wide $R=1.0$ jet with $p_T>250$ GeV and large mass, and finding narrow track-jets at least one of which $b$-tagged by a custom algorithm~\cite{htagboostatlas}. The number of $b$-tags (one or two) and the number of leptons (zero, one, or two) allowed to define six categories, where resonant signals were sought for by studying the four-body mass or the transverse mass (in the zero-lepton cases) distribution. The absence of any significant signal was turned into upper limits on the production cross section of the considered resonances  times branching fractions of $h \to b \bar{b}$ or $c \bar{c}$ (the ratio of branching fraction of $h$ to the two heavy quark pairs was considered fixed); for the $W'$ limits the 0- and 1-lepton categories were fit together, and for $Z'$ limits the 0- and 2-lepton categories were used. Using model A of the HVT scenario with $g_V=1$, ATLAS excluded at 95\% CL a $W'$ boson with mass in the 200-1750 GeV range and a $Z'$ of mass in the 200-1490 GeV range; using model B and $g_V=3$, the exclusions were respectively in the range $200<m_{W'}<2220$ GeV and $200<m_{Z'}<1580$ GeV; those limits were obtained by fixing the Higgs branching fraction to heavy quarks to its SM value~\cite{HSMBR}. Considering the two heavy bosons degenerate in mass, and fitting all data categories together, the limit was set at 1730 GeV (2310 GeV) for model A and B, respectively.

The search in Run 2 data collected in 2015 and 2016, corresponding to a total of 36.1 $fb^{-1}$ of integrated luminosity of 13-TeV collisions reported in~\cite{VHatlas3}, is the most recent result by ATLAS on $Vh$ resonances at the time of writing. The analysis considered the all-hadronic decay mode of the SM bosons. Events containing two $R=1.0$ anti-$k_T$ jets with transverse momentum thresholds  $p_T>450$ (250) GeV on the leading (sub-leading) jet were considered, searching for narrow $R=0.2$ $k_T$ jets within the wide jet cones. Jets were trimmed by discarding narrow $k_T$ jets with a $p_T$ of less than 5\% of the total $p_T$ of the wide jet; the jet mass was recomputed after trimming using both tracking and calorimeter information, to improve the mass resolution~\cite{jetmassatlasr2}. $b$-tagging optimized to boosted jets~\cite{htagboostatlas} was applied on track-jets inside the wide cones to identify the $h \to b \bar{b}$ decay candidate. Vector boson candidate jets were identified by 50\%-efficient substructure criteria~\cite{vtaggingatlasr2}. The events were finally divided in four classes depending on the number of $b$-tags and on whether the vector-boson wide jet were compatible with a $W$ or $Z$ decay candidate. A statistical analysis was used to test different mass hypotheses for heavy bosons in the HVT model. A signal with a local significance of 3.3 standard deviations was obtained at a mass of 3 TeV from the fit of the $Zh$ final state data categories; the global significance of the excess was estimated at $2.1 \sigma$. Fig.~\ref{f:vh_atlas3} shows the data in the four considered search categories. Finally, 95\% CL limits were set on the parameter space of the HVT, and for the considered model B $W'$ bosons were excluded in the mass region 1.1-2.5 TeV, and $Z'$ bosons in the mass region 1.1-2.6 TeV.

\begin{figure}[h!]
\begin{center}
\begin{minipage}{0.49\linewidth}
\includegraphics[scale=0.32]{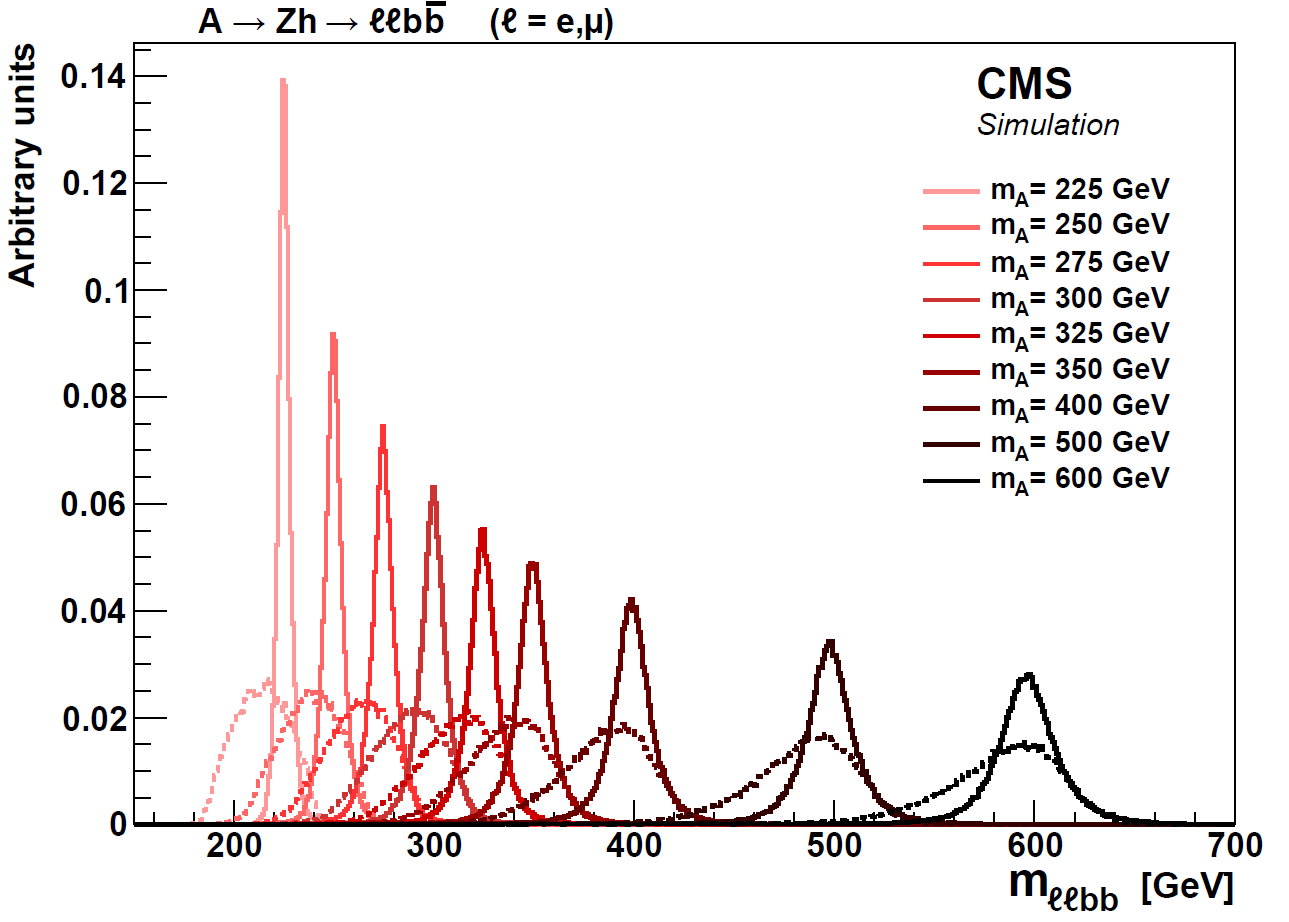}
\end{minipage}
\begin{minipage}{0.49\linewidth}
\includegraphics[scale=0.32]{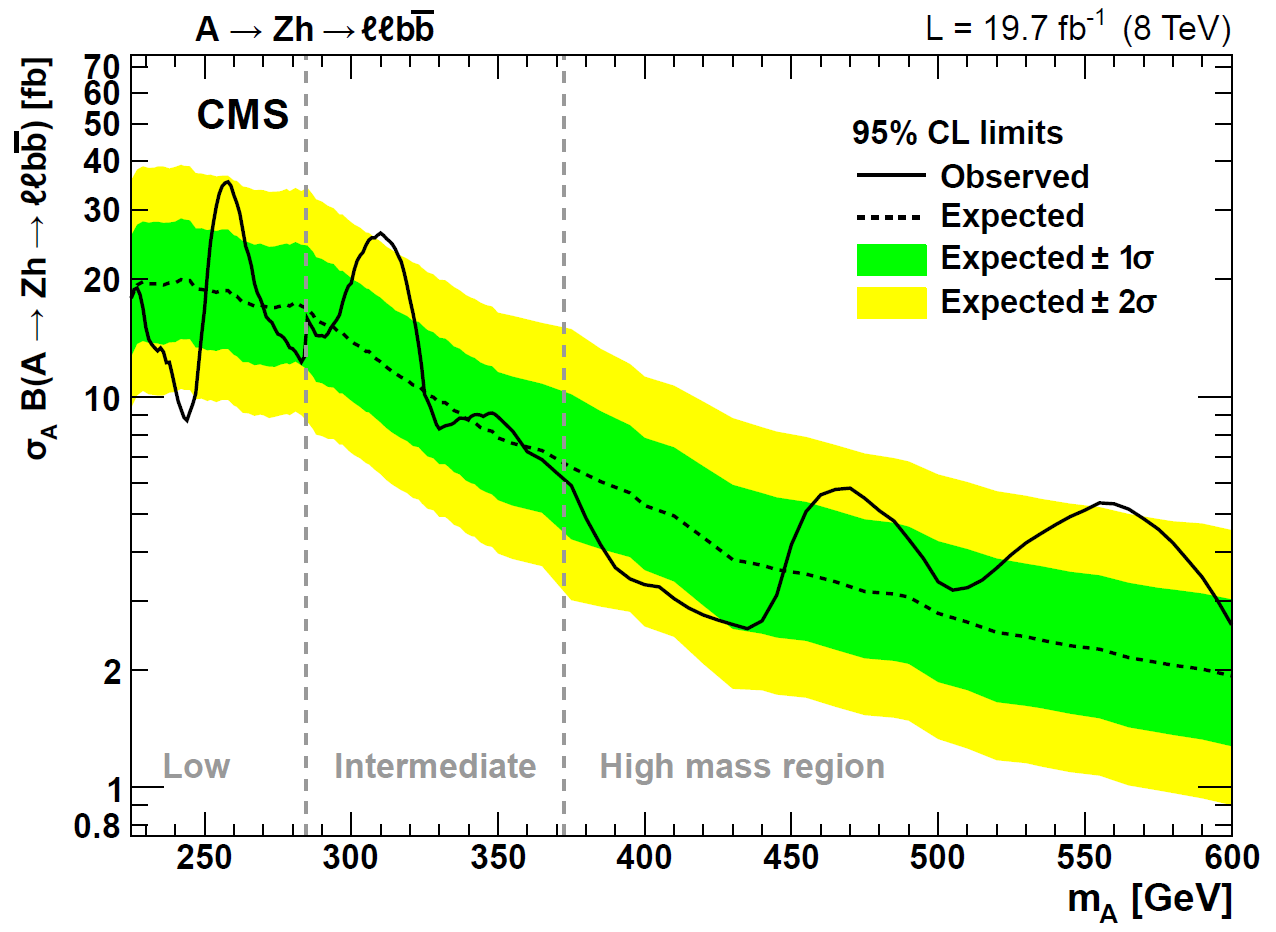}
\end{minipage}
\caption{\em Left: reconstructed $A$ mass templates from kinematic fits to the decay products of the CMS $A \to Z h$ search described in~\cite{azhcms1}. For each mass hypothesis, the $A$ mass distribution is drawn before and after the kinematic fit with dashed and continuous lines, respectively. Right: 95\% CL upper limit on the product of $A$ cross section times branching fraction to $Zh$ pairs as a function of $A$ boson mass obtained by CMS in the same search.}
\label{f:azh1}
\end{center}
\end{figure}

CMS produced a search for a heavy pseudoscalar $A$ boson in the decay chain $A \to Zh \to l^+ l^- b \bar{b}$, using an integrated luminosity of 19.7 $fb^{-1}$ of 8-TeV collisions~\cite{azhcms1}. On-shell $Z \to ee$ and $Z \to \mu \mu$ decays of the $Z$ boson were considered, while the Higgs boson was sought in its $b \bar{b}$ decay mode. In 2HDM theories the $A$ boson can be effectively sought for in the considered final state if its mass is above $m_h +m_Z \sim 216$ GeV and below the $A \to t \bar{t}$ threshold; in addition, within a wide range of parameter space the $h \to b \bar{b}$ decay has a large branching fraction~\cite{craig_galloway}. Events were selected to contain a leptonic $Z$ boson decay and two jets with a CSV $b$-tag, one above the tight working point of the algorithm and the other above the loose one (see Sec.~\ref{s:btagging}). A kinematic fit was performed to the signal hypothesis by constraining the $Z$ and $h$ masses to their known values; the fit produced a narrow four-body distribution for the signal templates, as shown in Fig.~\ref{f:azh1}. Background contributions were mainly due to electroweak production of vector bosons and jets, along with top pair-production; they were estimated from MC simulations and their rates validated with control samples enriched in each of the main components. Three disjunct mass regions were studied independently, and for each a BDT classifier was trained to separate the signal from backgrounds. The resulting mass distributions in high-BDT events did not show excesses above predicted backgrounds, and 95\% CL limits were set on the $A$ cross section (Fig.~\ref{f:azh1}, right). The limits were interpreted as exclusions of the $(\tan \beta, \cos(\beta-\alpha))$ parameter space of Type-I and Type-II 2HDM.

\begin{figure}[h!]
\begin{center}
\begin{minipage}{0.49\linewidth}
\includegraphics[scale=0.42]{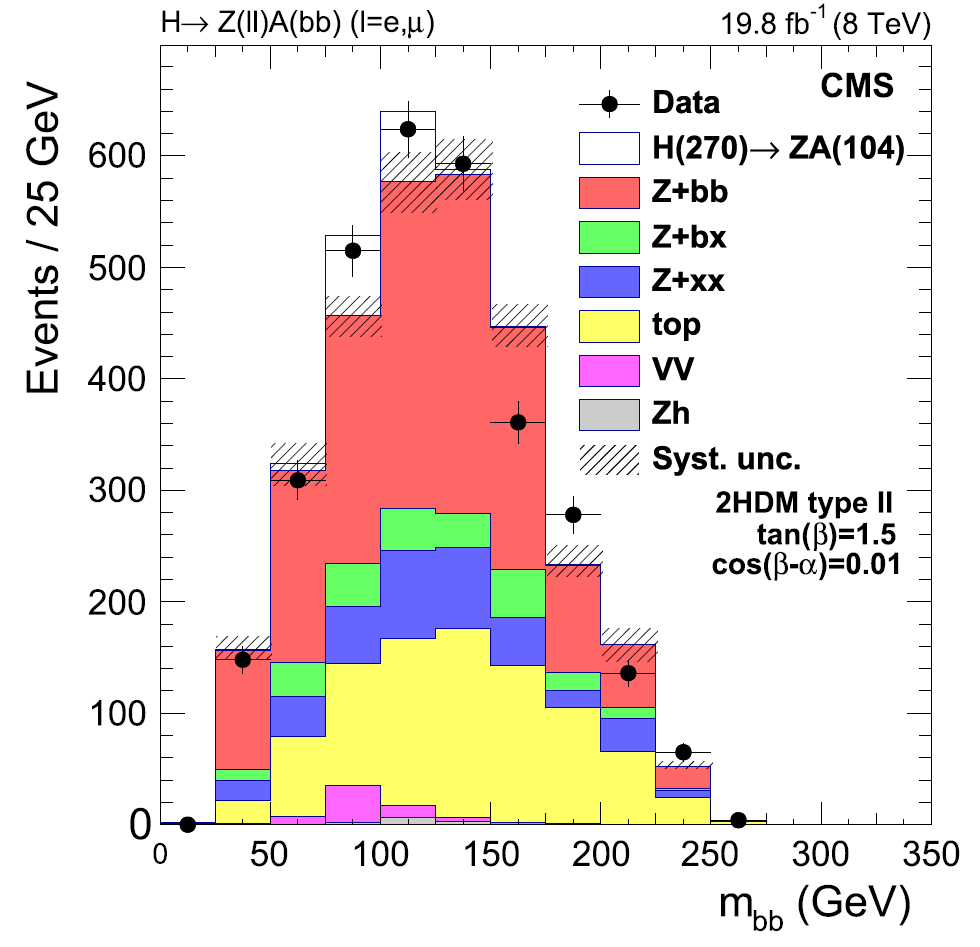}
\end{minipage}
\begin{minipage}{0.49\linewidth}
\includegraphics[scale=0.44]{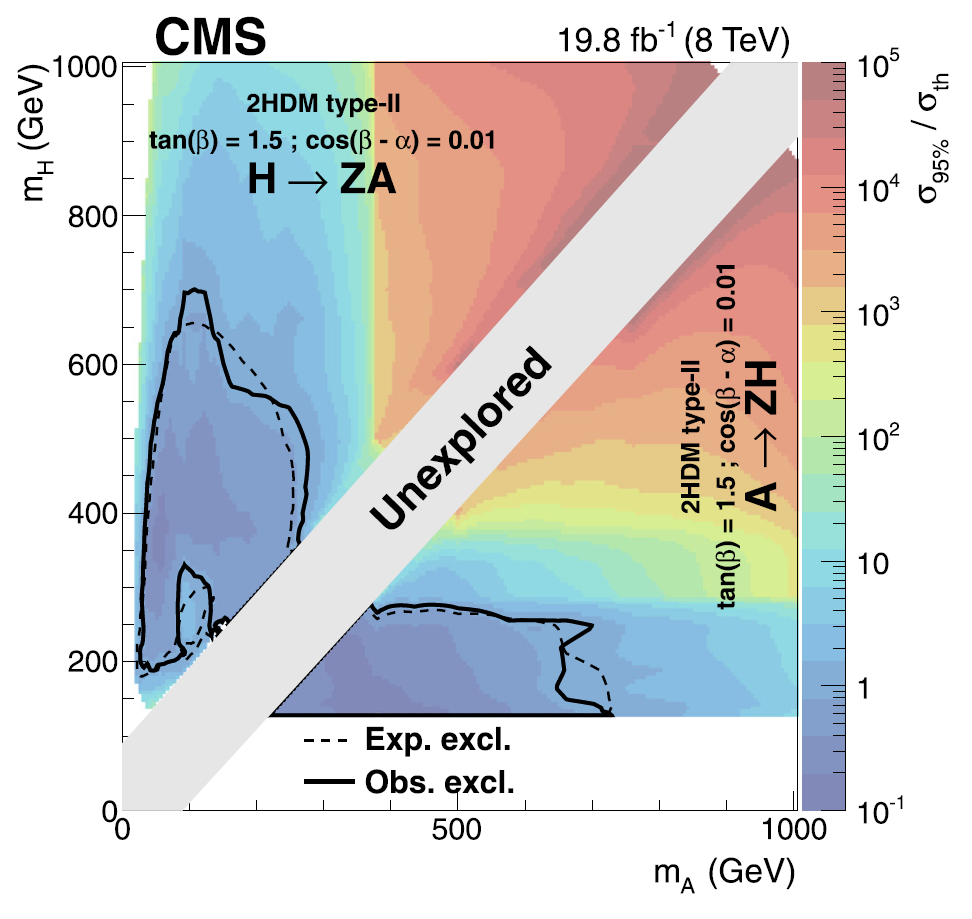}
\end{minipage}
\caption{\em Left: reconstructed $m_{b \bar {b}}$ distribution for $ll b \bar b$ candidates in the four-body mass range $222<m_{llb \bar {b}}<350$ GeV found by the CMS search for $A \to ZH$ and $H \to ZA$ decays~\cite{azhcms2} (points with vertical uncertainty bars). Full histograms show the composition of SM backgrounds; a 2HDM signal computed at NLO with the SusHi MC~\cite{sushi1,sushi2,sushi3}, corresponding to $m_H=270$ GeV and $m_A=104$ GeV, is shown by an empty histogram added to background ones.  Right: exclusion region obtained by the combination of the two searched final states considered in~\cite{azhcms2} in the $(m_H, m_A)$ plane, for a Type-II 2HDM with $\tan \beta=1.5$ and $\cos(\beta-\alpha)=0.01$. The excluded region is delimited by the full curve; the colour map shows the upper 95\% CL limit on the cross section in units of the cross section predicted by the 2HDM model. }
\label{f:azhlouvain}
\end{center}
\end{figure}

Another study targeting the $A \to ZH$ decay as well as the decay $H \to ZA$ was published by CMS in 2016 from the analysis of the same 8-TeV dataset of the previous one. In this case, the CP-even $H$ boson was not considered to necessarily coincide with the 125 GeV $h$ scalar; this violates the stipulation of the present review, which considers final states made of SM bosons, yet we include a brief discussion of that analysis here for its coherence with the other studies presented in this Section.  In both considered final states, the lighter scalar was sought for in its decay to a pair of $b$-quarks or a $\tau \tau$ pair, while the $Z$ boson was identified through its decay to an $ee$ or $\mu \mu$ pair.  The event selection for the $ll b \bar b$ decay mode required the presence of two $R=0.4$ anti-$k_T$ jets with a medium CSV $b$-tag. In the $ll \tau \tau$ mode the $\tau$ lepton pair was accepted if it produced the following signatures: $e \mu$, $e \tau_h$, $\mu \tau_h$, or $\tau_h \tau_h$, where $\tau_h$ indicates a hadronic decay of the $\tau$ lepton, and $e$, $\mu$ indicate the respective fully leptonic decays of the $\tau$. In the $ll \tau \tau$ selection the presence of $b$-tagged jets was vetoed to suppress the top pair production background. The mass of $\tau$ lepton pairs was reconstructed with the SVfit algorithm~\cite{svfit}. The shape of the mass distributions of the reconstructed resonance for backgrounds was determined with MC simulations, while their normalization was determined with data-driven techniques. In the $ll b \bar{b}$ selection an excess of events was observed in the region of $b \bar b$ mass versus 4-body mass around (95, 285) GeV (see Fig.~\ref{f:azhlouvain}, left), as well as in the region (575, 660) GeV. The local significances of these effects in the data were estimated at $2.6\sigma$ ($2.85\sigma$) respectively; the global significance was estimated at $1.6\sigma$ ($1.9 \sigma$). No significant excess was observed in the $ll \tau \tau$ decay mode. A combination of the searches in the two considered final states allowed to extract 95\% CL limits on the space spanned by the $(m_A, m_H)$ plane of a Type-II 2HDM with $\tan \beta=1.5$ and $\cos(\beta-\alpha)=0.01$ (Fig.~\ref{f:azhlouvain}, right). The exclusion region ranges from 200 to 700 GeV for $m_H$ when $m_A$ is between 20 and 270 GeV (for $m_H>m_A$), or for $m_A$ in the range 200-700 GeV when $m_H$ is in the 120-270 GeV range (for $m_A>m_H$).

A third CMS search using the same dataset of the ones above targeted events containing high-$p_T$ jets clustered with the CA algorithm~\cite{CA1,CA2} using a wide $R=0.8$ cone to identify the signal of a heavy resonance decaying to a $Wh$ pair, with a subsequent hadronic decay of the latter~\cite{VHcms2}. The pruned mass of the wide jets was used, along with the $\tau_{21}$ and $\tau_{42}$ subjettiness variables, to reduce the overwhelming QCD background. While $\tau_{21}$ can distinguish how well a single jet conforms to the hypothesis of being produced by the hadronization of two separate, energetic quarks emitted from a boosted vector boson decay, $\tau_{42}$ provides information on the validity of the hadronic $WW$ decay hypothesis altogether. The $h \to b \bar{b}$ decay was identified by requiring that the sub-jets associated to $h$ decay in a wide jet had a CSV $b$-tagging value above its loose working point when they had an angular separation above $\Delta R>0.3$; for narrower pairs, the $b$-tagging requirement was applied to the wide CA jet. The mass distribution of $Wh$ candidates in the data was fit by parametrizing the background, predominantly constituted by QCD multijet production, with a smooth four-parameter function, and the signal with a Gaussian shape of 5\%-10\% relative resolution. The absence of any significant signal in the investigated mass range was used to set upper limits on the cross section of a heavy $W'$ or $Z'$ boson in the context of the HVT model. The resulting 95\% CL excluded mass ranges were $1.0<m_{Z'}<1.1$ TeV and $1.3<m_{Z'}<1.5$ TeV for a $Z'$ boson, and $1.0<m_{W'}<1.6$ TeV for a $W'$ boson; a mass-degenerate state of $W'$ and $Z'$ bosons was excluded in the interval $1.0-1.7$ TeV.

\begin{figure}[h!]
\begin{center}
\begin{minipage}{0.49\linewidth}
\includegraphics[scale=0.34]{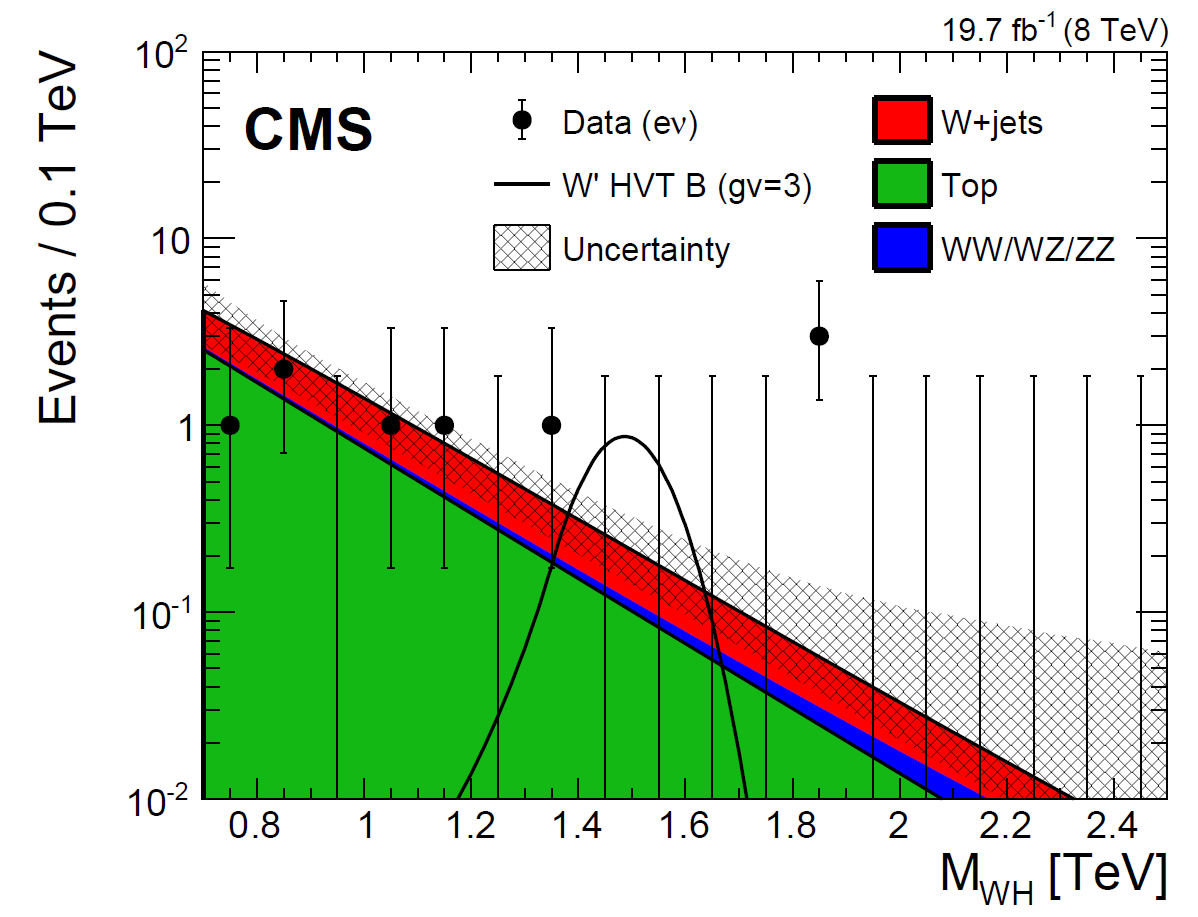}
\end{minipage}
\begin{minipage}{0.49\linewidth}
\includegraphics[scale=0.32]{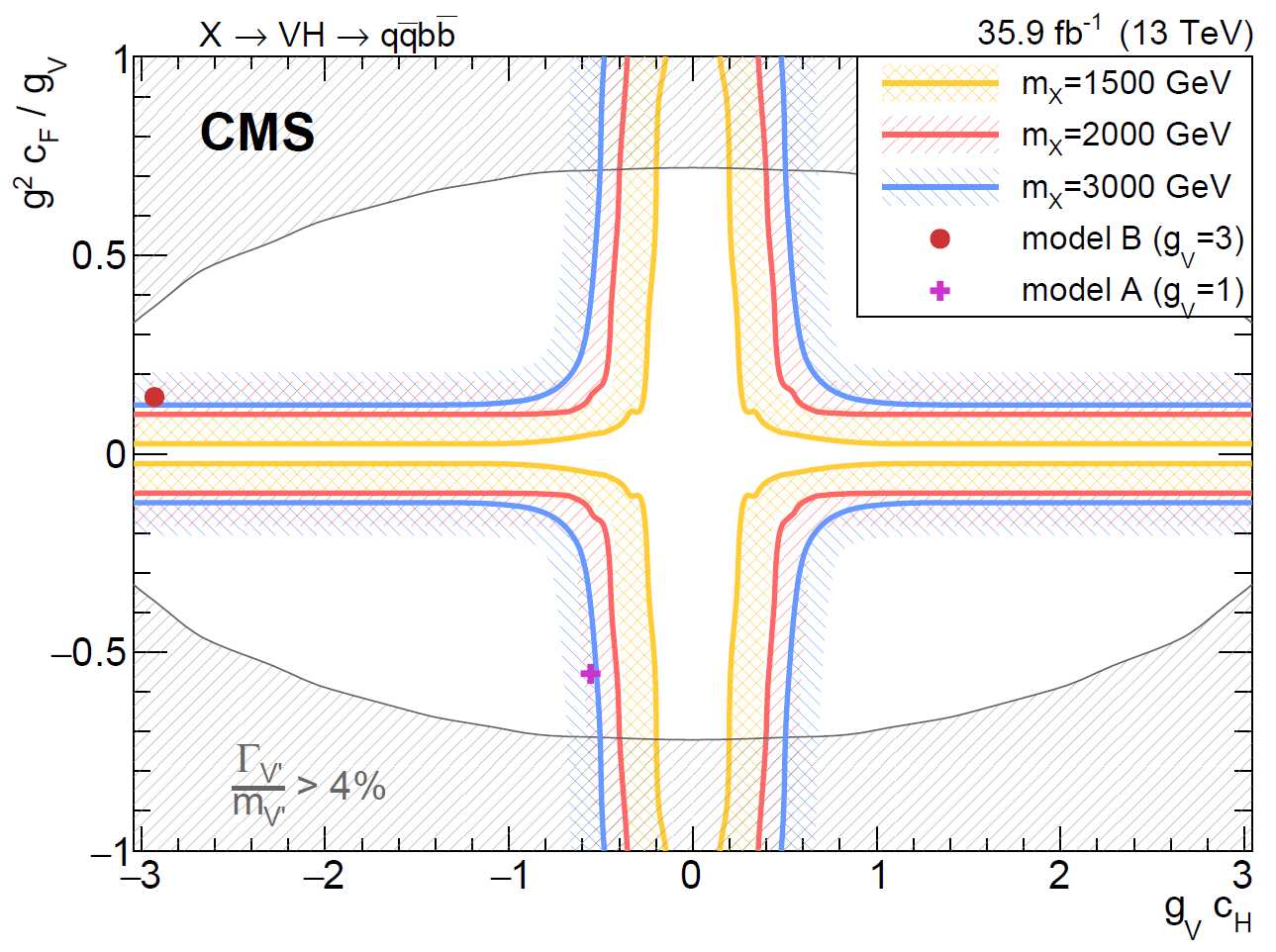}
\end{minipage}
\caption{\em Left: Reconstructed mass distribution of $Wh$ pairs in the $e \nu b \bar{b}$ decay mode obtained by CMS in 19.7 $fb^{-1}$ of 8-TeV collisions~\cite{VHcms2}. The data (points with 68\% CL uncertainty bars) are compared with estimated backgrounds from $W$+jets and $t \bar t$ production; the expected shape of a HVT $W'$ signal with a mass of 1.5 TeV is overlaid. Right: observed exclusion limit on the HVT parameter space for three different values of the heavy boson mass (1.5, 2.0, and 3.0 TeV) obtained by the CMS analysis in~\cite{VHcms5}. Models A and B are represented by a cross and a point, respectively. In the shaded area outside the oval in the graph, the narrow-width approximation is inapplicable. 
}
\label{f:whcms2}
\end{center}
\end{figure}

In a fourth study of the same 2012 dataset~\cite{WHcms3}, CMS considered the $W' \to Wh \to l \nu b \bar{b}$ decay chain, using the same analysis method of their search for high-mass $WW$ resonances already discussed {\em supra}~\cite{wwcms1}, and including $b$-tagging to identify the Higgs decay in the boosted jet as they did in their fully-hadronic $Wh$ search~\cite{VHcms2}. The study reported observing three events in the $e \nu b \bar{b}$ topology with a mass of about 1.8 TeV, where backgrounds yielded an estimated 0.3 events (see Fig.~\ref{f:whcms2}). No similar excess was observed in the $\mu \nu b \bar b$ channel, and the significance of the data was estimated at the level of $1.9 \sigma$ once accounting for the LEE. A limit was set at 95\% CL on the cross section of $W'$ production in HVT and LH models; once converted into lower mass limits, a $W'$ boson of mass below 1.4 TeV was excluded in the LH model, and below 1.5 TeV in HVT model B~\cite{HVT}, which mimics the properties of a composite Higgs scenario; in the latter case, a combination with the fully hadronic CMS search in the same dataset allowed to extend the lower mass limit to 1.8 TeV.

Two CMS publications reported on resonance searches in $Vh$ ($V=W,Z$) final states from the analysis of Run 2 data. A dataset corresponding to a luminosity of 2.2 to 2.5 $fb^{-1}$ of 13-TeV collisions acquired in 2015 was used~\cite{wwcms5} to search for leptonic $W$ and $Z$ decays accompanied by a high-$p_T$ wide jet with substructure, tagged to be originated by $h \to b \bar{b}$ decay. Both the $W \to l \nu$, the $Z \to \nu \nu$, and the $Z \to l l$ final states of vector boson decays were considered, with $l=e, \mu$. The search allowed to extend the lower mass limit on heavy vector bosons in HVT model B, considered in their previous publication~\cite{VHcms2}, to 2 TeV. A search in the fully-hadronic topology of $Vh$ decays was performed~\cite{VHcms5} using 35.9 $fb^{-1}$ of 13-TeV collisions, targeting the still untested high-mass region. At variance with previous analyses, jets were reconstructed with the anti-$k_T$ algorithm with a cone radius of $R=0.8$, and substructure was sought for by applying to jets the PUPPI algorithm~\cite{puppi} and the soft-drop algorithm~\cite{softdrop1,softdrop2} (see Sec.~\ref{s:jets}). To identify the Higgs-originated jet, a dedicated $b$-tagging algorithm based on a multivariate discriminant was used~\cite{boostbtag}.  No significant excess was observed over background predictions, and 95\% CL upper limits were obtained on the product of cross section for the production of a spin-1 resonance and the branching fraction of its decay to $Vh$ and the $h \to b \bar{b}$ decay; these range from 90 to 0.9 $fb$ for resonance masses  in the range 1500-4500 GeV, corresponding to an exclusion of resonance masses below 3.3 TeV. The limits were converted into exclusion regions of the parameter space of HVT model A and B (see Fig.~\ref{f:whcms2}, right).

\subsubsection{\it Searches for $hh$ Resonances}


\noindent
Non-resonant production of Higgs boson pairs occurs in the SM through triangle and box diagrams that interfere destructively, with a resulting cross section of less than 40 fb~\cite{hhxs} in 13-TeV proton-proton collisions. The interest of that process, which will likely require the full integrated luminosity collected during the future high-luminosity program of the LHC to be put in evidence, stems from the direct access that the triangle diagram provides to the measurement of the Higgs self-coupling  parameter $\lambda$ of the SM Lagrangian, and consequently to the test of possible SM extensions with anomalous couplings. Yet the final state of two Higgs bosons can also naturally arise in new physics models that predict the decay of a heavy narrow resonance (denoted by the letter $X$ in the following, when no specific model is referred to in particular). The cross section of that process could be large enough to be testable with the datasets currently available. Examples of classes of theories that can be probed are the singlet model~\cite{singlet1,singlet2,singlet3}, the 2HDM~\cite{2HDM}, the MSSM~\cite{mssm1,mssm2,mssm3,mssm4,mssm5,mssm6}, and models with warped extra dimensions~\cite{gbulkrevised1,wed2}. 

At variance with non-resonant Higgs pair production  searches, the searches for $X \to hh$ in ATLAS and CMS benefit from the application of kinematic fits to fully reconstruct the decay chain including the two-body decay of the $X$ particle and the subsequent Higgs boson decays in the selected events, as that strategy strongly reduces all otherwise irreducible backgrounds. Because of that, fully reconstructed final states like $b \bar {b} b \bar{b}$ and $b \bar{b} \gamma \gamma$ turn out to be comparatively more sensitive than partly-reconstructed ones, such as those including {\em e.g.} one $h \to WW$ decay with leptonic $W$ final states. In addition, jet substructure techniques have proven able to significantly increase the sensitivity by allowing the identification of $h \to b \bar{b}$ decays in high-momentum jets of high visible mass~\cite{boostsensitivitytohh}.The LHC experiments have produced a number of searches for resonances decaying into $hh$ pairs, leveraging on those ideas. Below we provide only a quick summary of the earlier searches, giving more space to the more recent ones, which provide the most stringent limits. 


The first study reported by ATLAS for $X \to hh$ signals involved the analysis of events with two photons and two $b$-tagged jets in 20 $fb^{-1}$ of 8-TeV collisions acquired in the 2012 data taking run~\cite{hhatlas_1}. They found an excess event rate above predicted backgrounds in the signal region with a local significance of 3.0 standard deviations. The impact of the LEE was estimated with pseudo-experiments, by considering the whole search region; this allowed the estimate of a global significance of 2.1 standard deviations. In the hypothesis of no signal in the data, ATLAS extracted upper limits at 95\% CL on the product of resonance cross section times branching fraction into Higgs boson pairs at 3.5 to 0.7 $pb$ for $X$ masses ranging from 260 to 500 GeV. They compared their upper limits on signal cross section to predictions by a representative Type-I 2HDM model~\cite{eriksson2hdm,branco2hdm} not excluded by the data, obtained by setting $\cos (\beta-\alpha)=-0.05$ and $\tan \beta=1$, with mass-degenerate heavy $H$ and $A$ bosons, and with the light state $h$ corresponding to the 125 GeV scalar.
A second ATLAS study~\cite{hhatlas_2} of the same 2012 data sample considered the fully hadronic final state of two wide jets of high momentum, searching for the signature of two narrow $b$-tagged jets (the ``resolved'' analysis) or a wide jet with track-based $b$-tags within its area (the ``boosted'' search); the resolved analysis considered resonances of mass in the range 500-1500 GeV, while the boosted analysis could extend the sensitivity to 2000 GeV. In both cases a constraint on the reconstructed mass of the $b$-jet pair (in the resolved case) or of the trimmed fat jet (in the boosted case) to the mass of the 125 GeV Higgs boson helped increase the resolution on the mass of the four-body system.  The absence of any significant signal in the data led to set constraints on several benchmark models: the bulk RS model with $\tilde{k}=1.0$, 1.5, or 2.0, and 2HDM models with  CP-conserving scalar potentials under the condition $m_H=m_A=m_H^+$, choosing a mixing parameter between the doublets such that $m_{12}= m_A^2 \tan \beta / (1+\tan^2 \beta)$~\cite{2HDMpheno}, and interpreting the results in the $(\tan \beta , \cos(\beta-\alpha))$ plane. 


The searches for $hh$ resonances described above were combined by ATLAS~\cite{hhatlas_3} with reported results of two other independent decay modes: the $\gamma \gamma \tau \tau$ and the $\gamma \gamma WW$ final states. In all cases an integrated luminosity of 20.3 $fb^{-1}$ of 8-TeV proton-proton collisions was used. Combined limits on the production cross section of a heavy scalar decaying to the considered final states through the production of a pair of 125-GeV Higgs boson were obtained in the mass range $260 - 1000$ GeV. These were turned into exclusion limits in the $(m_A , \tan \beta)$ plane in the context of two MSSM scenarios: the ``hMSSM''~\cite{hmssm1,hmssm2} and the ``low-tb-high'' scenario~\cite{lowtanbeta}. The natural width of the heavy scalar is always smaller than the mass resolution on the Higgs pair system in the considered models, so no accounting was made for it. The cross section of the heavy Higgs was computed using SusHi 1.4.1~\cite{sushi1,sushi2,sushi3}, and branching ratios were calculated with Hdecay~\cite{hdecay} for the hMSSM, and with FeynHiggs~\cite{feynhiggs1,feynhiggs2,feynhiggs3} for the low-tb-high scenario. 

In 2016 ATLAS published a new search for resonant $hh \to b \bar{b} b \bar{b}$ production in 3.2 $fb^{-1}$ of luminosity collected from 13-TeV collisions in 2015, again considering both a ``resolved'' final state when all the four $b$-quarks were reconstructed as independent jets, and a ``boosted'' topology where both Higgs bosons decayed into a single wide jet~\cite{hhatlas_4}. Three models were considered for the resonant search: spin-2 Kaluza-Klein gravitons with $\tilde{k}=2$ and $\tilde{k}=1$, and heavy Higgs-like spin-0 particles. The analysis employed a number of kinematical requirements on the four $b$-tagged jets selected for the resolved $hh$ search, reducing backgrounds from QCD multijet production (modeled using data failing the $b$-tag requirement in one jet in each Higgs decay pair) and $t \bar{t}$ production (with shape modeled with MC). In the boosted topology two wide jets were selected with $250<p_T<1500$ GeV; at least three $b$-tags in $R=0.2$ subjets within the $R=1.0$ wide jets were required, defining two orthogonal categories depending on the $b$-tag properties of the fourth sub-jet. Backgrounds were estimated similarly to the resolved analysis. The searches did not return any significant signal in the investigated range of Higgs-pair masses, and upper limits in the cross section of the considered resonances were extracted with a profile likelihood technique, using the $CL_s$ criterion. For heavy spin-0 Higgs-like particles $H$ a limit on the cross section times branching fraction squared of $h \to b \bar{b}$ was set in the range from 30 to 300 $fb$ for masses between 500 and 3000 GeV. For gravitons, the cross section limits were translated into restrictions on their mass ranges: in the case of  $\tilde{k}=2$, an exclusion of the range $480<m_G^*<965$ GeV; and assuming instead $\tilde{k}=1$ an exclusion of the range $480<m_G^*<770$ GeV. 

CMS reported searching for a signal of $H \to hh \to b \bar{b} \tau \tau$ as well as $A \to Z h \to ll \tau \tau$ decays using data corresponding to a luminosity of $19.7 fb^{-1}$ of 8-TeV proton-proton collisions acquired in 2012, reporting the null results as model-independent 95\% CL limit of $\sim 0.2 pb$ on $\sigma(gg \to H \to hh) \times B(hh \to b \bar b \tau \tau)$ for $260<m_H<350$ GeV and of 17 to 5 $fb$ on $\sigma(gg \to A \to Zh) \times B(Z \to ll) \times B(H \to \tau \tau)$ for $220<m_A<350$ GeV. These were turned into parameter space limits in the low-$\tan \beta$ scenario of the MSSM~\cite{lowtanbeta}, as well as in generic Type-II 2HDM models. 

Narrow resonances of mass in the 1.15-3.0 TeV range were the target of a search performed by CMS exploiting the large branching fraction of the $X \to hh \to b \bar{b} b \bar{b}$ decay, again using $19.7 fb^{-1}$ of 8-TeV collisions, employing the technique of jet pruning~\cite{ellissubstructure,ellissubstructure2,pruning3,pruning4} on high-$p_T$ dijet events with small pseudorapidity difference $|\Delta \eta_{jj}|<1.3$ and using a loose working point of the CSV $b$-tagger. The division of events in three categories depending on signal purity enhanced the sensitivity. The results allowed to exclude at 95\% CL a radion with scale parameter $\Lambda_R=1$ TeV for masses up to 1.55 TeV.

\begin{figure}[h!]
 \begin{center}
\begin{minipage}{0.49\linewidth}
\includegraphics[scale=0.4]{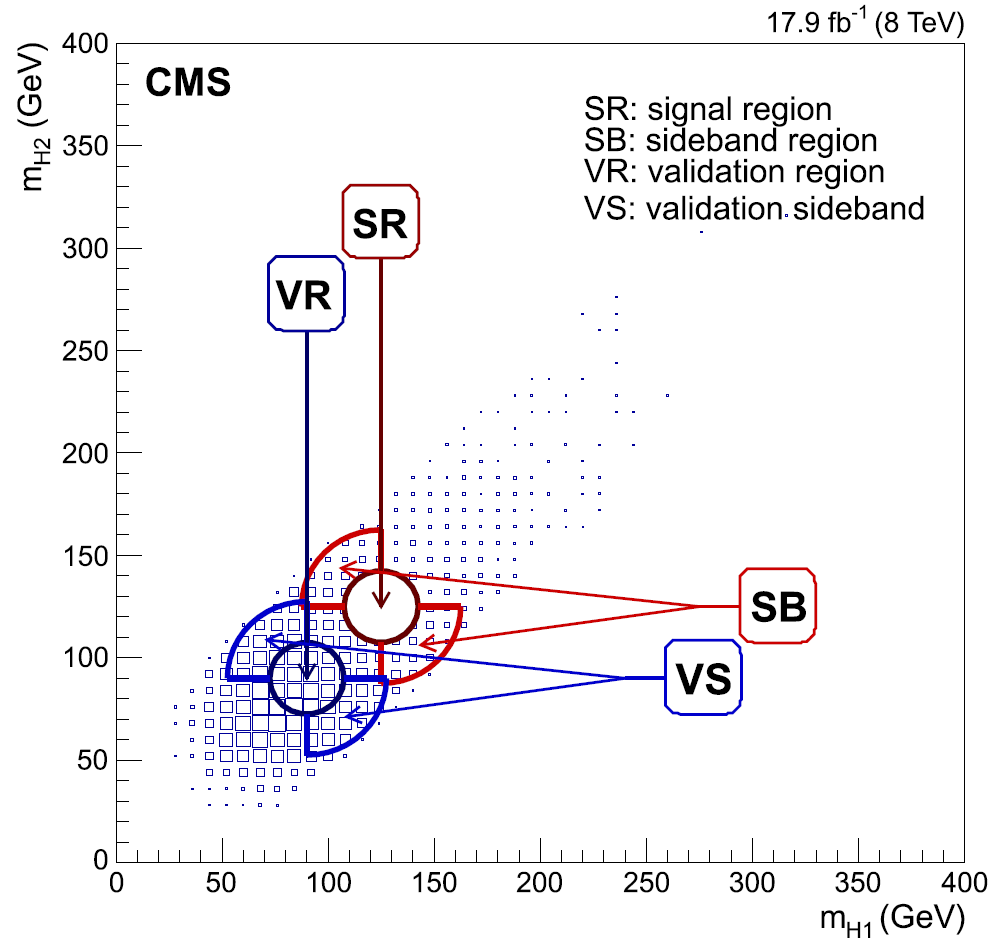}
\end{minipage}
\begin{minipage}{0.49\linewidth}
\includegraphics[scale=0.45]{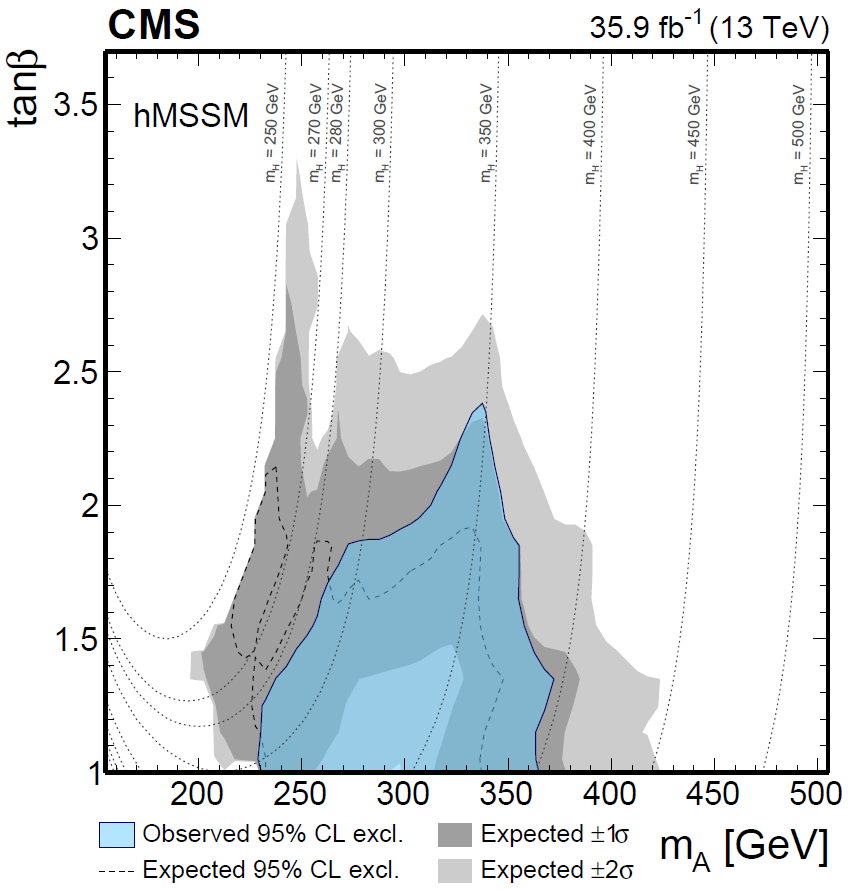}
\end{minipage}
\caption{\em Left: kinematic regions defined in the $(m_{h_1},m_{h_2})$ plane of the reconstructed masses of the two $h \to b \bar{b}$ decays, used to construct and validate the model of QCD backgrounds in the CMS search for $X \to hh \to b \bar{b} b \bar{b}$ search~\cite{hhbbbbr1cms}. The zone labeled ``SR'' is the signal region and data contained therein is not shown. ``VR'' indicates validation regions, ``SB'' is a sideband, and ``VS'' is the validation sideband. Right: Exclusion limits in the $(m_A,\tan \beta)$ plane of the hMSSM model obtained by the CMS search for Higgs pairs in the $bb \tau \tau$ final state with Run 2 data~\cite{cmshhbbtt}. The lighter neutral CP-even Higgs boson is assumed to be the observed 125 GeV scalar. Trajectories corresponding to equal values of $m_H$ are shown with dotted lines.  
}
\label{f:annulibbbb}
\end{center}
\end{figure}

\noindent
A search for resonances producing the  $hh \to b \bar {b} b \bar{b}$ final state was reported by CMS in a Run 1 dataset corresponding to $17.9 fb^{-1}$ of 8-TeV proton-proton collisions~\cite{hhbbbbr1cms}. Data were selected by a trigger collecting events containing at least four jets with $p_T>30$ GeV and $|\eta|<2.4$, two of which with $p_T>80$ GeV, and two jets $b$-tagged by a online CSV algorithm. Offline, the $p_T$ threshold was increased to 40 GeV, and events with four $b$-tagged jets were selected. Higgs decay candidates were reconstructed from pairs of $b$ jets if $|M_{b\bar{b}}-125|<35$ GeV. Since the kinematics of a heavy particle decay to four jets strongly depends on its mass, the selection criteria were further optimized with kinematic cuts in three separate search regions: $270<M_X<450$, $450<M_X<730$, and $M_X<1100$; for larger masses the $b$ quarks from $h$ decay merge into single jets, making the search method ineffective. CMS relied on a data-driven background-estimation technique to reduce the overwhelming background from QCD multijet processes, extrapolating the four-body mass distribution from events falling into quadrants of an annular region constructed around the signal region in the plane described by the two Higgs candidate masses (see Fig.~\ref{f:annulibbbb}, left). The procedure was validated using as a target events with both dijet masses around 90 GeV. The four-body mass distribution was then fit with a binned maximum likelihood technique as the sum of the data-driven background template, a top-pair background model extracted from MC,  and a radion signal template obtained from a Madgraph simulation~\cite{gouzevichradion}. Limits to the mass of the radion from 350 to 1100 GeV were set at 95\% CL  in the scenario of zero mixing with the Higgs boson, considering a product of the warp factor $k$ and the half-length of the extra dimension $L$ set to 35, and a radion decay constant $\Lambda_R=1$ TeV, using the theoretical cross section in~\cite{bargerradion}. For a spin-2 graviton a mass exclusion region was obtained between 380 and 830 GeV.

\begin{figure}[h!]
 \begin{center}
\includegraphics[scale=0.64]{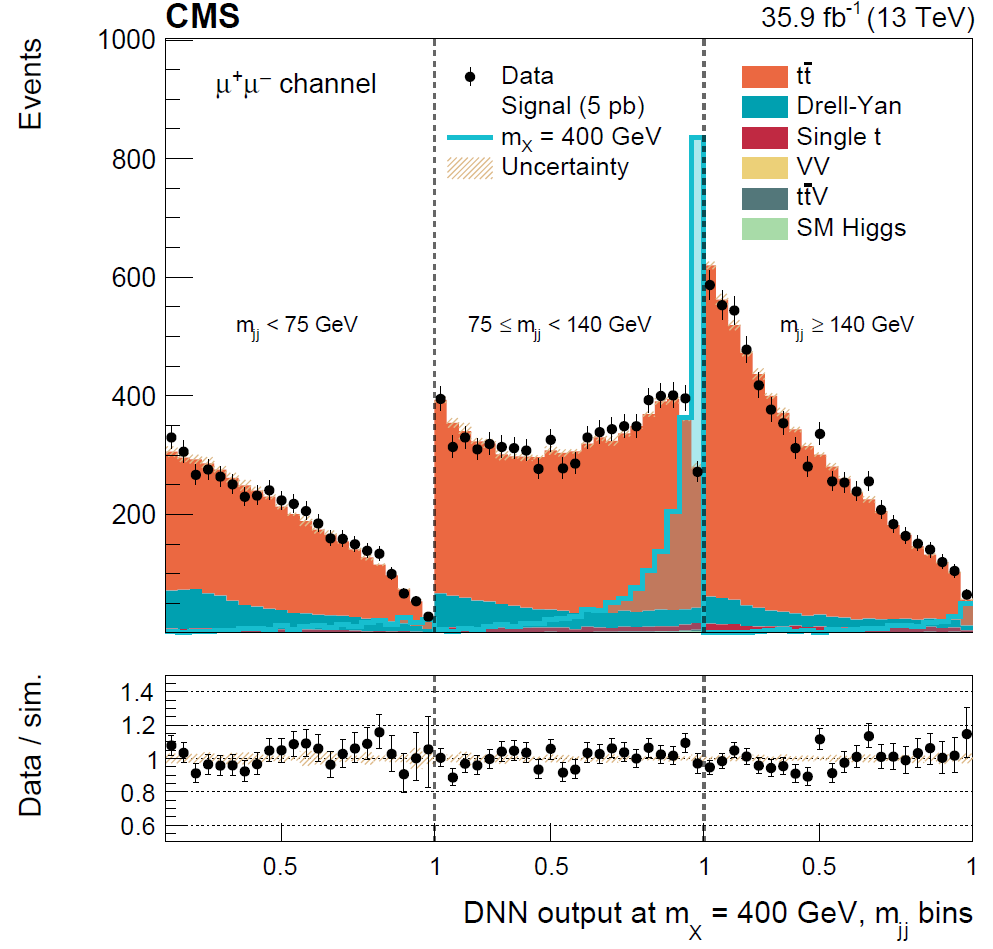}
\caption{\em Output of the DNN parametrized for $m_X=400$ GeV  in events with two muons for dijet masses $m_{jj}<75$ GeV (left), $75 \leq m_{jj} < 140$ GeV (center), and $m_{jj} \ge 140$ GeV (right) in the CMS analysis reported in~\cite{cmshhbbllnn}. The resonant signal was scaled to a cross section of 5 pb. Statistical uncertainties are displayed as vertical bars on the data points; shaded bands indicate the size of post-fit systematic uncertainties. }
\label{f:DNN}
\end{center}
\end{figure}

\begin{figure}[h!]
\begin{center}
\includegraphics[scale=0.42]{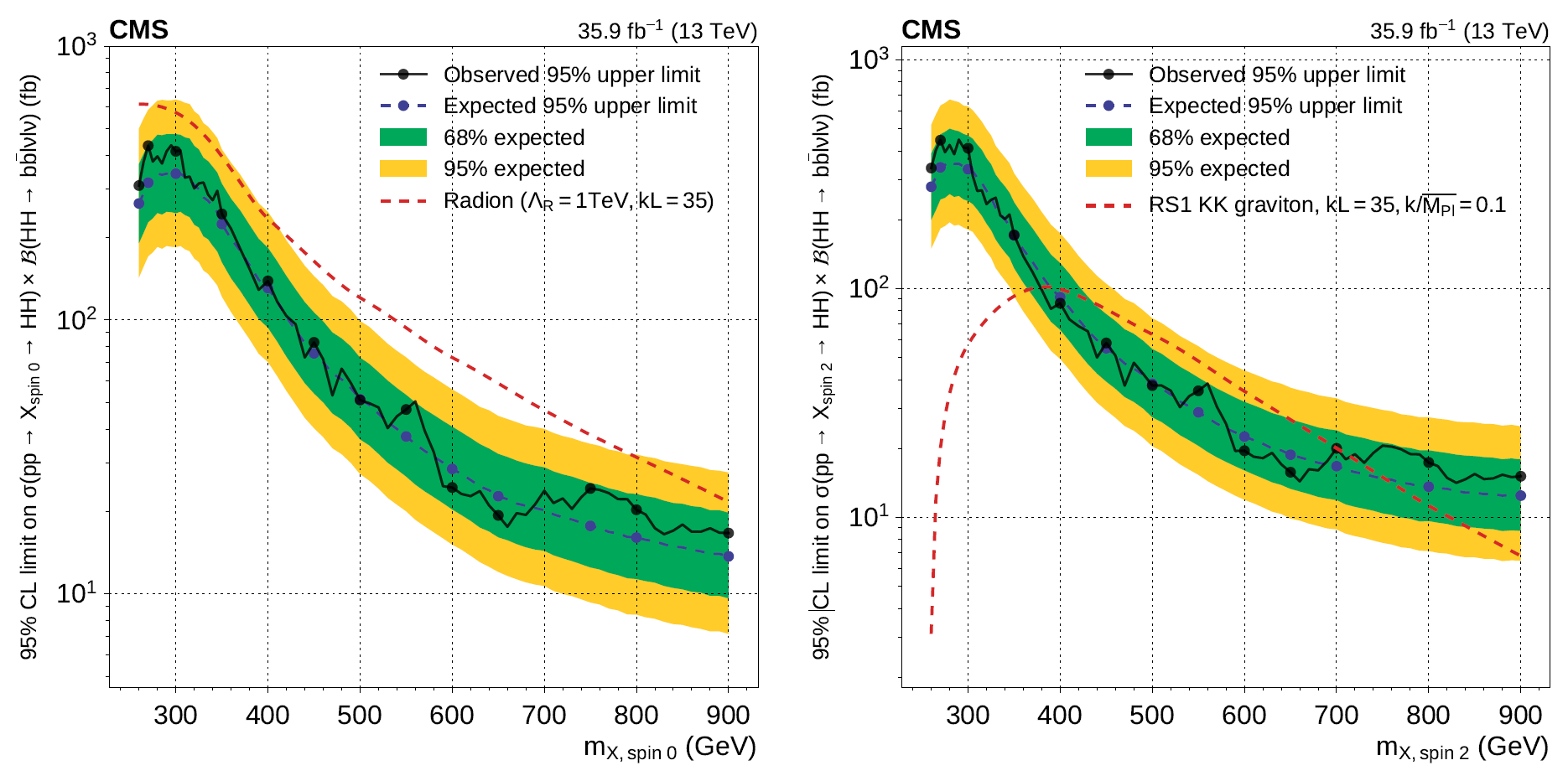}
\caption{\em 95 \% CL upper limits on the cross section times branching fraction of $X \to hh \to b \bar{b}VV \to b \bar{b} l \nu l \nu$ $(V=W,Z)$ as a function of the new resonance mass $m_X$ by the search reported in~\cite{cmshhbbllnn}. The dashed curves show predicted cross sections for radion (left) and KK graviton (right) in the absence of mixing with Higgs bosons~\cite{gbulkrevised2}.}
\label{f:cmshhbbllnn}
\end{center}
\end{figure}

Although still in preprint form at the time of writing, we include for completeness here a mention of two further results on resonances decaying to Higgs pairs. The first one was produced by CMS using the final state including two $b$ quarks, two charged leptons, and neutrinos, which may arise both from $hh \to b \bar {b} WW$ and from $hh \to b \bar {b} ZZ$ decays~\cite{cmshhbbllnn}. Events were collected from a dilepton trigger, and required to contain two opposite-charge electrons or muons with $12<m_{ll}<M_Z-15$ GeV, a criterion aimed at reducing quarkonia decays, semileptonic $B$-hadron decays, and the high-tails of $t \bar{t}$ and Drell-Yan backgrounds. The cut leaves $h \to WW$ candidates unaffected and retains a good part of $h \to ZZ \to ll \nu \nu$ candidates. Jets reconstructed with the anti-$k_T$ algorithm from PF candidates using a radius of 0.4  were required to have a CMVA value of $b$-tagging above the medium working point of the algorithm~\cite{cmva1,cmva2}. A deep neural network from the Keras package~\cite{keras} was trained to distinguish signal events from the irreducible background arising from top quark pair production. Binned-likelihood fits to the DNN output discriminant in three different ranges of dijet mass were fit to extract upper limits on the signal component (see Fig.~\ref{f:DNN}). The data allowed to set 95\% CL limits to the cross section times branching ratio of narrow spin-zero resonances as well as spin-two resonances with minimal gravity-like couplings, in the mass range $260<m_X<900$ GeV. The limits range respectively from 430 to $17 fb$ and from 450 to $14 fb$, as shown in Fig.~\ref{f:cmshhbbllnn}.

The other CMS result still in preprint form which we exceptionally report on here used the 2012 dataset and reported on a search in the $b \bar{b} \gamma \gamma$ final state~\cite{cmsbbgg}. The bulk RS model was considered by adopting the so-called ``elementary top hypothesis'', which includes right-handed top quarks localized on the TeV brane~\cite{wed2}.  Events were selected by a diphoton trigger path, and photon candidates were identified offline with a cut-based selection of ECAL clusters. Pairs of photons in the range $100<m_{\gamma_1 \gamma_2}<180$ GeV were selected, and further requirements $p_T^{\gamma_1} >m_{\gamma_1 \gamma_2}/3$, $p_T^{\gamma_2} > m_{\gamma_1 \gamma_2}/3$ on $p_T$-ordered photons $\gamma_1$ and $\gamma_2$ were applied to create a smooth background distribution in the mass of photon candidate pairs. Anti-$k_T$ jets were obtained by clustering PF candidates within a cone of radius $R=0.5$, and considered if they were well-separated in angle ($\Delta R>0.5$) from photon candidates. At least one jet must contain a medium CSV $b$-tag; the pair of jets with highest combined $p_T$ was retained as product of Higgs decay. Events were then categorized according to the number of $b$-tags in the Higgs candidate jets. A kinematic fit was applied to the four Higgs decay products to improve the resolution on $m_X$ in the $X \to hh \to b \bar{b} \gamma \gamma$ hypothesis. The search for the resonant signal distinguished a low-mass ($260<m_X<400$ GeV)  from a high-mass ($400<m_X<1100$ GeV) region. In the background-rich low-mass region, for each $m_X$ hypothesis events were selected in a narrow interval of the four-body mass distribution, and the $m_{\gamma_1 \gamma_2}$ and $m_{b \bar{b}}$ distributions were searched for a signal component. In the high-mass region, where backgrounds were smaller, the $m_{b \bar{b} \gamma_1 \gamma_2}$ distribution was searched directly for a resonant signal. No significant signal was found in the data. Using theoretical predictions for the cross sections of the new resonances~\cite{hiddensec_higgsium,bargerradion,ref76bbgg,ref77bbgg}, an exclusion at 95\% CL on the radion hypothesis with an ultraviolet cutoff scale $\Lambda_R=1$ TeV was obtained for masses below 980 GeV; for $\Lambda_R=3$ TeV the exclusion reduced to the mass region 200-300 GeV. For a spin-2 KK graviton, assuming a value of the dimensionless parameter $\tilde{k}=0.2$, an excluded region was found in the range 325-450 GeV. 

CMS recently started to produce results with Run 2 data on resonant Higgs pair production, with the search in the $b \bar{b} \tau \tau$ final state using $35.9 fb^{-1}$ of 13-TeV collisions~\cite{cmshhbbtt}. The $b \bar{b} \tau \tau$ decay mode has a combined branching fraction of 7.3\% in the SM; the considered decay modes of $\tau$ lepton pairs were ones with one $\tau$ decaying to $\nu_\tau$ plus hadrons, and the other decaying to either the same final state or to $e \nu_e \nu_\tau$ or $\mu \nu_\mu \nu_\tau$; overall these combinations cover 88\% of the possible decay modes of the $\tau$ lepton pairs. The mass of $\tau$ lepton pairs was computed with a dynamical likelihood technique~\cite{svfit}.  Jets were reconstructed by identifying particles with the PF algorithm and clustering them with the anti-$k_T$ algorithm using jet radii of $0.4$ and $0.8$. The jet mass was computed for the  larger cone radius jets using the soft-drop grooming algorithm~\cite{softdrop}. A wide jet was classified as a boosted dijet candidate if it had a mass is larger than 30 GeV and was made by two subjets each geometrically matched to jets reconstructed with the anti-$k_T$ algorithm using the radius parameter $R=0.4$. Events were divided in a high- and a low-purity category depending on having two or only one $b$-tagged jet. A four-body mass was finally obtained with a kinematic fit~\cite{kinfitbbtt}. Backgrounds arising from $Z \gamma \to ll$  and $W \to l \nu$ in association with jets (with $l = e, \mu, \tau$), diboson ($WW$, $ZZ$, and $WZ$), and SM single Higgs boson production were simulated with Madgraph 5~\cite{madgraph}, while the single top and $t\bar{t}$ backgrounds were simulated at NLO precision with POWHEG 2.0. The NNPDF3.0~\cite{nnpdf30} PDF set was used.  The search sets model-independent cross section limits on the production of a narrow resonance decaying to Higgs pairs. In the context of the hMSSM scenario these correspond to a 95\% CL exclusion of the region in parameter space $230<m_A<360$ GeV for $\tan \beta <2$ (see Fig.~\ref{f:annulibbbb}, right).

\subsubsection {\it Searches for $h \to aa$ Decays}

In 2HDM extensions of the SM such as the NMSSM~\cite{nmssm1,nmssm2,nmssm3,nmssm4}, the pseudoscalar Higgs boson $a$ can be assumed to be lighter than the 125 GeV Higgs, by having its mass protected by the Peccei-Quinn symmetry~\cite{pecceiquinn1,pecceiquinn2,pecceiquinn3}. Before the LHC, searches for these light pseudoscalars targeting masses above the $\tau$-pair production threshold were performed both at the Tevatron and at LEP II. ALEPH set a limit $m_H>107$ GeV searching for $H \to aa$ decays and assuming a branching fraction of unity for the decay~\cite{haaaleph}; DZERO also set limits on $H \to aa$ by searching in the $\mu \mu \tau \tau$ final state~\cite{haadzero}. At the LHC, the ATLAS collaboration sought for the signature of decays of the 125 GeV scalar into two $a$ bosons in $\tau \tau \mu \mu$ and in $b \bar{b} b \bar{b}$  final states; the present precision on the allowed branching fraction into undetected final states of Higgs decays still leaves considerable room for these possibilities. The first study was performed on 20.3 $fb^{-1}$ of 8-TeV proton-proton collisions data collected in the 2012 run~\cite{haaatlas_1}. Events were selected to contain a muon pair plus an additional signature of $\tau$ lepton decays; the mass distribution of the muon pair assigned to the decay of an $a$ boson was sought for narrow bumps. The absence of detected signals was turned into exclusion limits on the product of cross section of Higgs bosons times their branching fraction to $a$ boson pairs, normalized to the SM  production of Higgs bosons by gluon fusion; limits on the heavy-$H$ mass were also obtained, in the hypothesis of $m_a=5$ GeV.
A second search in a dataset corresponding to 3.2 $fb^{-1}$ of 13-TeV collisions considered Higgs bosons produced in association with a $W$ boson, and targeted the range $20 < m_a < 60$ GeV with both $a$ bosons decaying to $b$ quarks~\cite{haaatlas_2}. The experimental signature included an electron or muon and missing transverse energy from the decay of the $W$ boson, and multiple hadronic jets from the $a$ bosons decay; the leptons provided efficient triggers for the search. The data were selected to contain at least three jets, two or more $b$-tagged by a multi-variate discriminant using secondary vertex and soft lepton information; a further BDT discriminant was employed to separate the signal from the dominant background coming from $t \bar{t}$ decays. The search did not return any significant signal; ATLAS thus set 95\% CL limits on the cross section of $WH$ production  times the branching fraction of the Higgs boson to an $a$ pair and the squared branching fraction of $a \to b \bar{b}$. The limits range from 6.2 to 1.5 $pb$ in the investigated interval of $m_a$ values. 

\begin{figure}[h!]
\begin{center}
\begin{minipage}{0.49\linewidth}
\includegraphics[scale=0.45]{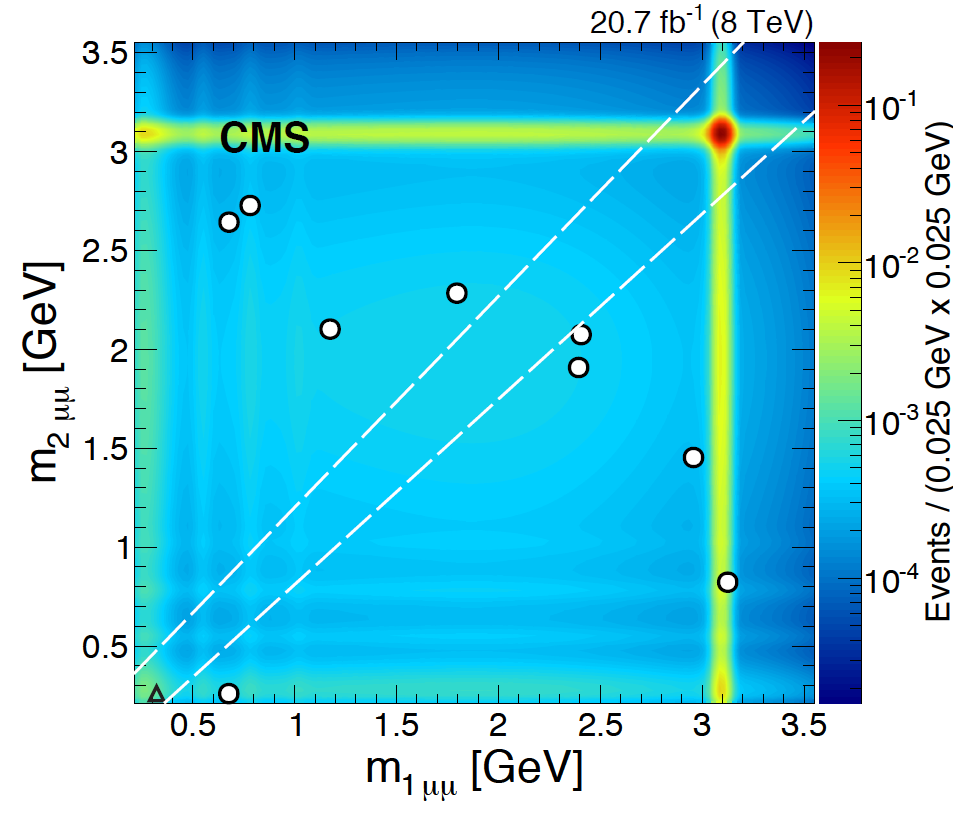}
\end{minipage}
\begin{minipage}{0.49\linewidth}
\includegraphics[scale=0.46]{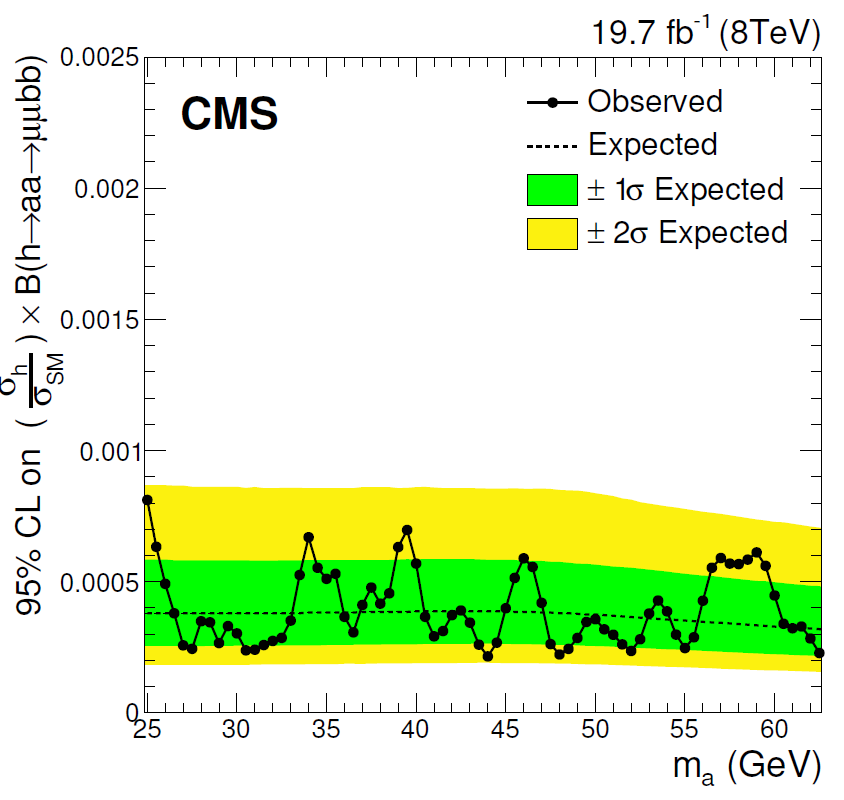}
\end{minipage}
\caption{\em Left: event candidates in the CMS search for $a_1 a_1 \to 4\mu$ decays~\cite{haa4mucms1}. The triangle in the lower left corner is the only event accepted in the signal region along the diagonal. Right: 95\% CL upper limits obtained by CMS in 8-TeV searches~\cite{haacms2} on the signal-strength modifier $\sigma_h / \sigma_h^{SM}$ multiplied by the branching fraction of $a_1 a_1$ to $\mu \mu b \bar{b}$ final state.}.
\label{f:haacms1}
\end{center}
\end{figure}

The CMS collaboration also produced several searches for $h \to aa$ decays. In a search on 19.7 $fb^{-1}$ of 8-TeV collisions~\cite{haa4mucms1}, they considered $a$ bosons in the $2m_{\mu} - 2m_{\tau}$ range in the context of the NMSSM and dark sector models~\cite{darksusy1,darksusy2,darksusy3,darksusy4}. Events with four muons were selected from data collected by a dimuon trigger, and a signal was sought in the plane defined by the invariant masses of muon pairs. Backgrounds from $b \bar{b}$ production and from prompt- and non-prompt-production of $J / \psi$ meson pairs were estimated with data-driven techniques. One event with the required characteristics was found in the search region, compatible with the estimated background of $2.2\pm 0.7$ events (see Fig.~\ref{f:haacms1}). The resulting rate limits were provided in a model-independent way and interpreted in the context of the NMSSM.
A second search in the 2012 dataset was also performed for pairs of NMSSM bosons $a_1$ or $h_1$ produced in the decay of the Higgs boson,  looking for the final state of  four $\tau$ leptons~\cite{haacms2}. This study focused on very light states, from 4 to 8 GeV. Events were selected by a dimuon trigger and required to contain a pair of same-sign muons, each accompanied by another isolated track identified as a one-prong $\tau$ decay candidate. The plane defined by the invariant masses of the two track-muon pairs, used as a discriminant of the diboson signal from QCD backgrounds, was fit with a binned likelihood technique to extract information on a possible signal component. The data were found to be well-modeled by backgrounds alone; this resulted in the setting of 95\% CL upper limits on the product of production cross section times branching fraction of the Higgs boson into the pair of light bosons, times the squared branching fraction of the latter into $\tau$ lepton pairs. These vary from 4.5 to 10.3 $pb$ in the considered mass range.
Finally, the most recent publication by CMS~\cite{haacms3} on the considered $h \to aa$ signature was based on the same dataset as the one above. It considered the higher mass range 5.0-62.5 GeV, and three different signatures of $aa$ decay: the $4 \tau$ mode, the $\mu \mu b \bar{b}$ mode, and the $\tau \tau \mu \mu$ mode. Upper limits at 95\% CL on the Higgs production cross section times branching fraction of $h \to aa$, divided by the SM prediction for the Higgs cross section, were obtained by the searches down to values of 17\%, 16\% and 4\% using the three decay modes. 

\subsection {\it Searches for New Resonances Decaying to Gluon Pairs}
\label{s:gg}

\noindent
Resonances decaying to jet pairs are attractive game at hadron colliders. A strongly-interacting object can be produced with comparatively high rates even at masses that are a significant fraction of the total collision energy, electing dijet final states to the role of the most effective direct probe of the highest reachable energy scales. After early searches at the SppS~\cite{dijetua1_1,dijetua1_2,dijetua2_1} and a long campaign of searches in Tevatron collisions~\cite{dijetcdf1,dijetd02,dijetcdf3,dijetcdf4,dijetd01,dijetcdf5,dijetd02}, the ATLAS and CMS experiments followed the same thread, producing a number of searches for dijet resonances in their Run 1 and Run 2 datasets. Here we however discuss only a subset of those results, focusing on  searches for resonances decaying to  gluon pairs, which rightfully belong to the class of diboson final states object of this review. While the signature of energetic quarks and gluons cannot be distinguished well enough to allow for an advantageous experimental categorization of dijet events as the result of the $qq$, $qg$, or $gg$ final states, all models that predict the existence of a new particle undergoing $X \to gg$ decay foresee that the natural width $\Gamma_X$ is sizable. Due to the steep decrease of parton distribution functions as a function of Bjorken-$x$ at  large $x$ values, the mass distribution of the resonance at particle level assumes a quite non-Gaussian shape, with long tails toward masses much smaller than $m_X$. This peculiarity makes the search for gluon-pair resonances rather different from the search of generic narrow states decaying to jet pairs, and the result of the latter not usable to extract precise information on models predicting $gg$ resonances.

The ATLAS collaboration performed a number of searches for the decay of heavy resonances in dijet final states. Most of them consider only models producing decays to quark pairs or $qg$ final states~\cite{dijetatlas_1,dijetatlas_3}; these are not relevant to the present review. In~\cite{dijetatlas_3} and later publications ATLAS also considered a simplified model by searching for Gaussian-shaped signals produced by resonances of natural width ranging from 3\% to 15\% of the resonance mass. We feel these limits cannot be easily converted into exclusion bounds for gluon-gluon resonances in the data, which would display different distributions due to more prominent initial and final state radiation effects than those present in processes involving resonances primarily coupling and decaying to quarks. 

\begin{figure}[h!]
 \begin{center}
\begin{minipage}{0.49\linewidth}
\includegraphics[scale=0.45]{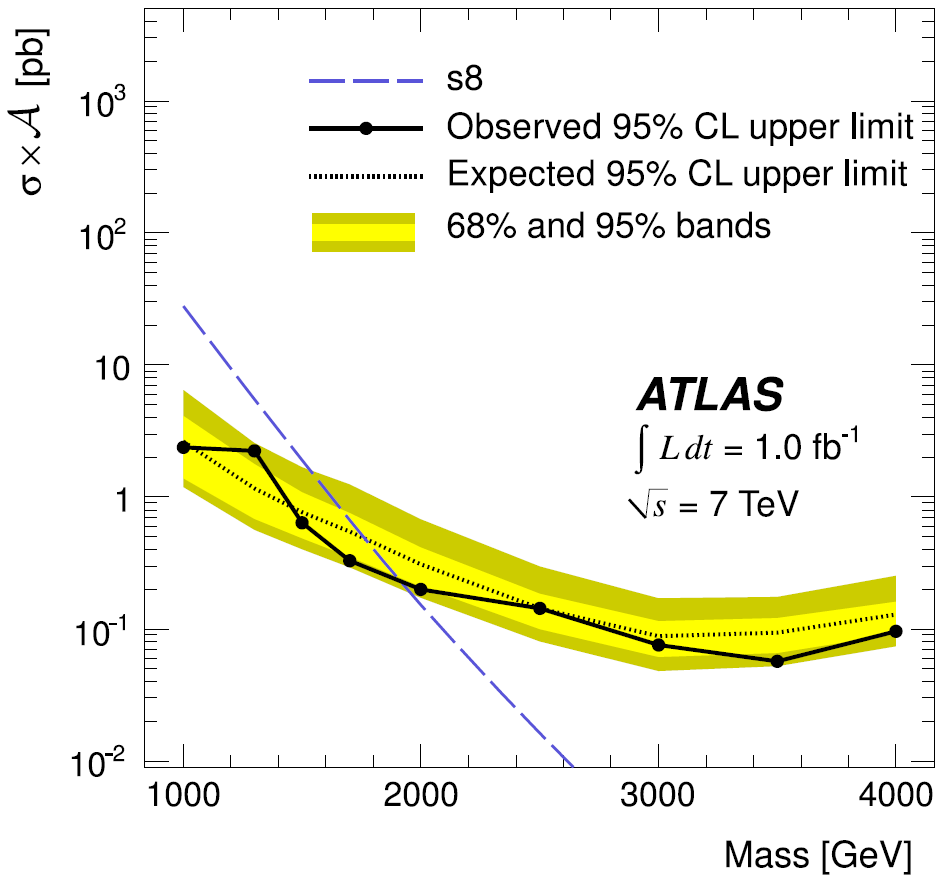}
\end{minipage}
\begin{minipage}{0.49\linewidth}
\includegraphics[scale=0.42]{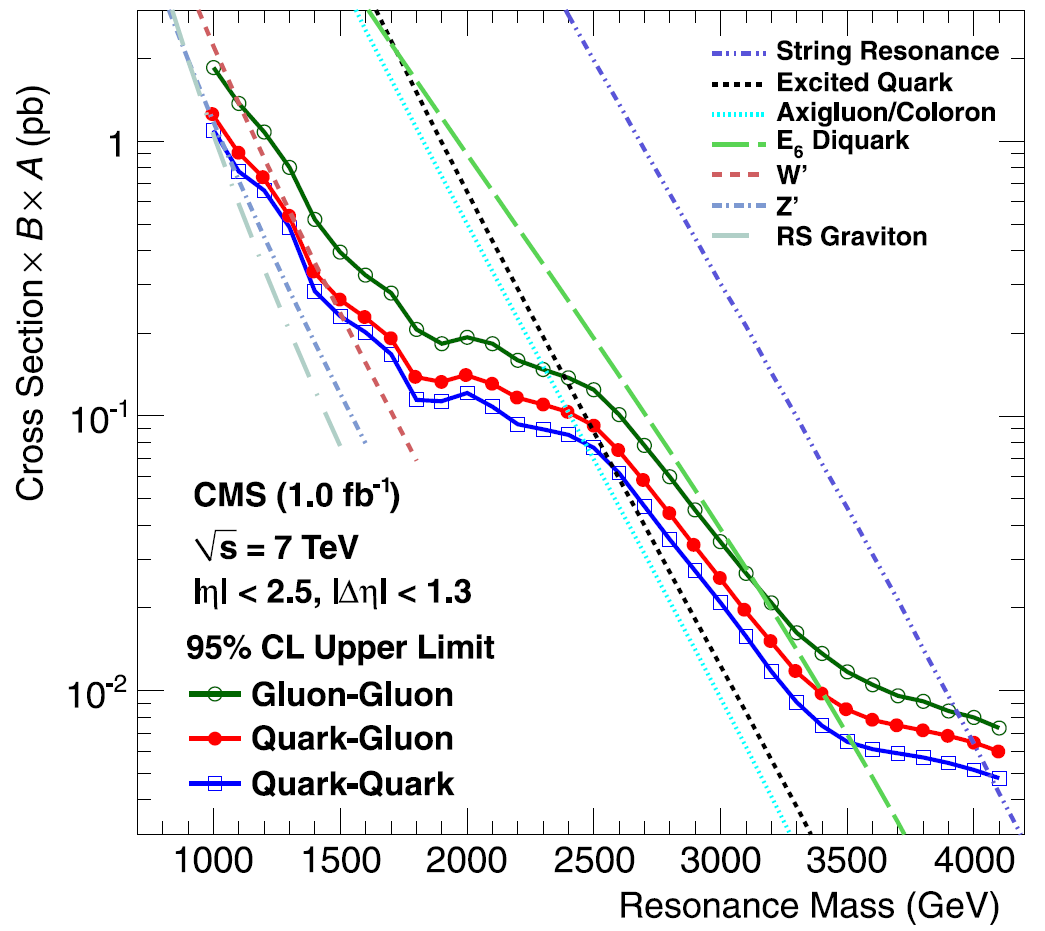}
\end{minipage}
\caption{\em Left: 95\% CL upper limit on the product of cross section and acceptance of a colour-octet scalar as a function of the particle mass obtained by ATLAS~\cite{dijetatlas_4}. The coloured bands indicate the expected range of upper limits at 68\% and 95\% given the search sensitivity; the points are the actual excluded values. The predicted cross section times acceptance of a S8 colour-octet scalar is also shown as a dashed curve. 
Right: 95\% CL upper limits obtained by the CMS search described in~\cite{dijetcms_4} on the product of cross section times branching fraction times acceptance for resonances decaying into $qq$ (open boxes), $qg$ (solid circles), and $gg$ (open circles) final states. A number of theoretical predictions are overlaid for comparison. }
\label{f:dijetatlas4}
\end{center}
\end{figure}

The search reported in~\cite{dijetatlas_4}, which employed $1 fb^{-1}$ of proton-proton luminosity of 7-TeV collisions, was the first ATLAS result considering the S8 model. They scanned with the Bumphunter technique~\cite{bumphunter1,bumphunter2} the mass distribution for a signal modeled with Madgraph 5 at leading order, interfaced to Pythia with CTEQ6L1 PDFs, with the ATLAS MC09 tune~\cite{mc09tune} and passed through a full detector simulation. The wider shape of the S8 model clearly impacted the sensitivity of the search, which reported cross section times acceptance limits that were indeed two to 10 times worse than those set on excited quarks in the mass range from 1 to 4 TeV (see Fig.~\ref{f:dijetatlas4}, left). The resulting 95\% CL lower limit on the mass of colour-octet scalars was set at 1.92 TeV. The study also  provided cross section times acceptance limits for generic signals producing a Gaussian mass distribution, assuming that the detector reconstructed the final state with a 4\% resolution. 
In a repetition of the same search, using the full 7 TeV dataset corresponding to $4.8 fb^{-1}$ of integrated luminosity collected in 2011~\cite{dijetatlas_5}, ATLAS could only report a limit at 1.86 TeV because of a 1-sigma upper fluctuation of the data for masses of about 2 TeV. In that study they also set for the first time 95\% CL limits on the mass of string resonances at 3.61 TeV. 
Successively, using 20.3 $fb^{-1}$ of 8-TeV collisions collected in 2012, ATLAS could significantly increase the lower limit on the mass of S8 resonances to 2.70 TeV~\cite{dijetatlas_6}. In that study a more complete treatment of generic dijet mass resonance shapes was also applied, using a non-relativistic approximation of the Breit-Wigner resonance shape convoluted with parton distribution functions, non-perturbative effects, and a Gaussian detector resolution model. 
The worsening of the sensitivity as the relative width $\Gamma/m$ increases was found significant for resonance masses of 2.0 (3.5) TeV, where expected upper limits on the cross section times acceptance increase by a factor of two (ten), respectively. 
The same generic search was also performed in~\cite{dijetatlas_8}, using data corresponding to 3.6 $fb^{-1}$ of luminosity collected in 2015 at 13-TeV proton-proton collisions, and in~\cite{dijetatlas_10}, which added to that dataset the one collected in 2016, searching in a total luminosity of 37 $fb^{-1}$ of Run 2 collisions. The last study did not set explicit limits on $gg$-decaying resonances but provided and improved folding procedure of detector and physics effects of the generic search for Gaussian signals. The extraction of precise information on the existence of $gg$-decaying resonances from those results is however not straightforward.

\begin{figure}[h!]
\begin{center}
\begin{minipage}{0.53\linewidth}
\includegraphics[scale=0.4]{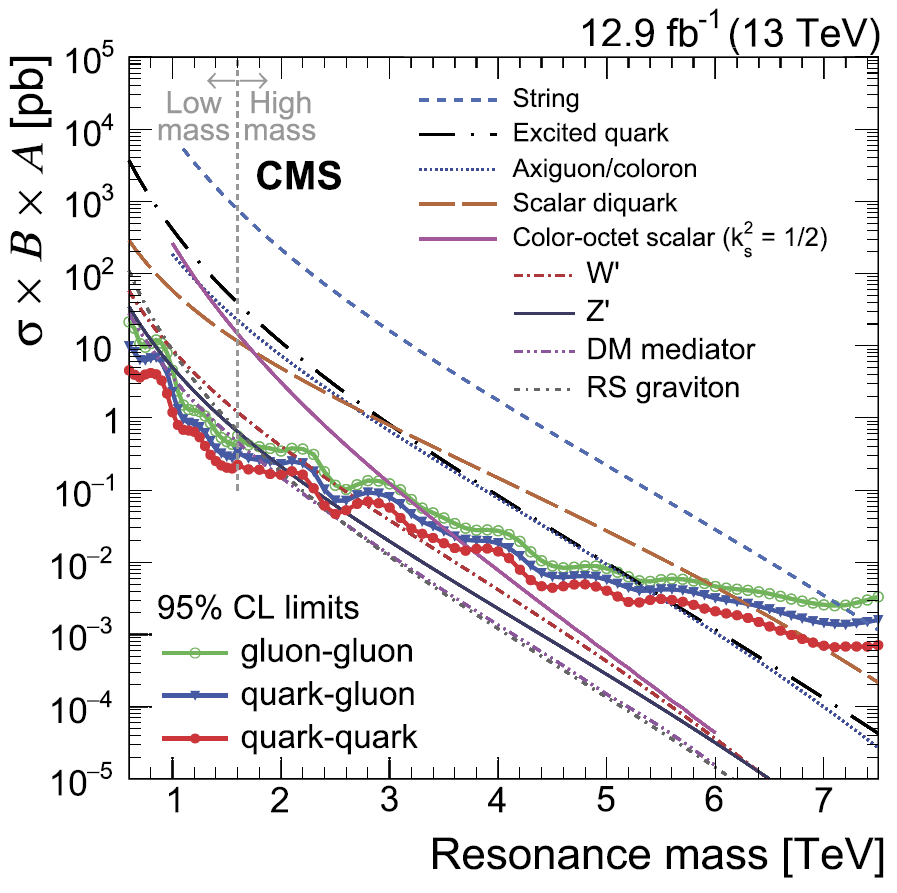}
\end{minipage}
\begin{minipage}{0.46\linewidth}
\includegraphics[scale=0.37]{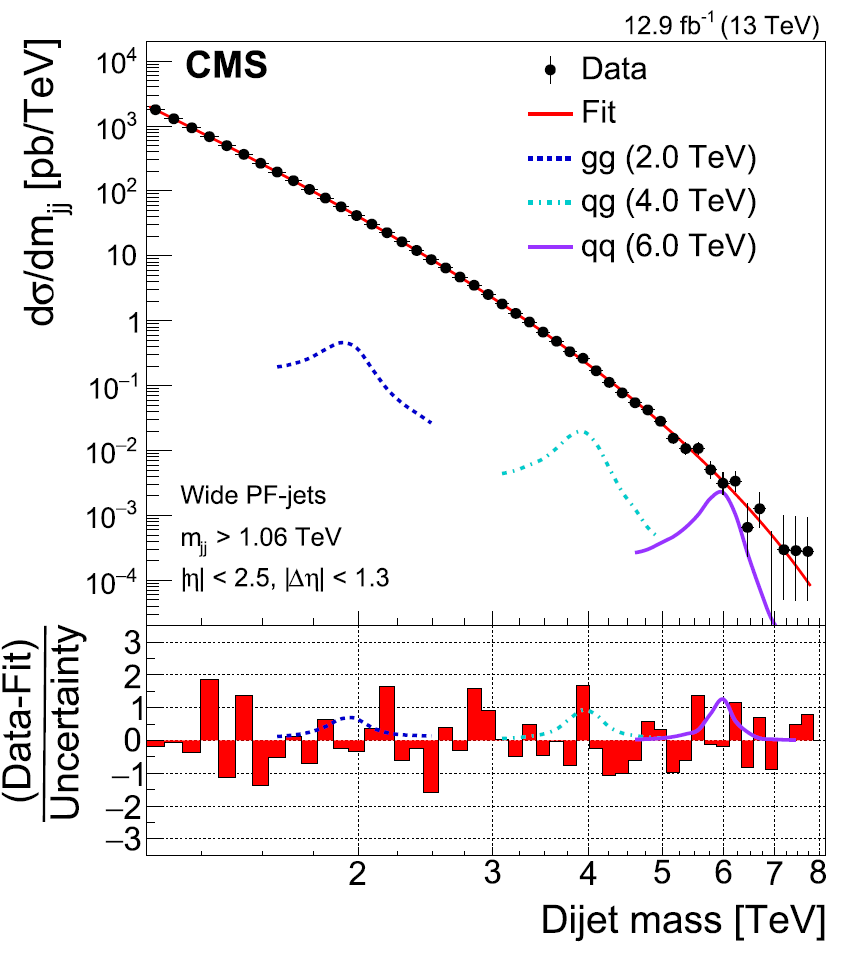}
\end{minipage}
\caption {\em Left: Mass distribution of jet pairs from the CMS analysis of data corresponding to 12.9 $fb^{-1}$ of 13-TeV proton-proton collisions reported in~\cite{dijetcms_7}. A background-only fit is overimposed to the data; sample signal shapes for $gg$, $qg$, and $qq$ resonances of mass respectively of 2.0, 4.0, and 6.0 TeV are also shown. The lower panel shows the data minus fit residuals, normalized by their statistical uncertainty.
Right: 95\% CL upper limits on the product of cross section times branching fraction times acceptance  ($\sigma B A$) of resonances decaying to jet pairs, obtained by CMS in the same analysis~\cite{dijetcms_7}. The three thick curves refer to $qq$-, $qg$-, and $gg$-decaying resonances; the graph includes the predicted dependence of the product $\sigma B A$ on resonance mass for a number of considered models. }
\label{f:dijetcms_78}
\end{center}
\end{figure}

\noindent
The CMS collaboration devoted great attention to the study of dijet events, producing a number of searches for high-mass states which could become manifest in that signature. Again, we report here only studies which resulted in the extraction of information on models predicting resonances decaying to gluon pairs. The experiment uses PF objects clustered in ``wide jets'' to perform these searches, as a larger cone radius reduces sensitivity to systematic effects and improves the mass reconstruction; a selection of the data to require a pseudorapidity separation $|\Delta \eta|<1.3$ between the two jets is also usually applied to reduce QCD backgrounds. The CMS collaboration produced an early search for dijet resonances employing only 2.9 $pb^{-1}$ of 7-TeV proton-proton collisions~\cite{dijetcms_1}, and a more sensitive result using a larger data sample corresponding to a luminosity of 1 $fb^{-1}$~\cite{dijetcms_4}, reporting upper limits on the product of cross section times branching fraction times acceptance for a number of dijet-decaying resonances (see Fig.~\ref{f:dijetatlas4}, right). Those searches could not limit the mass range of models predicting new $Z'$ or $G^*$ resonances decaying to $gg$ pairs; the search however set a lower limit on the string resonance mass at 4.0 TeV.  A search later reported in~\cite{dijetcms_5} considered 4.0 $fb^{-1}$ of 8-TeV collisions and searched for $qq$, $qg$, and $gg$ resonances in the mass range from 1.0 to 4.8 TeV. The data were found to be well described by the four-parameter function\par

\begin{equation}
\frac{d\sigma}{dx}=\frac{p_0 (1-x)^{p_1}}{x^{p_2+p_3 \log{x}}}
\end{equation}

\noindent
with $x=m_{jj}/\sqrt{s}$, and upper limits on the strength of the three different resonance shapes were extracted at each mass point with a Bayesian technique~\cite{statpdg}. Limits on the mass of a S8 colour-octet scalar were set at 2.79 TeV. For gravitons described by the Randall-Sundrum model, the limit was obtained by considering in the signal-extraction fit a weighted average of the expected signal shape of $gg$ and $qq$ decaying states, to account for the possible decay modes of the searched state; the lower limit was set at 1.45 TeV.
The study of~\cite{dijetcms_7} considered a luminosity of 12.9 $fb^{-1}$ of 13-TeV proton-proton collisions, extending a previous search~\cite{dijetcms_6} on a smaller Run 2 dataset. The search was performed separately in two different mass regions: together with regular high-energy dijet events collected by the HLT system, for which the trigger selection was fully efficient for dijet masses above 1.1 TeV, events were also collected in a separate data path by reconstructing jets online with simplified ``calorimeter-based'' algorithm, and storing only calorimeter information, thus allowing for lower $E_T$ thresholds. This way, the low-mass search could extend from 0.6 to 1.6 TeV, while the high-mass search covered the rest of the spectrum up to 7.5 TeV (the highest-mass event had $m_{jj}=7.7$ TeV). To test the scalar colour-octet model a value of the squared anomalous coupling $k^2_s=1/2$ was used~\cite{scalarcolourkonehalf}. This implies a reduction of both natural width and cross section by a factor of two with respect to the values assumed in the previous search~\cite{dijetcms_6}. The data were observed to be well described by a QCD-inspired four-parameter functional form, as shown in Fig.~\ref{f:dijetcms_78} (left), and limits were set on the mass of resonances for a number of models; in particular, for what concerns $gg$ resonances, 95\% CL lower limits on the new particle mass were set at 3.0 TeV for colour-octet scalars. A lower limit on the mass of RS gravitons was also set by considering a 60\% branching fraction of the graviton to quark pairs and 40\% to gluon pairs; that result excluded the region $m_G^*<1.9$ TeV, again at 95\% CL (Fig.~\ref{f:dijetcms_78}, right).
In the search described in~\cite{dijetcms_8} CMS exploited the technique called ``data scouting'' to collect events with hadronic jets at trigger level with a rate of 1 kHz, by storing a reduced amount of information on the online-reconstructed jets subjected to a  calibration procedure; this allowed the high-statistics investigation of the mass range from 400 to 1850 GeV, enhancing the sensitivity to resonances with reduced couplings to quarks and gluons. The data corresponded to an integrated luminosity of 18.8 $fb^{-1}$ and were collected at a center-of-mass energy of 8 TeV during 2012. 
The extracted cross section times acceptance limits on narrow resonances decaying to gluon pairs were used to obtain mass limits on RS gravitons, considered as a benchmark. The mass region $500 < m_{G^*} < 1000$ GeV was excluded by averaging the observed limits for $qq$ and $gg$ resonances, a technique which we stigmatize as improper although in the considered case it is probably a reasonable approximation.

\noindent
Table~\ref{t:dijetlimits} offers a summary of the results obtained by the LHC experiments in their searches for resonances decaying to gluon pairs. We omit to report the result of searches for generic Gaussian signals by ATLAS, which we consider not relevant for the $gg$ final state.
\begin{table}[h!]
\begin{center}
\caption{\em Summary of searches produced by the LHC experiments in dijet final states which
considered the signature of a S8 scalar or a RS graviton decay to gluon pairs, and 
lower limits set on the resonance masses, when applicable.}
\begin{tabular}{|l|c|c|c|c|c|}
\hline
Experiment & Luminosity & cm energy & S8 scalar & RS graviton & Reference \\
                   & ($fb^{-1}$) & (TeV) &  (TeV) & (TeV) & \\
\hline
ATLAS & 1.0 & 7.0 & $>1.92$ & - &~\cite{dijetatlas_4} \\
ATLAS & 4.8 & 7.0 & $>1.86$ & - &~\cite{dijetatlas_5} \\
ATLAS & 20.3 & 8.0 & $>2.70$ & - &~\cite{dijetatlas_6} \\
CMS  & 0.003 & 7.0 & - & - &~\cite{dijetcms_1}\\
CMS  & 1.0 & 7.0 & - & - &~\cite{dijetcms_4}\\
CMS  & 4.0 & 8.0 & $>2.79$ & $>1.45$ &~\cite{dijetcms_5}\\
CMS & 18.8 & 8.0 & - & $<0.5$ or $>1.0$ &~\cite{dijetcms_8}\\
CMS  & 2.4 & 13.0 & $>3.1$ & - &~\cite{dijetcms_6}\\
CMS & 12.9 & 13.0 & $>3.8$& $>1.9$ &~\cite{dijetcms_7}\\
\hline
\end{tabular}
\label{t:dijetlimits}
\end{center}
\end{table}

\clearpage

\section {Summary and Outlook \label{summary}}
\label{s:outlook}

\subsection {\it Summary of LHC Searches}

\noindent
The results described in the previous section are all negative, as none of the dozens of searches for new resonances in diboson final states has so far yielded a credible hint of the presence of new physics in the analyzed datasets. The amount of information they provide is however conspicuous, as they exclude large swaths of the parameter space of theories put forth to extend the SM. The results clarify that no new physics is at easy reach in TeV-scale collisions, exacerbating the naturalness problem and the other nagging questions we are facing in our attempts to understand the structure of the subnuclear world. 

We can summarize the results by grouping them in four main categories.\par

\begin{itemize}

\item The absence of new gauge bosons extending the group structure of the SM has been verified up to the scale of 2 TeV and above using diboson final states. This of course does not explicitly exclude the general idea of new U(1) groups or still larger structural additions, but the limits have started to make them look less appealing as a way to alleviate the shortcomings of the SM.

\item New Higgs bosons belonging to a 2HDM extension of the Higgs sector have been sought in a number of diboson final states. Due to the wide variety of models, the absence of any signal of additional Higgs bosons cannot be used to say that 2HDM theories are disfavored by the data; yet for some of the simplest versions, like the MSSM, this is certainly true.

\item Large extra dimensions have been probed with boson pairs in the original scenario of Randall and Sundrum, as well as in its bulk graviton version, and in the ADD model. These models have become less fashionable since the TeV region was unsuccessfully probed by LHC searches.

\item Resonances decaying to gluon pairs have also been excluded in a wide range of energies. The absence of clear indications on the mass scale of new such states makes the continuation of these searches an obligation as long as the LHC will continue to deliver new data.

\end{itemize}

\subsection {\it Outlook}

\noindent
With the collection of over 50 inverse femtobarns of integrated luminosity of 13-TeV proton-proton collisions in 2017, the ATLAS and CMS experiments both surpassed by a wide margin the mark of 100 inverse femtobarns of luminosity declared ``good for physics analysis'', most of which were acquired at the highest collision energy among those scanned by the LHC. The resulting datasets allow the production of a wealth of measurements. Most notably, the Higgs boson was studied in great detail, obtaining cross section measurements in all the main production modes and estimates of the Higgs branching fraction to five different final states, all of which were found in excellent agreement with SM predictions. The Higgs boson mass is now known with a precision of a fraction of a percent, and a vigorous campaign of measurements has started to close in on some of its most detailed properties, like the intrinsic width and the self-coupling constant. In parallel, detailed measurements of other SM observables allowed a qualitative step forward in the area of precision physics. From the viewpoint of experimental studies the standard model is in a quite healthy state, with measurement precisions of ${\cal{O}}(10^{-4})$ on the $W$ mass and ${\cal{O}}(10^{-3})$ on the top quark and Higgs boson masses still perfectly matching the indirect experimental inputs on electroweak observables produced at the $Z$ pole and in other precision measurements~\cite{gfitter}.


On the other hand, we know that the SM must break down at some point, as it is at best an effective theory, valid until we reach an energy scale where new effects start to contribute visibly to the phenomenology. The search for those new effects by the LHC experiments  has been arguably even more intense than the measurement campaign targeting the Higgs boson or other SM physics, as is testified by the fact that the exotica groups of the two collaborations are the most prolific ones  in terms of published results. Yet no anomaly has been detected yet. This dearth of hints that we might ever be getting close to the breaking point of the SM is a sobering input, as the LHC will only be able to increase its center-of-mass energy by 7\% in the near future, with a resulting limited increase of parton luminosities for very-high-energy collisions. The discovery potential of new massive states which constitute the most straightforward and cleanest way to detect physics beyond the SM is thus only going to be driven by increases in the total integrated luminosity. As a result, the lack of a first evidence of new effects in those 100 inverse femtobarns of luminosity collected so far guarantees that no new discovery is behind the corner. To this bleak picture one must add the notion that even in case an evidence of something new should appear in the latest analyzed data, consumers of LHC results should be advised to take a wait-and-see attitude rather than starting to produce tentative models that explain the effect in the context of beyond-the-standard-model concoctions. The hundreds of considered final states, the testing of dozens of differential distributions in wide ranges of the relevant variables, and the multiple iteration of the same testing procedures as more data is collected, all together in fact imply that occasional three-sigma, and even four- and five-sigma effects due to statistical fluctuations or imperfect assessment of systematic uncertainties are {\em bound} to arise sooner or later. A demonstration of this simple observation was given in December 2015 by the appearance of the infamous ``gamma-gamma'' resonance spotted at 750 GeV as a small bump in an otherwise smooth mass distribution of photon pairs by ATLAS and CMS. The resulting explosion of published theoretical models capable of incorporating the diphoton decay of a massive resonance without affecting the observed agreement of other observables with SM predictions is not necessarily nocuous; it has actually been argued to be beneficial for the field~\cite{giudiceprivcomm}; nonetheless, it might also be taken as an advice for the LHC experiments to be even more careful in the presentation of their results.

The detection of a diboson resonance is indeed the dream of experimental physicists working at the high-energy frontier, because of the unmistakable conclusions that it would enable on the existence of new interactions at the TeV scale. The excellent techniques developed by ATLAS and CMS to optimize their sensitivity to such signals have allowed the best use to be made of the collected data. Since the signatures to be considered to search for diboson-decaying resonances are not as complex and multiform as those {\em e.g.} required to investigate the parameter space of SUSY theories, they enable an economical investigation of multiple theoretical models. The recent approval of the future high-luminosity program of the Large Hadron Collider imply that its experiments studying the high-energy frontier will continue to search for resonant signals in their mass distributions for many years to come. In terms of increments in our knowledge of what lays beyond the SM the returns, however, run the concrete risk of being null. 

Our insufficient clues about possible new physics would demand a time-out to ponder on the situation, and yet the huge challenges posed by the construction of new high-energy colliders force us to move forward and plan for new accelerators that might see the light no earlier than in the 2040ies. Among the new machines proposed to extend our reach at the high-energy frontier, a few~\cite{fcc,helhc} promise to extend our investigation of hadron collisions to higher center-of-mass energies, and even to reach into the 100-TeV range; others suggest to return to electron-positron collisions to invest in the detailed measurement of Higgs boson couplings and do precision physics with Higgs bosons and top quark pairs~\cite{fccee,ilc4,ilc5,ilc6,ilc8}, or foresee a combination of the two ideas~\cite{fcc,fccee}. The even more ambitious project of producing and colliding muon beams in the TeV energy range~\cite{muoncollider1,muoncollider2} is also being studied, with the prospects of doing both precision Higgs physics (exploiting the large coupling of the Higgs to muon pairs) and new physics searches; but it is clear that such a machine would require an even longer gestation.

Whatever machine will inherit the LHC leadership at the high-energy frontier, it will highly benefit from the phenomenological and experimental studies and from the analysis techniques that have been developed there over the course of the past decade for the production of results such as those described in this review. We find of particular significance those belonging to five macro-areas of research.
The first is the development of new calculational tools that have allowed for more precise predictions of the phenomenology of parton collisions, in particular in complex final states with many hadronic jets. The baseline for most processes is now the next-to-leading order, with many processes computed and modeled to Next-to-Next-to-Leading Order (NNLO), and some even to three-loop accuracy~\cite{higgsnnnlo}; parton-level MC programs that include those calculations have greatly improved the understanding of experimental data and supported all BSM searches.
The second is constituted by statistical methods devised to extract inference on the presence of small localized signals from likelihood fits to smooth mass distributions, when account needs to be taken of a large number of nuisance parameters. The third area, which is undergoing an explosive expansion inside and outside HEP, is the one of machine-learning methods which lend themselves excellently to the challenging classification and regression use cases that are common in particle physics research. The fourth is represented by particle-flow-based reconstruction techniques of high-level objects, which has proven crucial for the precise measurement of hadronic jets in high-pile-up circumstances, as well as for the high-efficiency identification of difficult signatures such as hadronically-decaying $\tau$ leptons. The fifth area includes all procedures that allow the tagging of heavy objects within wide hadronic jets of very high energy, by pruning calorimeter clusters of their soft contributions and extracting information on their origin from their substructure. Guidance on the design requirements of the next generation of particle detectors will come in particular from the latter two macro-areas above, as they call for both very high granularity and redundant detection of energy deposits in the calorimeters, as well as a precise tracking in very strong magnetic fields capable of ``fanning out'' the soft components of hadronic jets.


\appendix

\section*{Acronyms}

To ease the task of correctly interpreting the many acronyms used in this review, whose definition is otherwise only given in the first occurrence of the term, we provide below a conversion table.

\begin{table}[h!]
\begin{center}
\caption{\em Definition of acronyms used in this document.}
\begin{tabular}{l|l|l}
Acronym & Meaning & Section \\
\hline
LHC & Large Hadron Collider &~\ref{s:lhc} \\
SM & Standard Model &~\ref{s:sm} \\
SUSY & Supersymmetry &~\ref{s:susy} \\
LH & Little Higgs &~\ref{s:lh}\\
2HDM & Two-Higgs Doublet Model &~\ref{s:2hdm} \\
GUT & Grand-Unification Theories &~\ref{s:gut}\\ 
GLR & Generalized Left-Right symmetric models &~\ref{s:glr}\\
GSM & Generalized Sequential Model &~\ref{s:gsm}\\
SSM & Sequential Standard Model &~\ref{s:ssm}\\
VBF & Vector-Boson Fusion &~\ref{s:vbf} \\
HVT & Heavy Vector Triplet &~\ref{s:hvt} \\
EGM & Extended Gauge Model &~\ref{s:egm}\\
RS & Randall-Sundrum &~\ref{s:rs} \\
FCNC & Flavour-Changing Neutral Current &~\ref{s:fcnc} \\
KK & Kaluza-Klein &~\ref{s:kk} \\
MSSM & Minimal Supersymmetric Standard Model &~\ref{s:mssm}\\
hMSSM & habemus MSSM &~\ref{s:hmssm}\\
NMSSM & Next-to-Minimal Supersymmetric Standard Model &~\ref{s:nmssm}\\
QCD & Quantum Chromo-Dynamics &~\ref{s:qcd}\\
LSTC & Low-Scale TechniColour &~\ref{s:lstc}\\
PDF & Parton Distribution Functions &~\ref{s:pdf}\\
ATLAS & A large ToroidaL ApparatuS &~\ref{s:atlas} \\
CMS & Compact Muon Solenoid &~\ref{s:cms} \\
ECAL & Electromagnetic Calorimeter &~\ref{s:ecal}\\
HCAL & Hadronic CALorimeter &~\ref{s:hcal}\\
HLT & High-Level Trigger &~\ref{s:hlt}\\
EM & ElectroMagnetic &~\ref{s:em}\\
PF & Particle Flow &~\ref{s:pf} \\
CA & Cambridge-Aachen &~\ref{s:ca} \\
CSV & Combined Secondary Vertex &~\ref{s:csv}\\
CMVA & Combined Multi-Variate Algorithm &~\ref{s:cmva} \\
MVA & Multi-Variate Algorithm &~\ref{s:mva}\\
NN & Neural Network &~\ref{s:nn} \\
BDT & Boosted Decision Tree &~\ref{s:bdt}\\
DNN & Deep Neural Network &~\ref{s:dnn} \\
HEP & High-Energy Physics &~\ref{s:hep} \\
LEE & Look-Elsewhere Effect &~\ref{s:lee} \\
CL & Confidence Level &~\ref{s:cl} \\
BSM & Beyond the Standard Model &~\ref{s:bsm} \\
CPS & Complex-Pole System &~\ref{s:cps}\\
HTM & Higgs Triplet Model &~\ref{s:htm} \\
MWT & Minimal Walking Technicolour &~\ref{s:mwt}\\
\hline
\end{tabular}
\end{center}
\label{t:conversion}
\end{table}

\clearpage

\end{document}